\documentstyle[epsfig,mathbbol]{lamuphys}
\makeatletter
\input{chicago-sp.sty}
\let\chapter\hid@chapter
\makeatother
\newcommand{\halfwidth}{0.495\textwidth}

\newcommand{\fullwidth}{\textwidth}

\newcommand{\be}{\begin{equation}}
\newcommand{\ee}{\end{equation}}
\newcommand{\bea}{\begin{eqnarray}}
\newcommand{\eea}{\end{eqnarray}}\newcommand{\bml}{\begin{mathletters}
\baselineskip 10pt}
\newcommand{\eml}{\baselineskip 12pt \end{mathletters}}
\newcommand{\nn}{\nonumber}

\newcommand{\m}{{\scriptscriptstyle -}}
\newcommand{\p}{{\scriptscriptstyle +}}

\newcommand{\intl}{\int\limits_{-L}^L}

\newcommand{\bra}{\langle}
\newcommand{\ket}{\rangle}
\newcommand{\cond}{\bra 0 | \bar \psi \psi | 0 \ket}

\newcommand{\simleq}{\scriptstyle{\stackrel{<}{\sim}}}
\newcommand{\simgeq}{\scriptstyle{\stackrel{>}{\sim}}}
\newcommand{\lca}{\phi_{\mathrm{LC}}}
\newcommand{\bsa}{\chi_{\mathrm{BS}}}

\newcommand{\sfrac}[2]{{\textstyle \frac{#1}{#2}}}
\newcommand{\pad}[2]{\frac{\partial #1}{\partial #2}}
\newcommand{\fud}[2]{\frac{\delta #1}{\delta #2}}

\newcommand{\vc}[1]{\mbox{\bf #1}}
\newcommand{\svc}[1]{\mbox{\footnotesize\bf #1}}
\newcommand{\vcg}[1]{\mbox{\boldmath$#1$}}
\newcommand{\svcg}[1]{\mbox{\footnotesize\boldmath$#1$}}

\newcommand{\pb}[2]{\left\{#1 \, , \, #2 \right\}}

\newcommand{\sgn}{{\mbox{sgn}}}
\newcommand{\tr}{\mbox{tr}}
\newcommand{\FP}{\mbox{FP}}

\begin{document}
\pagenumbering{arabic}
\title{Light--Cone Quantization: Foundations and Applications}

\author{Thomas Heinzl}

\institute{Friedrich--Schiller--Universit\"at Jena,
Theoretisch--Physikalisches Institut, Max--Wien--Platz 1, D-07743 Jena}

\maketitle

\begin{abstract}
  These lecture notes review the foundations and some applications of
  light--cone quantization. First I explain how to choose a time in
  special relativity. Inclusion of Poincar\'e invariance naturally
  leads to Dirac's forms of relativistic dynamics. Among these, the
  front form, being the basis for light--cone quantization, is my main
  focus. I explain a few of its peculiar features such as boost and
  Galilei invariance or separation of relative and center--of--mass
  motion. Combining light--cone dynamics and field quantization
  results in light--cone quantum field theory. As the latter
  represents a first--order system, quantization is somewhat
  nonstandard. I address this issue using Schwinger's quantum action
  principle, the method of Faddeev and Jackiw, and the functional
  Schr\"odinger picture. A finite--volume formulation, discretized
  light--cone quantization, is analysed in detail. I point out some
  problems with causality, which are absent in infinite volume.
  Finally, the triviality of the light--cone vacuum is established.
  Coming to applications, I introduce the notion of light--cone wave
  functions as the solutions of the light--cone Schr\"odinger
  equation. I discuss some examples, among them nonrelativistic
  Coulomb systems and model field theories in two dimensions. Vacuum
  properties (like chiral condensates) are reconstructed from the
  particle spectrum obtained by solving the light--cone Schr\"odinger
  equation. In a last application, I make contact with phenomenology
  by calculating the pion wave function within the Nambu and
  Jona-Lasinio model. I am thus able to predict a number of
  observables like the pion charge and core radius, the
  r.m.s.~transverse momentum, the pion structure function and the pion
  distribution amplitude. The latter turns out to be the asymptotic
  one.
\end{abstract}

\section{Introduction}

The nature of elementary particles calls for a synthesis of relativity
and quantum mechanics. The necessity of a quantum treatment is quite
evident in view of the microscopic scales involved which are several
orders of magnitude smaller than in atomic physics. These very scales,
however, also require a relativistic formulation. A typical hadronic
scale of 1 fm, for instance, corresponds to momenta of the order of $p
\sim \hbar c / 1 \mathrm{fm} \, \simeq 200$ MeV. For particles with masses $M$
$\simleq$ 1 GeV, this implies sizable velocities $v \simeq p/M \,
\simgeq$ 0.2 $c$.

It turns out that the task of unifying the principles of quantum
mechanics and relativity is not a straightforward one. One can neither
simply extend ordinary quantum mechanics to include relativistic
physics nor quantize relativistic mechanics using the ordinary
correspondence rules. Nevertheless, Dirac and others have succeeded in
formulating what is called ``relativistic quantum mechanics'', which
has become a subject of text books since --- see e.g.~
\cite{bjorken:64,yndurain:96}. It should, however, be pointed out that
this formulation, which is based on the concept of single-particle
wave-functions and equations, is not really consistent.  It does not
correctly account for relativistic causality (retardation effects
etc.)  and the existence of antiparticles. As a result, one has to
struggle with issues like the Klein paradox\footnote{For a nice recent
discussion, see \cite{holstein:98}.}, the definition of position
operators \cite{newton:49} and the like.

The well--known solution to these problems is provided by quantum {\em
field} theory, with an inherently correct description of
antiparticles that entails relativistic causality. In contrast to
single-particle wave mechanics, quantum field theory is a
(relativistic) many body formulation that necessarily involves
(anti-)particle creation and annihilation. Physical particle states
are typically a superposition of an infinite number of `bare' states,
as any particle has a finite probability to emit or absorb other
particles at any moment of time.  A pion, for example, would be
represented in terms of the following Fock expansion,
\be
  | \pi \ket = \psi_2 | q \bar q \ket + \psi_3 | q \bar q g \ket +
  \psi_4 | q \bar q q \bar q \ket + \ldots \; ,
\ee
where the $\psi_n$ are the probability amplitudes to find $n$
particles (quarks $q$, antiquarks $\bar q$ or gluons $g$) in the pion.
With the advent of QCD, however, a conceptual difficulty concerning
this many--particle picture has appeared. At low energy or momentum
transfer, hadrons, the bound states of QCD, are reasonably described
in terms of two or three constituent quarks and thus as {\em
few--body} systems. These `effective' quarks $Q$ are dressed
so that they gain an effective mass of the order of 300--400 MeV.
They are used as the basic degrees of freedom in the `constituent
quark model'. This model yields a reasonable mass spectroscopy of
hadrons \cite{lucha:91,mukherjee:93}, but its foundations are
not very well established theoretically. First, a nonrelativistic treatment
of light hadrons is not justified (see above). Second, the model
violates many symmetries of QCD (in particular \textit{chiral
symmetry}). Third, it is rather unclear how a constituent picture
can arise in a quantum field theory such as QCD.

In principle, in order to confirm the constituent quark model, one
would have to solve the `QCD Schr\"odinger equation' for hadron states
$| \mbox{hadron} \ket$ of mass $M_{\mathrm{h}}$,
\be
  H_{\mathrm{QCD}} |\mbox{hadron} \ket = M_{\mathrm{h}} |\mbox{hadron}
  \ket  \; ,
\ee
and check whether the eigenstates are reasonably well described in
terms of the constituent valence states $| Q\bar Q \ket$ or $| QQQ
\ket$. This is a very hard problem. A more moderate goal
would be to `relativize' the constituent quark model, ideally in such
a way that it respects the symmetries of QCD. I will discuss this
attempt in detail at the end of these lectures.

To arrive at this point, there is, of course, some way to go. Let me
start with the following claim. A particularly useful approach for our
purposes is based on a somewhat unorthodox choice of the `time arrow'
within special relativity: instead of the ordinary `Galileian time'
$t$, I choose `light--cone time', $x^\p \equiv t - z/c$. In the course
of these lectures\footnote{Naturally, these lectures cannot cover all
aspects of light--cone field theory. For a comprehensive recent review
on the subject see \cite{brodsky:98}.}, this claim will be
substantiated step by step.

I will begin with some general remarks on relativistic dynamics
(Section~2). As a paradigm example I discuss the free relativistic
particle which is the prototype of a reparametrization invariant
system.  I show that the choice of the time parameter is not unique as
it corresponds to a gauge fixing, the purpose of which is to get rid
of the reparametrization redundancies. By considering the stability
subgroups of the Poincar\'e group, one finds that there are
essentially three reasonable choices of `time' for a relativistic
system, corresponding to Dirac's `instant', `point' and `front' form,
respectively.  The latter choice is the basis of light--cone dynamics,
the main features of which will be discussed in the last part of
Section~2.

Section~3 is devoted to light--cone \emph{field} quantization. I show
how the Poincar\'e generators are defined in this case and utilize
Schwinger's quantum action principle to derive the canonical
commutators. This is the first method of quantization to be discussed.
The relation between equal--time commutators, the field equations and
their solutions for different initial and/or boundary conditions is
clarified. 

It turns out that light--cone field theories, being of first order in
the velocity, generally are constrained systems which require a
special treatment. I rederive the canonical light--cone commutators
using a second method of quantization (based on phase space reduction)
due to Faddeev and Jackiw. I extend this discussion to light--cone
quantization in finite volume and point out possible problems with
causality in this approach.  Going back to infinite volume, I
introduce a third method of quantization, the functional Schr\"odinger
picture, and combine it with the light--cone formalism.  I close this
section with a discussion of the presumably most spectacular feature
of light--cone quantum field theory, the triviality of the vacuum.

As a prelude to the applications I introduce the notion of light--cone
wave functions in Section~4. I show how light--cone wave functions
can be obtained by solving the light--cone Schr\"odinger equation. As
examples, I discuss nonrelativistic wave functions as they occur in
hydrogen--like systems, some model field theory in 1+1 dimensions and a
simple Gaussian model.
 
In Section~5, I finally make contact with phenomenology. I calculate
the light--cone wave function of the pion within the Nambu and
Jona-Lasinio model. This model is known to provide a good description
of spontaneous chiral symmetry breaking, as it is governed by the same
symmetry group as low--energy QCD. With the pion wave function at hand
I derive a number of observables like the pion charge and core
radius, the electromagnetic form factor, the r.m.s.~transverse
momentum and the pion structure function. I conclude with a
calculation of the pion distribution amplitude. 

\vfill
\eject

\section{Relativistic Particle Dynamics}

The physically motivated desire to describe hadrons as bound states of a
small, fixed number of  constituents is our rationale to go back
and reanalyze the relation between Hamiltonian quantum mechanics and
relativistic quantum field theory.

Quite generally, bound states are obtained by solving the Schr\"odinger
equation,
\begin{equation}
  i \hbar \pad{}{\tau} |\psi(\tau) \ket = H | \psi(\tau) \ket \; , 
\end{equation}
for {\em normalized, stationary states},  
\begin{equation}
  |\psi(\tau) \ket = e^{-i E \tau} | \psi(0) \ket .  
\end{equation}
This leads to the bound--state equation
\begin{equation}
  H |\psi(0) \ket = E | \psi(0) \ket \; ,
\end{equation}
where $E$ is the bound state energy. We would like to make this
Hamiltonian formalism consistent with the requirements of relativity.
It is, however, obvious from the outset that this procedure will not
be manifestly covariant as it singles out a time $\tau$ (and an energy
$E$, respectively).  Furthermore, it is not even clear what the time
$\tau$ really is as it does not have an invariant meaning.

\subsection{The Free Relativistic Point Particle}

To see what is involved it is sufficient to consider the classical
dynamics of a free relativistic particle. We want to find the associated
canonical formulation as a basis for subsequent quantization. We will
proceed by analogy with the treatment of classical free strings which is
described in a number of textbooks \cite{green:87,polyakov:87}.
Accordingly, the relativistic point particle may be viewed as an
infinitely short string.  

Recall that the action for a relativistic particle is essentially
given by the arc length of its trajectory
\begin{equation}
  \label{REL_ACT1}
  S = - m s_{12} \equiv - m \int_1^2 ds  \; .
\end{equation}
This action\footnote{We work in natural units, $\hbar = c = 1$.} is a
Lorentz scalar as
\begin{equation}
  ds = \sqrt{g_{\mu \nu} x^\mu x^\nu} \; 
\end{equation}
is the (infinitesimal) invariant distance. We can rewrite the action
(\ref{REL_ACT1}) as 
\begin{equation}
  \label{REL_ACT2}
  S = -m \int_1^2 ds \, \sqrt{\dot x_\mu \dot x^\mu} \equiv 
  \int_1^2 ds \, L(s) \; ,
\end{equation}
in order to introduce a Lagrangian $L(s)$ and the four velocity $\dot
x^\mu \equiv dx^\mu/ds $. The latter obeys 
\begin{equation}
  \label{V2}
  \dot x^2 \equiv \dot x_\mu \dot x^\mu = 1 \; , 
\end{equation}
as the arc length provides a natural parametrization. Thus, $\dot x^\mu$
is a time-like vector, and we assume in addition that it points into the
future, $\dot x^0 > 0$. In this way we guarantee relativistic causality
ensuring that a real particle passing through a point $P$ will always
propagate into the future light cone based at $P$.

We proceed with the canonical formalism by calculating the canonical
momenta as
\begin{equation}
  \label{CAN_MOM1}
  p^\mu = - \pad{L}{\dot x_\mu} = m \dot x^\mu \; .
\end{equation}
These are not independent, as can be seen by calculating the square
using (\ref{V2}),
\begin{equation}
  p^2 = m^2 \dot x^2 = m^2 \; , 
\end{equation}
which, of course, is the usual mass--shell constraint. This constraint
indicates that the Lagrangian $L(s)$ defined in (\ref{REL_ACT2}) is {\em
  singular}, so that its Hessian $W^{\mu \nu}$ with respect to
the velocities,
\begin{equation}
  W^{\mu \nu} \equiv \frac{\partial^2 L}{\partial \dot x_\mu \partial \dot
  x_\nu} = - \frac{m}{\sqrt{\dot x^2}} \left( g^{\mu \nu} - \frac{\dot
  x^\mu \dot x^\nu}{\dot x^2} \right) =  - m \left( g^{\mu \nu} - \dot
  x^\mu \dot x^\nu \right)\; ,  
\end{equation}
is degenerate. It has a zero mode given by the velocity itself,
\begin{equation}
  \label{SING_LAG}
  W^{\mu \nu} \dot x_\nu = 0 \; . 
\end{equation}
The Lagrangian being singular implies that the velocities cannot  be
uniquely expressed in terms of the canonical momenta. This, however, is
not obvious from (\ref{CAN_MOM1}), as we can easily solve for the
velocities, 
\begin{equation}
  \dot x^\mu = p^\mu /m \; .
\end{equation}
But if one now calculates the canonical Hamiltonian,
\begin{equation}
  \label{VAN_HAM}
  H_c = - p_\mu \dot x^\mu - L = -m \dot x^2 + m \dot x^2 = 0 \; ,
\end{equation}
one finds that it is vanishing! It therefore seems that we do not have a
generator for the  time evolution of our dynamical system. In the following,
we will analyze the reasons for this peculiar finding.

First of all we note that the Lagrangian is homogeneous of first degree
in the velocity,    
\begin{equation}
  L (\alpha \dot x^\mu) = \alpha L(\dot x^\mu) \; .
\end{equation}
Thus, under a  reparametrization of the world-line,
\begin{equation}
  \label{REPAR} 
  s \mapsto s^\prime \; , \quad x^\mu (s) \mapsto x^\mu \big(s^\prime (s)\big) \; ,
\end{equation}
where the mapping $s \mapsto s^\prime$ is one--to--one with $ds^\prime /
ds > 0$ (orientation conserving), the Lagrangian changes according to
\begin{equation}
  L(dx^\mu/ds) = L\Big( (dx^\mu/ds^\prime) (ds^\prime/ds) \Big) = (ds^\prime/ds)
  L(dx^\mu/ds^\prime)\; .
\end{equation}
This is sufficient to guarantee  that the action is invariant under
(\ref{REPAR}), that is, {\em reparametrization invariant}, 
\begin{equation}
  S =  \int_{s_1}^{s_2}ds \, L(dx^\mu/ds) = 
  \int_{s_1^\prime}^{s_2^\prime}ds^\prime \, \frac{ds}{ds^\prime}\frac{ds^\prime}{ds}
  L(dx^\mu/ds^\prime) \equiv S^\prime \; , 
\end{equation}
if the endpoints remain unchanged, $s_{1,2} = s_{1,2}^\prime$. On the
other hand, $L$ is homogeneous of the first degree if and only if
Euler's formula holds, namely
\begin{equation}
  \label{EULER}
  L = \pad{L}{\dot x^\mu} \dot x^\mu = - p_\mu \dot x^\mu \; .
\end{equation}
This is exactly the statement (\ref{VAN_HAM}), the vanishing of the
Hamiltonian. Furthermore, if we differentiate (\ref{EULER}) with respect
to $\dot x^\mu$, we recover (\ref{SING_LAG}) expressing the singular
nature of the Lagrangian. Summarizing, we have found the general result
\cite{hanson:76,sundermeyer:82} that if a Lagrangian is homogeneous of
degree one in the velocities, the action is reparametrization invariant,
and the Hamiltonian vanishes. In this case, the momenta are homogeneous
of degree zero, which renders the Lagrangian singular.

The reparametrization invariance is generated by the first class constraint
\cite{dirac:64,sundermeyer:82}, 
\begin{equation}
\label{SR_CONSTR}
  \theta \equiv p^2 - m^2 = 0 \; , 
\end{equation}
as can be seen as follows. From the canonical one form $- g_{\mu \nu}
p^\mu dx^\nu$ we read off the Poisson bracket
\begin{equation}
\label{SR_PB}
  \pb{x^\mu}{p^\nu} = - g^{\mu\nu} \; , 
\end{equation}
and calculate the change of the coordinate $x^\mu$,
\bea
  \delta x^\mu &=& \pb{x^\mu}{\theta \delta \epsilon} = - 2 p^\mu \delta
  \epsilon = - 2m \, \dot x^\mu \delta \epsilon \equiv \dot x^\mu \delta
  \tau \nn \\
  &=& x^\mu (\tau + \delta \tau) - x^\mu (\tau) = x^\mu (\tau^\prime)
  - x^\mu (\tau) \; . 
\eea
Thus, the reparametrization (\ref{REPAR}) is indeed generated by the
constraint (\ref{SR_CONSTR}). 

Reparametrization invariance can be viewed as a gauge or
\textit{redundancy} symmetry. The redundancy consists in the fact that a
single trajectory (world--line) can be described by an infinite number
of different parametrizations. The physical objects, the trajectories,
are therefore equivalence classes obtained by identifying (`dividing
out') all reparametrizations. The method to do so is well known,
namely gauge fixing. For the case at hand, this corresponds to
a particular choice of parametrization, or, more physically, to the
choice of a time parameter $\tau$. This amounts to choosing a
foliation of space--time into space \textit{and} time.  Minkowski
space is thus decomposed into hypersurfaces of equal time, $\tau =
const$, which in general are three-dimensional objects, and the time
direction `orthogonal' to them. The time development thus continuously
evolves the hypersurface $\Sigma_0: \tau = \tau_0$ into $\Sigma_1:
\tau = \tau_1 > \tau_0$. Put differently, initial conditions provided
on $\Sigma_0$ together with the dynamical equations (being
differential equations in $\tau$) determine the state of the dynamical
system on $\Sigma_1$.

Practically, the (3+1)--foliation is done as follows. We introduce
some arbitrary coordinates, $\xi^\alpha = \xi^\alpha (x)$, which may
be curvilinear. We imagine that three of these, say $\xi^i$,
$i=1,2,3$, parametrize the three--dimensional hypersurface $\Sigma$,
so that the remaining one, $\xi^0$, represents the time variable,
i.e.~$\tau = \xi^0 (x)$. This equation can equivalently be viewed as a
gauge fixing condition.

The first question to be addressed is: what is a `good' choice of
time? There are two criteria to be met, namely existence and
uniqueness. Existence means that the equal--time hypersurface $\Sigma$
should intersect \emph{any} possible world--line, while uniqueness
requires that it does so once and only once.  Mathematically,
uniqueness can be analysed in terms of the Faddeev--Popov (FP)
`operator', which is given by the Poisson bracket of the gauge fixing
condition with the constraint (evaluated on $\Sigma$),
\be
\label{FP}
  \FP \equiv \pb{\xi^0 (x)}{\theta} = \pb{\xi^0 (x)}{p^2} = - 2
  \, \pad{\xi^0}{x^\mu} \, p^\mu \equiv -2 \, N \cdot p \; .
\ee
Here, we have introduced the normal $N$ on $\Sigma$,
\be
  N^\mu (x) = \left. \pad{\xi^0 (x)}{x^\mu} \right|_{\Sigma} \; ,
\ee
which will be important later on. The statement now is that uniqueness
is achieved (for a single degree of freedom) if the FP operator does
not vanish, i.e. if $N \cdot p \ne 0$. Generically, this means that
the particle trajectory must not be parallel to the hypersurface
$\Sigma$ of equal time. 

As an aside we remark that this is completely analogous to the
reasoning in standard gauge (field) theory. There, the constraint
$\theta$ is given by Gauss's law which generates gauge transformations
$A \to A + D \omega$, $D$ denoting the covariant derivative. For a
gauge fixing $\chi[A] = 0$, the equation corresponding to (\ref{FP})
becomes
\be
  \FP  = \pb{\chi}{\theta} = \fud{\chi}{\omega} = \fud{\chi}{A}
  \fud{A}{\omega} = N \cdot D \; ,
\ee
where all (functional) derivatives are to be evaluated on the gauge
fixing hypersurface, $\chi = 0$. 

Let us now perform an analysis of the canonical formalism for a
general choice of hypersurface $\Sigma$. For this we need some
notation. We write the line element as
\begin{equation}
  \label{H_METRIC}
  ds^2 = g_{\mu\nu}dx^\mu dx^\nu =
  g_{\mu\nu}\pad{x^\mu}{\xi^\alpha} \pad{x^\nu}{\xi^\beta} d\xi^\alpha
  d\xi^\beta \equiv h_{\alpha\beta}(\xi) d\xi^\alpha d\xi^\beta \; . 
\end{equation}
Introducing a vierbein $e^\mu_{\;\alpha} (\xi)$, the metric
$h_{\alpha\beta}(\xi)$ is alternatively given by
\begin{equation}
  h_{\alpha\beta} (\xi) = g_{\mu\nu} \, e^\mu_{\;\; \alpha} (\xi) \, 
  e^\nu_{\;\; \beta} (\xi) \; .
\end{equation}
The transformation $x \to \xi$ is well known from general relativity,
where it corresponds to the transformation from a local inertial frame
described by the flat metric $g_{\mu\nu}$ to a noninertial frame with
coordinate dependent metric $h_{\alpha\beta}(\xi)$. For our purposes
we write this metric in a (3+1)-notation as follows,
\begin{equation}
  \label{(3+1)_METRIC}
  h_{\alpha\beta} = \left( \begin{array}{cc}
                            h_{00} & h_{0i} \\
                            h_{i0} & h_{ij} 
                      \end{array}        \right) \equiv
                    \left( \begin{array}{cr}
                            h_{00}  & \vcg{h}^T \\
                            \vcg{h} & -H 
                      \end{array}        \right)    \; .
\end{equation}
Of particular interest is the component $h_{00}$, which  explicitly
reads 
\begin{equation}
  h_{00} = g_{\mu\nu} \pad{x^\mu}{\xi^0} \pad{x^\nu}{\xi^0} = g_{\mu\nu}
  e^\mu_{\;\; 0} e^\nu_{\;\; 0} \equiv n^2 \; ,
\end{equation}
where we have defined the unit vector in $\xi^0$-direction
\begin{equation}
  n^\mu = \pad{x^\mu}{\xi^0} =  e^\mu_{\;\; 0} \equiv \dot x^\mu \; ,
\end{equation}
which thus is the new four--velocity. It is related to the normal
vector $N^\mu$ via
\begin{equation}
  n \cdot N = e^\mu_{\;\; 0} e^{\;\; 0}_\mu = \pad{\xi^0}{x^\mu} \pad
  {x^\mu}{\xi^0} = 1 \; .
\end{equation}
The normal vector $N$ enters the inverse metric which we write as
follows, 
\be
   \label{INV_METRIC}
   h^{\alpha\beta} = \left( \begin{array}{cc}
                            g_{00} & g_{0i} \\
                            g_{i0} & g_{ij} 
                      \end{array}        \right) = 
                      \left( \begin{array}{cr}
                            N^2     & \vcg{g}^T \\
                            \vcg{g} &  -G 
                      \end{array}        \right)\; .
\ee
The $h_{ij}$ are the metric components associated with the
hypersurface. The invariant distance element (\ref{H_METRIC}) thus
becomes (with $h_{0i} \equiv h_i$),
\begin{eqnarray}
  ds^2 &=& h_{00} d\xi^0 d\xi^0 + 2 h_{0i} d\xi^0 d\xi^i + h_{ij} d\xi^i
  d\xi^j \; , \nn \\
  &=& \left( n^2 + 2 h_i \frac{d\xi^i}{d\tau} +
  h_{ij}\frac{d\xi^i}{d\tau} \frac{d\xi^j}{d\tau} \right) d\tau^2
  \equiv h(\tau) d\tau^2 \; , 
\end{eqnarray}
where, in the second step, we have used that $\xi^0 = \tau$. In the
last identity we have defined a world--line metric or einbein
\be
\label{WL_METRIC}
  h (\tau) \equiv \dot x^2 = h_{\alpha \beta} \dot \xi^\alpha \dot
  \xi^\beta \; ,
\ee
which expresses the arbitrariness in choosing a time by providing an
(arbitrary) `scale' for the velocity.  Introducing the velocities
expressed in the new coordinates, $w^i \equiv d\xi^i/d\tau$, the
world--line metric can be written as
\begin{equation}
  h(\tau) = n^2 + 2 h_i w^i + h_{ij} w^i w^j \; .
\end{equation}
Let us develop a canonical formalism for a general choice of the
einbein $h(\tau)$ corresponding to the gauge fixing $\tau = \xi^0 (x) =
0$. The Lagrangian becomes
\be
  L(\tau) = - m \sqrt{h(\tau)} \; , 
\ee
leading to the canonical momenta
\be
  \pi_\alpha = - \pad{L}{\dot \xi^\alpha} = \frac{m}{\sqrt{h}}
  \, h_{\alpha\beta} \, \dot \xi^\alpha = e_\alpha^\mu \, p_\mu \; .
\ee
We see that the einbein $h$ is appearing all over the place. The
canonical Hamiltonian is expressed in terms of the inverse metric
$h^{\alpha\beta}$,
\be
  H_c = - \pi_\alpha \dot \xi^\alpha - L = - \frac{\sqrt{h}}{m}
  (h^{\alpha\beta} \pi_\alpha \pi_\beta - m^2) = 0 \; .
\ee
It vanishes (as it should) as it is proportional to the constraint,
\be
\label{H_CONSTRAINT}
  \theta = h^{\alpha\beta} \pi_\alpha \pi_\beta - m^2= p^2 - m^2 = 0 \; .
\ee
The FP operator also depends on the entries of the inverse metric
(\ref{INV_METRIC}), in particular the normal vector $N$,
\be
  \FP = N^2 \pi_0 + g^i \pi_i \; .
\ee
After gauge fixing, the generator of $\tau$--evolution, $H_\tau \equiv
\pi_0$, is obtained by solving the constraint (\ref{H_CONSTRAINT}) for
$\pi_0$ which assumes the explicit form,
\be
\label{QUADRATIC}
  N^2 \pi_0^2 + 2 g^i \pi_i \pi_0 - G^{ij} \pi_i \pi_j - m^2 = 0 \; .
\ee
Depending on the value of $N^2$, we thus have to consider two
different cases. The generic one is that the normal $N$ on $\Sigma$ is
\textit{time--like}, $N^2 > 0$. In this case, the mass--shell
constraint is of \textit{second order} in $\pi_0$, so that there
are two distinct solutions,
\be
\label{2ND_ORDER}
  \pi_0 = \frac{1}{N^2} \left\{ - (\vcg{g}, \vcg{\pi}) \pm
  \sqrt{(\vcg{g}, \vcg{\pi})^2 + N^2 [(\vcg{\pi}, G \vcg{\pi}) + m^2]}
  \right\} \; .
\ee  
Not unexpectedly, the `problem' of two different signs in front of the
square root arises \cite{gitman:90}. Within quantum mechanics, this is
somewhat difficult to interpret. Upon `second quantization', i.e.~in
the context of quantum field theory, one has, of course, the natural
explanation in terms of antiparticles. As we will not quantize the
relativistic point particle, the sign `problem' is of no concern to
us. A possible arbitrariness will be removed ad hoc by demanding
$\pi_0 > 0$. With this additional condition the FP operator becomes
\be
  \FP = - 2 \sqrt{(\vcg{g}, \vcg{\pi})^2 + N^2 [(\vcg{\pi}, G \vcg{\pi})
  + m^2]} \; ,
\ee
which is clearly nonvanishing for a massive particle, $m \ne 0$. A
gauge fixing with $N^2 > 0$ is thus unique and leads to a
well--defined description of the $\tau$--evolution. 

The second case to be considered is in a sense degenerate. It
corresponds to a \textit{light--like} normal, $N^2 = 0$. In this case,
the constraint (\ref{QUADRATIC}) is only of 
\textit{first order} in $\pi_0$ leading to a single solution,
\be
\label{1ST_ORDER}  
  \pi_0 = \frac{(\vcg{\pi}, G \vcg{\pi}) + m^2}{(\vcg{g}, \vcg{\pi})}
  \; .
\ee
As a result, there is no `sign problem' and no `ugly' square
root. Conservation of difficulties, however, is at work, because it
is no longer obvious whether the FP operator,
\be
  \FP = - 2 (\vcg{g}, \vcg{\pi})  \; ,
\ee
is different from zero. Clearly, this is absolutely necessary for
(\ref{1ST_ORDER}) to represent a well--defined solution.

At this point, it should be mentioned that the results
(\ref{2ND_ORDER}) and (\ref{1ST_ORDER}) are not yet the full story.
The entries of the inverse metric, $N^2$, $\vcg{g}$ and $G$ should
actually be expressed in terms of the quantities $n^2$, $\vcg{h}$ and
$H$ defining the induced metric on $\Sigma$. So far, it is also not
completely clear which choices of these parameters actually make sense
physically. Of course, the normal $N$ should not be space--like as
this would imply that $\Sigma$ contains time--like directions and thus
possible particle trajectories. In the next subsection I will give
some criteria for reasonable choices of time.

Before we come to that let us apply the general formalism to the
standard choice of `Galileian' time, $\tau = \xi^0 (x) = x^0 = t$. In
this case, the surface $\Sigma: t = 0$ is an entirely space--like
hyperplane with constant normal vector $N = (1, \vc{0}) = n$. The
other metric entries are $\vcg{h} = \vcg{g} = 0$ and $H = G =
\Eins$. The world--line metric (\ref{WL_METRIC}) thus becomes
\be
  h(t) = \dot x^2 = 1 - v^2 \equiv 1/\gamma^2 \; , 
\ee
where $\gamma$ is the usual Lorentz contraction factor. The
Hamiltonian is obtained in line with the second--order case above,
\be
\label{H_T}
  H_t = N \cdot p = p^0 = \sqrt{\vc{p}^2 + m^2} \sim \FP \; .
\ee
It generates the dynamics via the basic Poisson bracket $\pb{x^i}{p^j}
= \delta^{ij}$ leading to
\be
\label{IF_DYNAMICS}
  \dot x^i = \pb{x^i}{H_t} = p^i / p^0 \; , 
\ee
with $p^0$ given by (\ref{H_T}). Note that a well--defined time
evolution requires a nonvanishing FP operator (which is proportional to
$p^0$). 

As already announced, we will discuss alternatives to this standard
choice of time in the next subsection.

\subsection{Dirac's Forms of Relativistic Dynamics}

To address this issue it is not sufficient to consider only the
$\tau$--development and the associated generator of time translations
(i.e.~the Hamiltonian) $H_\tau$. Instead, one has to refer to the full
Poincar\'e group to be able to guarantee full relativistic invariance.
The generators of the Poincar\'e group are the four momenta $P^\mu$ and
the six operators $M^{\mu \nu}$ which combine  angular momenta and
boosts according to
\begin{eqnarray}
  L^i &=& \frac{1}{2}\epsilon^{ijk}M^{jk} \; , \label{LK1} \\
  K^i &=& M^{0i} \; , \label{LK2}
\end{eqnarray}
with $i$, $j$, $k$ = 1,2,3. These generators are elements of the
Poincar\'e algebra which is defined by the Poisson bracket relations,
\begin{eqnarray}
  \label{PC_ALGEBRA}
  \pb{P^\mu}{P^\nu}  &=& 0 \; , \nn \\
  \pb{M^{\mu \nu}}{P^{\rho}}  &=& g^{\nu \rho}P^\mu - g^{\mu \rho}
  P^\nu  \; , \\
  \pb{ M^{\mu \nu}}{M^{\rho \sigma}}  &=& g^{\mu \sigma} M^{\nu \rho} -
  g^{\mu \rho} M^{\nu \sigma} - g^{\nu \sigma} M^{\mu \rho} + g^{\nu
  \rho} M^{\mu \sigma} \; . \nn
\end{eqnarray}
It is well known that the momenta $P^\mu$ generate space--time
translations and the $M^{\mu \nu}$ rotations and Lorentz boosts,
cf.~(\ref{LK1}, \ref{LK2}). In the following we will only consider
proper and orthochronous Lorentz transformations, i.e.~we exclude
space and time reflections.

Any Poincar\'e invariant dynamical theory describing e.g.~the
interaction of particles should provide a particular realization of the
Poincar\'e algebra. For this purpose, the Poincar\'e generators are
constructed out of the fundamental dynamical variables like positions,
momenta, spins etc. An elementary realization of (\ref{PC_ALGEBRA}) is
given as follows. Choose the space-time point $x^\mu$ and its conjugate
momentum $p^\mu$ as canonical variables, i.e.~adopt (\ref{SR_PB}),
\begin{equation}
  \label{PB_COV2}
  \pb{x^\mu}{p^\nu}  = -  g^{\mu \nu} \; .
\end{equation}
The Poincar\'e generators are then found to be
\begin{equation}
  \label{ELEM_PC_GEN}
  P^\mu = p^\mu \; , \quad M^{\mu \nu} = x^\mu p^\nu - x^\nu p^\mu \; , 
\end{equation}
as is easily confirmed by checking (\ref{PC_ALGEBRA}) using
(\ref{PB_COV2}).  An infinitesimal Poincar\'e transformation is thus
generated by
\begin{equation}
  \delta G = - \sfrac{1}{2}\delta\omega_{\mu\nu} M^{\mu\nu} 
  + \delta a_\mu P^\mu \; 
\end{equation}
in the following way,
\begin{equation}
  \delta x^\mu = \pb{x^\mu}{\delta G} = \delta \omega^{\mu\nu} x_\nu +
  \delta a^\mu \; , \quad \delta\omega^{\mu\nu} = - \delta\omega^{\nu\mu} \; .  
\end{equation}
The action of the Poincar\'e group on some scalar function $F(x)$ is
thus given by
\begin{equation}
  \label{PC_ACTION}
  \delta F = \pb{F}{\delta G} = \partial^\mu F \, \delta a_\mu -
  \sfrac{1}{2}  (x^\mu
  \partial^\nu - x^\nu \partial^\mu)F \, \delta \omega_{\mu\nu} \; .
\end{equation}
Though the realization (\ref{ELEM_PC_GEN}) is covariant, it has
several shortcomings. It does not describe any interaction; for
several particles the generators are simply the sum of the single
particle generators. This point, however, is of minor importance to
us, and will only be touched upon at the end of this subsection. The
solution of the problem, as already mentioned in the introduction, is
the framework of local quantum field theory. More importantly, the
representation (\ref{ELEM_PC_GEN}) does not take into account the
mass--shell constraint, $p^2 = m^2$, which we already know to guarantee
relativistic causality as it generates the dynamics upon solving for
$H_\tau$.

To remedy the situation we proceed as before by choosing a time
variable $\tau$, i.e.~a foliation of space--time into essentially
space--like hypersurfaces $\Sigma$ with time--like or light--like
normals $N$.  We have seen that $\Sigma$ should be chosen in such a
way that it intersects all possible world--lines once and only once
(existence and uniqueness).  Apart from this necessary consistency
with causality the foliation appears quite arbitrary. However, given
a particular foliation one can ask the question which of the
Poincar\'e generators will leave the hypersurface $\Sigma$
invariant. The set of all such generators defines a subgroup of the
Poincar\'e group called the stability group $G_\Sigma$ of
$\Sigma$. The associated generators are called {\em kinematical}, the
others {\em dynamical}. The latter map $\Sigma$ onto another
hypersurface $\Sigma^\prime$ and thus involve the development in
$\tau$. One thus expects that the dynamical generators will depend on
the Hamiltonian (and, therefore, the interaction) which, by
definition, is a dynamical quantity.

It is clear, however, that the stability group corresponding to a
particular foliation will be empty if the associated hypersurface looks
very irregular and thus does not have a high degree of symmetry. One
therefore demands in addition that the stability group acts transitively
on $\Sigma$: any two points on $\Sigma$ can be connected by a
transformation from $G_\Sigma$. With this additional requirement there 
are exactly five inequivalent classes of hypersurfaces
\cite{leutwyler:78} which are listed in Table \ref{T4}.

\begin{table}
\begin{center}
\renewcommand{\arraystretch}{1.3}
\caption{\label{T4} \textsl{All possible choices of hypersurfaces $\Sigma$:
  $\tau = const$ with transitive action of the  stability group
  $G_\Sigma$. $d$ denotes the dimension of $G_\Sigma$, that is, the
  number of kinematical Poincar\'e generators; $\vc{x}_\perp \equiv (x^1,
  x^2)$.}}
\vspace{.5cm}

\begin{tabular*}{\textwidth}[h]{ @{\extracolsep\fill} l  l  l  c }
\hline\hline
name        & $\Sigma$        & $\tau$          & $d$ \\
\hline
instant     & $x^0 = 0$       & $t$             & 6 \\
light front & $x^0 + x^3 = 0$ & $t + x^3/c$     & 7 \\ 
hyperboloid & $x_0^2 - \vc{x}^2 = a^2 > 0$, $x^0 > 0$ & $(t^2 -
              \vc{x}^2/c^2  - a^2/c^2)^{1/2}$  & 6 \\
hyperboloid & $x_0^2 - \vc{x}_\perp^2 = a^2 > 0$, $x^0 > 0$ & $(t^2 -
              \vc{x}_\perp^2 /c^2 -  a^2/c^2)^{1/2}$ & 4 \\
hyperboloid & $x_0^2 - x_1^2  = a^2 > 0$, $x^0 > 0$ & $(t^2 -
              x_1^2 /c^2 -  a^2/c^2)^{1/2}$ & 4 \\
\hline\hline
\end{tabular*}
\end{center}
\end{table}

The first three choices have already been found by \cite{dirac:49} in
his seminal paper on `forms of relativistic dynamics'.  He called the
associated forms the `instant', `front' and `point' forms,
respectively. These are the most important choices as the remaining
two forms have a rather small stability group and thus are not very
useful.  We have only listed them for the sake of completeness.

It is important to note that for all forms one has $\lim_{c \to \infty}
\tau = t$, which means that in the nonrelativistic case there is only
one possible foliation leading to the absolute Galileian time
$t$. This is consistent with the fact that there is no limiting
velocity in this case implying that particle trajectories can have
arbitrary slope (tangent vector). Therefore, the hypersurface
$\Sigma_{nr}: t = const$ is the only one intersecting all possible
world-lines. For other choices, the criterion of existence introduced
in the last subsection would be violated.

To decide which of the Poincar\'e generators are kinematical, we use
the general formula (\ref{PC_ACTION}) describing their action. Imagine
that $\Sigma$ is given in the form $\Sigma: \tau = \xi^0 (x) \equiv
F(x)$ as in Table~\ref{T4}. If $P^\mu$ or $M^{\mu\nu}$ are kinematical
for some $\mu$ or $\nu$, then, for these particular superscripts, the
components of the gradient and rotor of $F$ have to vanish on
$\Sigma$,
\begin{equation}
  \label{KINEMATIC1}
  \partial^\mu F = 0 \; , \quad (x^\mu \partial^\nu - x^\nu
  \partial^\mu)F = 0 \; .
\end{equation}
In terms of the normal vector $N$ these equations become
\be
\label{KINEMATIC2} 
  N^\mu = 0 \; , \quad  x^\mu N^\nu - x^\nu N^\mu = 0 \; , 
\ee
which again will hold for \emph{some} of the superscripts $\mu$ and/or
$\nu$, if $\Sigma$ has nontrivial stabilizer. The distinction between
kinematical and dynamical is thus completely encoded in the normal
vector $N$. 

The choice of Galileian time $\tau = t$ is of course the most common
one also in the relativistic case, and we have discussed it briefly at
the end of the last subsection. To complete this discussion, we
construct the associated representation of the Poincar\'e generators
on $\Sigma: t =0$.  The idea is again to explicitly saturate the
constraint $p^2 = m^2$ by solving for $H_t = p^0 = N \cdot p =
(\vc{p}^2 + m^2)^{1/2}$, and setting $x^0 = 0$ in (\ref{ELEM_PC_GEN}).

As a result, we obtain the following (3+1)--representation of the
Poincar\'e generators,
\begin{equation}
  \label{PC_ALG_IF}
  \begin{array}{l c l l c l}
    P^i & = & p^i \; ,     & M^{ij} & = & x^i p^j - x^j p^i \; ,   \\
    P^0 & = & H_t \; , & M^{i0} & = & x^i H_t 
    \; . 
  \end{array} 
\end{equation}
This outcome is as expected: Compared to (\ref{ELEM_PC_GEN}), $p^0$ has
been replaced by $H_t$, and $x^0$ has been set to zero. It should,
however, be pointed out that for non--Cartesian coordinates the
construction of the Poincar\'e generators is less straightforward. 

Let us address the question of kinematical versus dynamical
generators.  In agreement with (\ref{KINEMATIC1}) and
(\ref{KINEMATIC2}), one has
\begin{equation}
  N^i  = 0 = x^i N^j - x^j N^i  \; , \quad i,j = 1,2,3 \; , 
\end{equation}
so that $\Sigma$ is both translationally and rotationally invariant
confirming that the dimension of its stability group is six
(cf.~Table~\ref{T4}). On the other hand,
\begin{eqnarray}
  N^0  = 1 &\ne& 0 \; , \\
  x^0 N^i - x^i N^0  = - x^i &\ne& 0 \; ,
\end{eqnarray}
from which we read off that, apart from the Hamiltonian, also the boosts
are dynamical, i.e., $\Sigma$ is not boost invariant. The latter fact
is, of course, well known because the boosts  mix space and time. Under a
boost along the $\vc{n}$-direction with velocity $\vc{v}$, $t$ transforms as
\begin{equation}
  \label{BOOST}
  t \to t^\prime = t \cosh \omega + (\vc{n} \cdot \vc{x}) \sinh \omega \; ,  
\end{equation}
where $\vc{n} = \vc{v}/v$ and $\omega$ is the rapidity, defined
through $\tanh \omega = v$. From (\ref{BOOST}) it is evident that the
hypersurface $\Sigma: t=0$ is not boost invariant.

In obtaining the representation (\ref{PC_ALG_IF}), we make the
Poincar\'e algebra compatible with the instant--form gauge--fixing
constraint, $x^0 = 0$. An elementary calculation, using $\pb{x^i}{p^j}
= \delta^{ij}$, indeed shows that the generators (\ref{PC_ALG_IF})
really obey the bracket relations (\ref{PC_ALGEBRA}). We have already
seen in (\ref{IF_DYNAMICS}) that the Hamiltonian $P^0 = H_t$
generates the correct dynamics.

At this point it is getting time to really consider an alternative to
the instant form in some detail.

\subsection{The Front Form}

\begin{figure}
   \caption{\label{fig-front} \textsl{The hypersurface $\Sigma : x^\p =
        0$ defining the front form. It is a null-plane tangential to the
        light--cone, $x^2 =0$.}}
   \begin{center} 
     \vspace{1cm}
     \epsfig{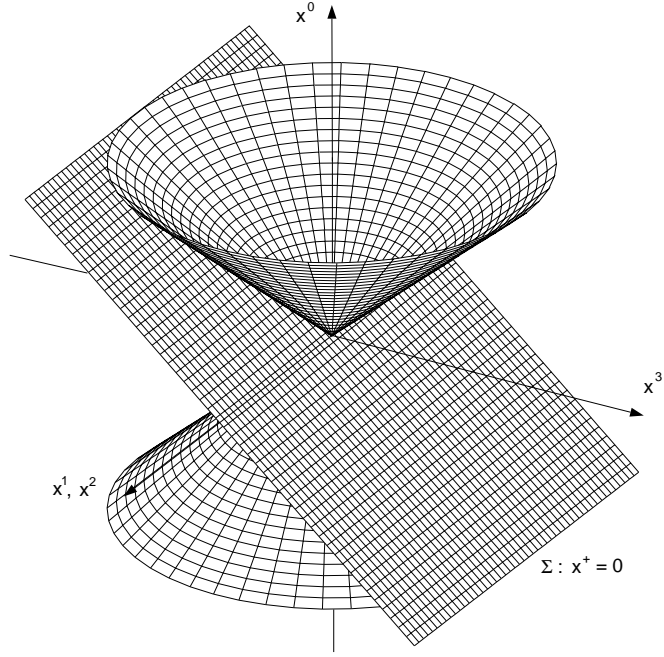}
   \end{center}    
\end{figure}

For an arbitrary four-vector $a$ we perform the following transformation
to {\em light--cone} coordinates,
\begin{equation}
  (a^0, a^1, a^2 , a^3 ) \mapsto (a^\p , a^1 , a^2 , a^\m) \; ,
\end{equation}
where we have defined
\begin{equation}
  a^\p = a^0 + a^3 \; , \quad a^\m = a^0 - a^3 \; .
\end{equation}
We also introduce the transverse vector part of $a$ as
\begin{equation}
  \vc{a}_\perp = (a^1 , a^2) \; .
\end{equation}
The metric tensor (\ref{(3+1)_METRIC}) becomes
\begin{equation}
\label{LC_METRIC}
  h_{\alpha\beta} = \left( \begin{array}{cr}
                            n^2     & \vcg{h}^T \\
                            \vcg{h} & -H 
                      \end{array}        \right) 
                  = \left( \begin{array}{crrc}
                         0   &  0 &  0 & 1/2 \\ 
                         0   & -1 &  0 &  0  \\
                         0   &  0 & -1 &  0  \\
                        1/2  &  0 &  0 &  0
                      \end{array} \right) 
\end{equation}
The entries 1/2 imply nonvanishing $\vcg{h}$ and thus a slightly
unusual scalar product, 
\begin{equation}
  \label{SCAL_PROD} 
  a \cdot b = g_{\mu\nu}a^\mu b^\nu =
  \sfrac{1}{2}a^\p b^\m + \sfrac{1}{2} a^\m b^\p - a^i b^i \; , \quad
  i = 1,2 \; .
\end{equation}
According to Table \ref{T4}, the front form is defined by choosing the
hypersurface $\Sigma: x^\p = 0$, which is a plane tangent to the
light--cone. It can equivalently be viewed as the wave front of a plane
light wave traveling towards the positive $z$--direction. Therefore,
$\Sigma$ is also called a \emph{light--front}. The normal vector is
\begin{equation}
  \label{N_FRONT}
  N = (1, 0,0, -1) \; , \quad N^2 = 0 \; ,  
\end{equation}
where $N$ has been written in ordinary coordinates. We see that $N^\p =
N^0 + N^3 = 0$ which implies that the normal $N$ to the hypersurface
lies {\em within} the hypersurface \cite{neville:71a,rohrlich:71}. As
$N$ is a light--like or null vector, $\Sigma$ is often referred to as a
\emph{null--plane} \cite{neville:71a,coester:92}. We have depicted the
front--form hypersurface $\Sigma$ together with the light--cone in
Fig.~\ref{fig-front}.

As is already obvious from (\ref{LC_METRIC}), the unit vector in
$x^\p$--direction is another null--vector, 
\begin{equation}
  n^\mu = \pad{x^\mu}{x^\p} = \sfrac{1}{2} (1 , 0 , 0, 1) \; ,
\end{equation}
so that $n \cdot N = 1$ as it should.  Given the scalar product
(\ref{SCAL_PROD}), we infer the invariant distance element
\begin{equation}
  ds^2 = g_{\mu\nu}dx^\mu dx^\nu = dx^\p dx^\m - dx^i dx^i
  = \left( \frac{dx^\m}{dx^\p} - \frac{dx^i}{dx^\p} \frac{dx^i}{dx^\p} 
  \right) dx^\p dx^\p
\end{equation}
from which the einbein $h$ can be read off as
\begin{equation}
  h(x^\p) = \dot x^\m - \dot x^i \dot x^i  \equiv
  v^\m - \vc{v}_\perp^2 \; .
\end{equation}
Note that velocities are dimensionless, so that despite appearance
the result is consistent (if you do not like it as it stands, just
insert the appropriate factors of $c$).

The Hamiltonian is obtained by solving the constraint $p^\p p^\m -
p_\perp^2 - m^2 = 0$, which is now \emph{linear} in $p^\m$. The result
is
\begin{equation}
  \label{FF_HAM}
  H_{x^\p} = n \cdot p = p^\m / 2 = \frac{p_\perp^2 + m^2}{2p^\p} \; .
\end{equation}
Let me reemphasize that this Hamiltonian does not contain a square
root as already pointed out by Dirac. However, now it is crucial that
the FP operator is nonvanishing,
\be
  \FP = - 2 N \cdot p = -2 p^\p \ne 0 \; .
\ee
While this is always true for  massive particles, it is violated for
massless `left--movers', i.e. for particles travelling in the negative
$z$--direction at the speed of light. In this case, we have a `Gribov
problem' \cite{gribov:78}, as the particles move \emph{within} our
gauge--fixing hyperplane, $x^\p = 0$. We will return to this issue
later on. 

For massive particles, the dynamics is consistently generated by means
of the Poisson brackets
\begin{equation}
  v^\m = \dot x^\m = \pb{x^\m}{H_{x^\p}} = \frac{p^\m}{p^\p} \; , \quad
  v^i = \dot x^i = \pb{x^i}{H_{x^\p}} = \frac{p^i}{p^\p} \; .
\end{equation}
Note finally, that the Hamiltonian (\ref{FF_HAM}) is \emph{not} the
normal projection $N \cdot p$ of the momentum, because $N \cdot p$ lies
within $\Sigma$ and thus corresponds to a kinematical direction.  

As for the instant form, the light--cone representation of the
Poincar\'e generators can be obtained by solving the constraint $p^2 =
m^2$ for $p^-$, inserting the result into the elementary
representation (\ref{ELEM_PC_GEN}) of the generators and setting $x^\p
= 0$.  The kinematical generators are
\bea
  P^i &=& p^i \; , \quad P^\p = p^\p \; , \nn \\
  M^{\p i} &=& - x^i p^\p \; ,
  \quad M^{12} = x^1 p^2 - x^2 p^1 \; , \quad M^{\p \m} = - x^\m p^\p
  \; . \label{FF_KINEMATIC} 
\eea
They correspond to transverse and longitudinal translations within
$\Sigma$ ($P^i$ and $P^\p$, respectively), transverse boosts and
rotations ($ M^{\p i}$), rotations around the $z$--axis ($ M^{12}$)
and boosts (!) in the $z$--direction ($ M^{\p \m}$). The latter will
be further analysed in a moment. We thus have found \emph{seven}
kinematical generators, so that the front form leads to the largest
stability group among Dirac's forms of dynamics (cf.~Table~1). 

The dynamical generators  are given by
\begin{equation}
  P^\m = \frac{p_\perp^2 + m^2}{p^\p} \; , \quad M^{\m i} = x^\m
  p^i - x^i p^\m \; .
\end{equation}
As expected, the $M^{\m i}$ depend on the Hamiltonian, $p^\m$.  If we
now consider rotations around the $x$-- or $y$--axis, generated by
\bea
  L^1 &=& M^{23} = \sfrac{1}{2} (M^{2\p} - M^{2\m}) \; , \\
  L^2 &=& M^{31} = \sfrac{1}{2} (M^{\p 1} - M^{\m 1}) \; , 
\eea
we note that they correspond to \emph{dynamical} operations due to the
appearance of $M^{\m i}$. This leads to the notorious `problem of
angular momentum' within the front form, see
e.g.~\cite{fuda:90}. Except for the free theory, it is very hard to
write down states with good angular momentum as diagonalizing
$\vc{L}^2$ is as difficult as solving the Schr\"odinger equation. A
similar problem arises for parity. This exchanges light--cone space
and time and thus also becomes dynamical \cite{burkardt:96c}. For the
kinematical component of the angular momentum, $L_z = M^{12}$, these
difficulties do not arise.

Consider now the following boost in $z$--direction with rapidity
$\omega$ written in instant--form coordinates,
\begin{eqnarray}
  t &\to&  t \cosh \omega + z \sinh \omega \; , \\
  z &\to&  t \sinh \omega + z \cosh \omega \; .
\end{eqnarray}
As stated before, such a boost mixes space and time coordinates $z$
and $t$. If we add and subtract these equations, we obtain the action
of the boost for the front form,
\begin{eqnarray}
  x^\p &\to& e^\omega x^\p \; , \label{XP_BOOST}\\
  x^\m &\to& e^{-\omega} x^\m \; . \label{XM_BOOST}
\end{eqnarray}
We thus find the important result that a boost in $z$--direction does not
mix light--cone space and time but rather rescales the coordinates! Note
that $x^\p$ and $x^\m$ are rescaled inversely with respect to each
other. The scaling factor can be written as
\begin{equation}
  e^\omega = \sqrt{\frac{1-v}{1+v}} \; ,
\end{equation}
if the rapidity $\omega$ is defined in the usual manner in terms of
the boost velocity $v$, $\tanh \omega = v$. One should note in
particular, that one has the fixed point hypersurface $\Sigma: x^\p =
0$ which is mapped onto itself, so that the relevant generator,
$M^{\p\m}= 2 M^{30}= - 2 K^3$, is kinematical, in agreeement with
(\ref{FF_KINEMATIC}). However, we see explicitly that this is no
longer true for $x^\p \ne 0$, where we get a rescaling of
$x^\p$. Stated differently, the transformation to light--cone
coordinates diagonalizes the boosts in $z$--direction. Therefore, the
behavior under such boosts becomes especially simple. A pedagogical
discussion and some elementary applications can be found in
\cite{parker:70a}.

We are actually more interested in the transformation properties of the
momenta, as these, being Poincar\'e generators, are more fundamental
quantities than the coordinates, in particular in the quantum theory
\cite{leutwyler:78}. As $P^\mu$ transforms as a four--vector we just have
to replace $x^\mu$ by $P^\mu$ in the boost transformations
(\ref{XP_BOOST}, \ref{XM_BOOST}) and obtain,
\begin{eqnarray}
  P^\p &\to& e^\omega P^\p \; , \label{P_BOOST} \\
  P^\m &\to& e^{-\omega} P^\m \; , \label{M_BOOST} \; .
\end{eqnarray}
We remark that $P^\p = 0$ is a fixed point under longitudinal boosts. In
quantum field theory, it corresponds to the vacuum. For the transverse
momentum, $P^i$, one finds a transformation law reminiscent of a
Galilei boost,
\be
\label{PERP_BOOST}
  P^i \to P^i + v^i P^\p \; .
\ee
In this identity, describing the action of $M^{\p i}$, longitudinal
and transverse momenta (which are both kinematical) get mixed.

We can now ask the question how to boost from $(P^\p , P^i)$ to momenta
$(Q^\p , Q^i)$. This can be done by fixing the boost parameters $\omega$
and $v^i$ as
\begin{equation}
  \omega = - \log \frac{Q^\p}{P^\p} \; , \quad v^i = \frac{Q^i -
  P^i}{P^\p} \; .
\end{equation}
Obviously, this is only possible for $P^\p \ne 0$. We emphasize that in
the construction above there is no dynamics involved. For the quantum
theory, this means that we can build states of arbitrary light--cone
momenta with very little effort. All we have to do is applying some
kinematical boost operators. The simple behavior of light--cone momenta
under boosts will be important for the discussion of bound states in
Section~4.  

The similarity between (\ref{PERP_BOOST}) and Galilei boosts is not
accidental. This is exhibited by the following subalgebra of the
light--cone Poincar\'e algebra \cite{susskind:68,bardakci:68}. Consider 
the Poisson bracket relations of the seven generators $P^\mu$, $M^{12}$, 
$M^{\p i}$,

\parbox{10cm}{
\begin{eqnarray*}
  \pb{M^{12}}{M^{\p i}} &=& \epsilon^{ij} M^{\p j} \; , \\
  \pb{M^{12}}{P^i}      &=& \epsilon^{ij} P^j \; , \\
  \pb{M^{\p i}}{P^\m}   &=& - 2 P^i \; , \\
  \pb{M^{\p i}}{P^j}    &=& - \delta^{ij} P^\p
\end{eqnarray*}}
\hfill
\parbox{1.5cm}{\begin{eqnarray}\label{GALILEI_SUB}\end{eqnarray}}

All other brackets of these generators vanish. Compare now with the
two-dimensio\-nal Galilei group. Its generators (for a free particle of
mass $\mu$) are: two momenta $k^i$, one angular momentum $L =
\epsilon^{ij}x^i k^j$, two Galilei boosts $G^i = \mu x^i$, the Hamiltonian
$H = k^i k^i /2\mu$ and the mass $\mu$, which is the Casimir
generator. Upon using $\pb{x^i}{k^j} = \delta^{ij}$ and identifying
$P^i \leftrightarrow k^i$, $M^{12} \leftrightarrow L$, $M^{\p i}
\leftrightarrow -2 G^i$, $P^\p \leftrightarrow 2\mu$ and $P^\m
\leftrightarrow H$, one easily finds that (\ref{GALILEI_SUB}) forms a
subalgebra of the Poincar\'e algebra which is isomorphic to the Lie
algebra of the two-dimensional Galilei group. (A second isomorphic
subalgebra is obtained via identifying $M^{\m i} \leftrightarrow 2
G^i$ and exchanging $P^\p$ with $P^\m$.) The first two identities in
(\ref{GALILEI_SUB}), for instance, state that $M^{\p i}$ and $P^i$
transform as ordinary two-dimensional vectors. $P^\p$ can be
interpreted as a variable Galilei mass which is also obvious from the
nonrelativistic appearance of the light--cone Hamiltonian, $P^\m  =
(P_\perp^2 + m^2)/P^\p$ and the Galilei boost (\ref{PERP_BOOST}). 

One thus expects that light--cone kinematics will partly show a
nonrelativistic behavior which is associated with the transverse
dimensions and governed by the two--dimensional Galilei group. This
expectation is indeed realized and leads, for instance, to a
separation of center--of--mass and relative dynamics in multi-particle
systems. This will be discussed at length in the beginning of
Section~4.

So far, our discussion of the Poincar\'e algebra was restricted to the
free case. With the inclusion of interactions, one expects all
dynamical Poincar\'e generators to differ from their free counterpart
by some `potential' term $W$. This has already been pointed out by
\cite{dirac:49}, who also stated that finding potentials which are
consistent with the commutation relations of the Poincar\'e algebra is
the ``real difficulty in the construction of a theory of a
relativistic dynamical system'' with a fixed number of particles.

It has turned out, however, that Poincar\'e invariance alone is not
sufficient to guarantee a reasonable Hamiltonian formulation. There are
no--go theorems both for the instant \cite{leutwyler:65} and the front
form \cite{jaen:84}, which state that the inclusion of any potential
into the Poincar\'e generators, even if consistent with the commutation
relations, spoils relativistic {\em covariance}. The latter is a
stronger requirement as it enforces particular transformation laws for
the particle coordinates. Thus, covariance imposes rather
severe restrictions on the dynamical system \cite{leutwyler:78}.

The  physical reason for these problems is  that potentials imply an
instantaneous interaction--at--a--distance which is in conflict with the
existence of a limiting velocity and retardation effects. Relativistic
causality is thus violated. This is equivalently obvious from the fact
that a \emph{fixed} number of particles is in conflict with the necessity of
particle creation and annihilation and the appearance of
antiparticles. 

Nevertheless, with considerable effort, it is possible to construct
dynamical quantum systems with a fixed number of constituents which are
consistent with the requirements of Poincar\'e invariance and
relativistic covariance \cite{leutwyler:78,sokolov:79,coester:82}.

At this point one might finally ask whether the different forms of
relativistic dynamics are physically equivalent. From the point of
view that different time choices correspond to different gauge fixings
it is clear that equivalence must hold.  After all, we are just
dealing with different coordinate systems. People have tried to make
this equivalence more explicit by working with coordinates which
smoothly interpolate between the instant and the front form
\cite{prokhvatilov:88,prokhvatilov:89,lenz:91,hornbostel:92}. In the
context of relativistic quantum mechanics, it has been shown that the
Poincar\'e generators for different forms are unitarily equivalent
\cite{sokolov:79}.

We are, however, more interested in what might be called a `top--down
approach'. Our aim is to describe few--body systems not within quantum
mechanics but quantum field theory to which we now turn.

\vfill
\eject

\section{Light--Cone Quantization of Fields}

\subsection{Construction of the Poincar\'e Generators}

We want to derive the representation of the Poincar\'e generators within
field theory and their dependence on the hypersurface $\Sigma$ chosen to
define the time evolution. To this end we follow \cite{fubini:73} and
describe the hypersurface mathematically through the equation
\begin{equation}
  \Sigma:  F(x) = \tau  \; .
\end{equation}
The surface element on $\Sigma$ is implicitly defined via
\begin{equation}
  \int_\Sigma d\sigma _\mu u(x) = \int d^4 x \, N^\mu  
  \delta(F(x) - \tau) u(x) \; ,
\end{equation}
where, as before, $N^\mu = \partial^\mu F(x)$ is the normal on
$\Sigma$ and $u$ some integrable function. We will write this
expression symbolically as
\begin{equation}
  \label{SURF_EL} 
  d\sigma _\mu = d^4 x \, N^\mu  \delta(F(x) - \tau) \; .
\end{equation}
The central object of this subsection will be the energy-momentum tensor, 
\begin{equation}
\label{ENMOM_TENSOR}
  T^{\mu \nu} = \pad{\mathcal{L}}{(\partial_\mu \phi)} \partial^\nu \phi -
  g^{\mu \nu} \mathcal{L} \; ,
\end{equation}
with $\mathcal{L}$ being the Lagrangian depending on fields that are
collectively denoted by $\phi$. With the help of the energy-momentum
tensor (\ref{ENMOM_TENSOR}) we can define a generator
\begin{equation}
  \label{GENA}
  A[f] = \int_\Sigma d \sigma_\mu f_\nu (x) T^{\mu \nu}(x) \; ,
\end{equation}
where $A$ and $f$ can be tensorial quantities carrying dummy indices
$\alpha$, $\beta, \ldots$ which we have suppressed.  $A[f]$ generates the
infinitesimal transformations
\begin{eqnarray}
  \delta_f B(x) &=& f_\mu (x) \, \partial^\mu B (x) \; , \\
  \delta_f x^\mu &=& f^\mu (x) \; ,
\end{eqnarray}
where $f$ is now understood as being infinitesimal. In the same way as
for a finite number of degrees of freedom, the generator $A$ is called
kinematical, if it leaves $\Sigma$ invariant, that is,
\begin{equation}
  \label{AKIN}
  \delta_f F = f_\mu \partial^\mu F = f \cdot N = 0 \; .
\end{equation}
Otherwise, $A$ is dynamical. With the energy-momentum tensor
$T^{\mu\nu}$ at hand, we can easily show that kinematical generators
are interaction independent. We decompose $T^{\mu \nu}$,
\begin{equation}
  T^{\mu \nu} = T_0^{\mu \nu} - g^{\mu \nu} \mathcal{L}_{\mathrm{int}}
  \; ,
\end{equation}
into a free part $T_0^{\mu \nu} = T^{\mu \nu}(g= 0)$, $g$ denoting the
coupling, and an interacting part (we exclude the case of derivative
coupling). If $A$ is kinematical, we have from (\ref{SURF_EL},
\ref{GENA}, \ref{AKIN}),
\begin{equation}
  A_{\mathrm{int}}[f] = - \int d^4 x \, \delta(F -
  \tau) {\cal L}_{\mathrm{int}} f^\mu \partial_\mu F
    = - \int d^4 x \, \delta(F -\tau) {\cal L}_{\mathrm{int}}
  \delta_f F  = 0 \; ,
\end{equation}
which indeed shows that $A$ does not depend on the
interaction. Dynamical operators, on the other hand, \emph{will}
contain interaction dependent pieces. Of course, we are particularly
interested in the Poincar\'e generators, $P^\alpha$ and $M^{\alpha
\beta}$. They correspond to the choices $f_\mu^\alpha = g_\mu^\alpha$
and $f_\mu^{\alpha \beta} = x^\alpha g_\mu^\beta - x^\beta
g_\mu^\alpha$, respectively, so that, from (\ref{GENA}), they are
given in terms of $T^{\mu\nu}$ as
\begin{eqnarray}
  P^\alpha &=& \int_\Sigma d\sigma_\mu T^{\mu\alpha} \; , \label{QFT_POINC_P}\\
  M^{\alpha\beta} &=& \int_\Sigma d\sigma_\mu (x^\alpha T^{\mu\beta} -
  x^\beta T^{\mu\alpha}) \; .\label{QFT_POINC_M}
\end{eqnarray}
From (\ref{AKIN}) it is easily seen that the Poincar\'e generators
defined in (\ref{QFT_POINC_P}, \ref{QFT_POINC_M}) act on $F(x) = \tau$
exactly as described in (\ref{PC_ACTION}). The remarks of Section~2 on
the kinematical or dynamical nature of the generators in the different
forms are therefore equally valid in quantum field theory. 

Let us first discuss the instant form.  We recall the hypersurface of
equal time, $\Sigma: F(x) \equiv N \cdot x \equiv x^0 = \tau$, which
leads to a surface element
\begin{equation}
  d \sigma^\mu = d^4 x \, N^\mu \delta(x^0 - \tau) \; , \quad N^\mu =
  (1, \vc{0}) \; . 
\end{equation}
Using (\ref{QFT_POINC_P}, \ref{QFT_POINC_M}), the Poincar\'e
generators are obtained as
\begin{eqnarray}
  P^\mu &=&  \int_\Sigma d^3 x \, T^{0 \mu} \; ,   \\
  M^{\mu\nu} &=& \int_\Sigma d^3 x \, \big(x^\mu T^{0 \nu} - x^\nu
  T^{0 \mu}\big) \; .
\end{eqnarray}
For the front form, quantization hypersurface and surface element are
given by
\begin{equation}
  \Sigma: F(x) \equiv N \cdot x \equiv x^\p = \tau \; , \quad
  d\sigma^\mu = d^4 x \, N^\mu \, \delta(x^\p - \tau)  \; ,
\end{equation}
where $N$ is the light--like four-vector of (\ref{N_FRONT}).  In terms
of $T^{\mu\nu}$, the Poincar\'e generators are
\begin{eqnarray}
  P^\mu &=& \sfrac{1}{2} \int_\Sigma dx^\m d^2 x_\perp \, T^{\p \mu} \; ,
  \label{FF_P} \\
  M^{\mu\nu} &=& \sfrac{1}{2} \int_\Sigma dx^\m d^2 x_\perp \, 
  \big(x^\mu T^{\p \nu} - x^\nu  T^{\p \mu}\big) \; . \label{FF_M}
\end{eqnarray}
The somewhat peculiar factor 1/2 is the Jacobian which arises upon
transforming to light--cone coordinates.

\subsection{Schwinger's (Quantum) Action Principle}

Our next task is to actually quantize the fields on the hypersurfaces
$\Sigma: \tau = F(x)$ of equal time $\tau$. There is more than one
possibility to do so, and we will explain a few of these. We begin
with a method that is essentially due to Schwinger
\cite{schwinger:51,schwinger:53a,schwinger:53b}. We define a
four-momentum density
\begin{equation}
  \Pi^\mu = \pad{\mathcal{L}}{(\partial_\mu \phi)} \; ,
\end{equation}
so that the energy-momentum tensor $T^{\mu\nu}$ can be written as 
\begin{equation}
  T^{\mu\nu} = \Pi^\mu \partial^\nu \phi - g^{\mu\nu} \mathcal{L} \; .
\end{equation}
In some sense, this can be viewed as a covariant generalization of the
usual Legendre transformation between Hamiltonian and Lagrangian. Using
the normal $N^\mu$ of the hypersurface $\Sigma$, we define the canonical
momentum (density) as the projection of $\Pi^\mu$,
\begin{equation}
  \pi \equiv N \cdot \Pi \; .
\end{equation}
Schwinger's action principle states that, upon variation, the action $S =
\int d^d x \, \mathcal{L}$ changes at most by a surface term which (if
$\Sigma$ is not varied, i.e.~$\delta x^\mu = 0$) is given by
\begin{equation}
\label{DELTA_G}
  \delta G (\tau) = \int_\Sigma d\sigma_\mu \, \Pi^\mu \, \delta\phi =
  \int d^d x \, \delta   (\tau - F) \, \pi \,  \delta \phi \; . 
\end{equation}
The quantity $\delta G$ is interpreted as the generator of field
transformations, so that we have
\begin{equation}
  \label{DELTA_G_SL}
  \delta \phi = \pb{\phi}{\delta G} \; ,
\end{equation}
in case that $\Sigma$ is entirely space--like (with time--like normal)
\cite{schwinger:51,schwinger:53a}.  We note in passing that the
generator $\delta G$ is a field theoretic generalization of the
canonical one-form $dG \equiv p_i dq^i$ used in analytical mechanics.

As in the preceding section we have to distinguish two cases depending
on whether the normal vector $N$ is time--like or space--like. For
time--like $N$, the associated hypersurface is space--like. The basic
example for this case is the instant form, to which we immediately
specialize. The canonical momentum density is given by the velocity,
$\pi = \dot \phi$, and the Lagrangian is \emph{quadratic} in $\dot
\phi$. The canonical Poisson bracket is derived from Schwinger's
action principle using (\ref{DELTA_G_SL}),
\bea
  \delta \phi (x) &=& \pb{\phi(x)}{\delta G (\tau)} \nn \\
  &=& \int dy^0 \int d^3 y \, \delta (x^0 - y^0) \pb{\phi(x)}{\pi(y)}
  \delta \phi(y) \nn \\
  &=& \int d^3 y \, \left. \pb{\phi(x)}{\phi(y)} \delta
  \phi(y) \right|_{x^0 = y^0 = \tau} \; .
\eea
The canonical Poisson bracket, therefore, must be 
\be
  \pb{\phi(x)}{\phi(y)}_{x,y \in \Sigma} = \delta^3 (\vc{x} - \vc{y})
  \; ,
\ee
which, of course, is the standard result. The second case, $N$
light--like, corresponds to the front form. With minor modifications,
Schwinger's approach can also be used here, resulting in what is
interchangably called light--cone, light-front or null--plane
quantization. The canonical light--cone momentum is
\begin{equation}
  \pi = N \cdot \Pi = N \cdot \partial \phi = \partial^\p \phi \equiv 2
  \pad{}{x^\m} \phi \; ,
\end{equation}
which is peculiar to the extent that it does not involve a
(light--cone) time derivative. Therefore, $\pi$ is a dependent
quantity which does not provide additional information, being known on
$\Sigma$ when the field is known there. Thus, $\pi$ is merely an
abbreviation for $\partial^\p \phi$ which is a \emph{spatial}
derivative. Again, the reason is that the normal $N^\mu$ of the
null--plane $\Sigma$ lies within $\Sigma$. This leads to the important
consequence that the light--cone Lagrangian is \emph{linear} in the
velocity $\partial^\m \phi$, or, put differently, that light--cone
field theories are \emph{first--order systems}. As a result, $\phi$
and $\partial^\p \phi$ have to be treated on the same footing within
Schwinger's approach which leads to an additional factor 1/2 compared
to (\ref{DELTA_G_SL}),
\begin{equation}
  \label{DELTA_PHI_FF}
  \sfrac{1}{2} \delta \phi = \pb{\phi}{\delta G} \; ,
\end{equation}
with a front-form generator
\begin{equation}
  \label{DELTA_G_FF}
  \delta G (x^\p) = \sfrac{1}{2} \int_\Sigma dx^\m d^2 x_\perp
  \partial^\p \phi \, \delta \phi \; .
\end{equation}
The appearance of the peculiar factor 1/2 in (\ref{DELTA_PHI_FF}) has
been discussed at length by \cite{schwinger:53b} --- see also
\cite{chang:73a}. Roughly speaking it stems from the fact that the
independent field content within the front form is only one half of
that in the instant form. The factor 1/2 cancels the light--cone
Jacobian $J = 1/2$ in (\ref{DELTA_G_FF}), so that we are left with the
Poisson bracket,
\be
\label{LC_PB}
  \pb{\phi (x)}{\pi (y)}_{x,y \in \Sigma} = \delta (x^\m -
  y^\m) \delta^2 (\vc{x}_\perp - \vc{y}_\perp) \; .
\ee
As usual, commutators are inferred from  Poisson brackets by
invoking Dirac's correspondence principle, that is, by replacing the
bracket by $i$ times the commutator. For arbitrary classical
observables, $A$, $B$, this means explicitly,
\begin{eqnarray}
\label{DIRAC_CORR} 
 [\hat A , \hat B] = i \widehat{\pb{A}{B}} \; .
\end{eqnarray}
We do not address the question of operator--ordering ambiguities at
this point, as these will not be an issue in the applications to be
discussed later on. One should, however, be aware of this problem, as
it indeed \emph{can} arise within the framework of light--cone
quantization \cite{heinzl:96a}.  

Using (\ref{DIRAC_CORR}), the bracket (\ref{LC_PB}) leads to the
following commutator,
\begin{equation}
  \label{NON_FUND_COMM}
  [\phi (x), \pi (y)]_{x^\p = y^\p = \tau} = i \delta (x^\m -
  y^\m) \delta^2 (\vc{x}_\perp - \vc{y}_\perp) 
\end{equation}
As the independent quantities are the fields themselves, we invert the
derivative $\partial^\p$ and obtain the more fundamental commutator
\begin{equation}
  \label{FUND_COMM}
  [\phi (x) , \phi (y)]_{x^\p = y^\p = \tau} = - \sfrac{i}{4}\sgn(x^\m -
  y^\m) \delta^2 (\vc{x}_\perp - \vc{y}_\perp) \; .
\end{equation}
In deriving (\ref{FUND_COMM}) we have chosen the anti-symmetric Green
function $\sgn(x^\m)$ satisfying
\begin{equation}
\label{AS_GREEN}
  \pad{}{x^\m} \sgn (x^\m) = 2 \delta (x^\m) \; ,
\end{equation}
so that (\ref{NON_FUND_COMM}) is reobtained upon differentiating
(\ref{FUND_COMM}) with respect to $y^\m$.  We will see later that the
\emph{field} commutator (\ref{FUND_COMM}) can be derived directly
within Schwinger's method. Before that, however, let us study the
relation between the choice of initializing hypersurfaces, the problem
of field quantization and the solutions of the dynamical equations.

\subsection{Quantization as  an Initial-- and/or Boundary--Value Problem}
\label{subs:IBVP}

As a prototype field theory we consider a massive scalar field $\phi$
in 1+1 dimensions. Its dynamics is encoded in the action
\begin{equation}
  \label{SCAL_ACTION}
  S[\phi] = \int d^2 x \, {\cal L} =  \int d^2 x \left( \sfrac{1}{2}
  \partial_\mu  \phi \,  \partial^\mu
  \phi - \sfrac{1}{2} m^2 \phi^2 - {\cal V}[\phi] \right)\; ,
\end{equation}
where $\mathcal{V}$ is some interaction term like e.g.~$\lambda \phi^4$
and ${\cal L} = {\cal L}_0 + {\cal V}$. By varying the free
action in the standard way we obtain
\begin{eqnarray}
  \delta S = \int_{\partial M} d\sigma_\mu
  \Pi^\mu  \delta \phi + \int_M \left[ 
  \pad{{\cal L}_0}{\phi} - \partial_\mu \pad{{\cal L}_0}{(\partial_\mu
  \phi)} \right] \delta \phi \; .
\end{eqnarray}
If we do not vary on the boundary of our integration region $M$, $\delta
\phi|_{\partial M} = 0$, the surface term in $\delta S$ (which is
closely related to $\delta G$ from (\ref{DELTA_G})), vanishes and we end up
with the (massive) Klein--Gordon equation in 1+1 dimensions,
\begin{equation}
  \label{KG}
  (\Box + m^2) \phi = 0 \; .
\end{equation}
In this subsection, we will solve this equation by specifying initial
and/or boundary conditions for the scalar field $\phi$ on different
hypersurfaces $\Sigma$.  In addition, we will clarify the relation
between the associated initial value problems and the determination of
`equal--time' commutators.

It may look rather trivial to consider just the free theory, but this
is not entirely true.  Let us analyze what quantization of a field
theory means in the light of the different forms of relativistic
dynamics.  One specifies canonical commutators like $[\phi (x), \phi
(y)]_{x,y \in \Sigma}$, where the hypersurface $\Sigma$: $\tau =
const$ defines the evolution parameter $\tau$. As both $x$ and $y$ lie
in $\Sigma$, the commutator is evaluated at `equal time', which
implies that it is a \emph{kinematical} quantity. Therefore, it is the
same for the free \emph{and} the interacting theory.

\begin{figure} 
  \caption{\sl The Pauli-Jordan function as a function of $T = mx^0 /2$ and
  $X = mx^1 / 2$. It vanishes outside the light--cone and oscillates inside.}
    \begin{center}
    \epsfig{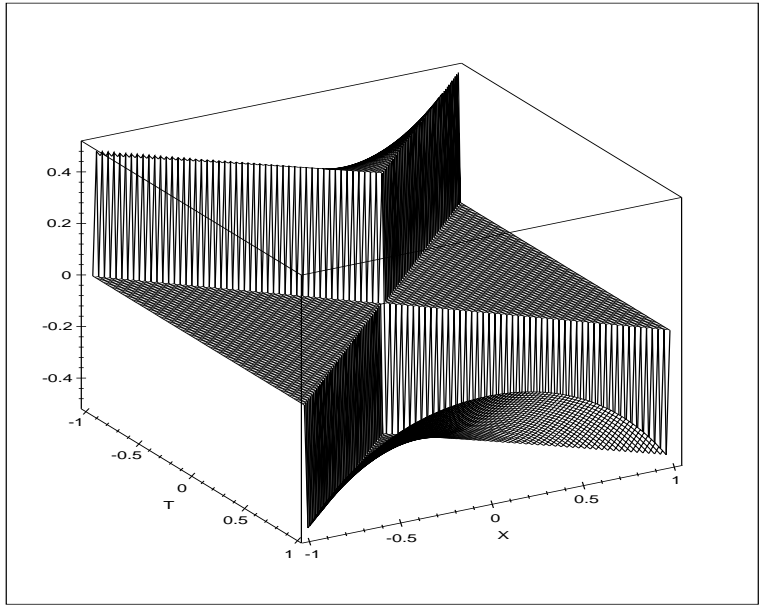}
    \vspace{10pt}
    \end{center}
  \label{fig:delta}
\end{figure}

Now, if $\phi$ \emph{is} a \emph{free} field, the commutator,
\begin{equation}
 [\phi (x) , \phi(0)] = i \Delta (x) \; ,
\end{equation}
is exactly known: it is the Pauli--Jordan or Schwinger function
$\Delta$ \cite{jordan:28,schwinger:49} which is a special solution of
the Klein--Gordon equation (\ref{KG}). It can be obtained directly from
the action in a covariant manner as a Peierls bracket
\cite{peierls:52,dewitt:83}. Alternatively, one can find it 
by evaluating the Fourier integral,   
\begin{eqnarray}
  \label{PJ}
  \Delta (x) &=& - \frac{i}{2\pi} \int d^2 p \, \delta (p^2 - m^2) \,
  \sgn(p^0) e^{- i p \cdot x} \nn \\
  &=& - \sfrac{1}{2} \sgn(x^0) \, \theta (x^2) \, J_0 (m \sqrt{x^2}) \nn \\
  &=& - \sfrac{1}{4} \big[ \sgn(x^\p) + \sgn(x^\m) \big] \; J_0 (m
  \sqrt{x^\p x^\m}) \; , 
\end{eqnarray}
where I have given both the instant and front form representation
\cite{heinzl:94a}. We note that $\Delta$ is antisymmetric, $\Delta (x)
= - \Delta (-x)$ and Lorentz invariant (under proper orthochronous
transformations). Most important, it is causal, i.e.~it vanishes
outside the light--cone, $x^2 < 0$ (see Fig.~\ref{fig:delta}).

If $\phi$ is an {\em interacting} field, causality, of course, must
still hold. If $x$ and $y$ are space--like with respect to each other,
the commutator thus still vanishes,
\begin{equation}
\label{COMM_CAUSAL}
  [\phi (x) , \phi (y)]_{(x-y)^2 < 0} = 0 \; . 
\end{equation}
This expresses the fact that fields which are separated by a
space--like distance cannot communicate with each other.  For the
front form, with the hypersurface $\Sigma: x^\p = 0$,
(\ref{COMM_CAUSAL}) cannot be used to obtain the canonical
commutators: In 1+1 dimensions, $\Sigma$ is part of the light--cone and
therefore entirely light--like. In higher dimensions, $\Sigma$ still
contains light--like directions namely where $x^\m = \vc{x}_\perp =
0$. For this reason, the light--front commutator (\ref{FUND_COMM}) of
two free fields does \emph{not} vanish identically.

Let me now discuss the explicit relation between the choice of
equal--time commutators and the classical initial/boundary for the
Klein--Gordon equation. Three examples are of interest.

\subsubsection{Cauchy Data: Instant Form.}

Conventional quantization on a space--like surface (based on the
instant form) corresponds to a Cauchy problem: if one specifies the
field $\phi$ and its time derivative $\dot \phi$ on $\Sigma: x^0 = t  = 0$,
\bea
  \phi(t=0, x)      &=& f(x) \; , \\
  \dot \phi(t=0, x) &=& g(x) \; , 
\eea
where the functions $f$ and $g$ denote the initial data (depending on
$x \equiv x^1$), the solution of the Klein--Gordon equation is
uniquely determined. This can be checked by considering the Taylor
expansion around $(t,x) \in \Sigma$, i.e.~$t=0$,
\be
  \phi(t,x) = \phi(0,x) + t \dot \phi(0,x) + \sfrac{1}{2} t^2 \ddot
  \phi(0, x) + \ldots \; ,
\ee
with the overdot denoting the time derivative. From this we see that
one has to know \emph{all} time derivatives of $\phi$ on $\Sigma$ 
once the data $f$, $g$ are given. If we calculate these, 
\bea
  \phi &=& f \; , \nn \\ \dot \phi &=& g \; , \nn \\ \ddot \phi &=&
  \phi^{\prime\prime} - m^2 \phi = f^{\prime\prime} - m^2 f \; , \nn
  \\ 
  \frac{\partial^3 \phi}{\partial t^3} &=& \dot \phi^{\prime\prime} -
  m^2  \dot \phi =
  g^{\prime\prime} - m^2 g \; , \nn \\ \vdots && \; .
\eea
we find that indeed all time derivatives are given in terms of $f$ and
$g$ and their known \emph{spatial} derivatives, denoted by the prime.
In the last two identities, we have made use of the equation of
motion. As a result, we see that the Cauchy problem is well posed: the
solution of the Klein--Gordon equation is uniquely determined by the
data on $\Sigma$.

Upon quantization, this translates into the fact that the Fock
operators can be expressed in terms of the data,
\begin{equation}
  \label{FOCK_IF} a (p^1) = \int dx^1 e^{-i p^1 x^1} \Big[ \omega_p \,
  \phi(x^0 = 0 , x^1) + i \dot \phi (x^0 = 0 , x^1) \Big] \; ,
\end{equation}
with $\omega_p = (p_1^2 + m^2)^{1/2}$. In addition, the canonical
commutators can be viewed as the Cauchy data for the Pauli-Jordan
function $\Delta$, 
\begin{eqnarray}
  \Big [\phi   (x),  \phi  (0) \Big]_{x^0   =  0}  &=&  i  \Delta 
  (x)\vert_{x^0 = 0} = 0 \; , \label{VANISHING} \\  
  \Big [\dot \phi   (x),  \phi  (0) \Big]_{x^0   =  0}  &=&  i \dot \Delta 
  (x)\vert_{x^0 = 0} = -i \delta (x^1) \; .  \label{CAUCHY_DATA}
\end{eqnarray}  
As stated above, the vanishing of the commutator (\ref{VANISHING}) is
due to causality.

\subsubsection{The  Characteristic Initial-Value Problem.} 

In the following I will perform an analogous discussion for the
hypersurfaces $\Sigma: x^\pm = const$, which, in $d=1+1$, constitute
the entire light--cone, $x^2 = 0$. In Dirac's classification, the
light--cone corresponds to a degenerate point form with parameter $a
=0$ (see Table~1). One thus does not have transitivity as points on
different `legs' of the cone are not related by a kinematical
operation. Still, it turns out that the associated initial--value
problem is well posed
\cite{neville:71a,rohrlich:71}

The light--fronts $x^\pm = 0$ are characteristics of the Klein--Gordon
equation \cite{domokos:71}.  Therefore, one is dealing with a characteristic
initial-value problem \cite{courant:62}, for which one has to provide the
data
\bea
  \phi(x^\p = 0 , x^\m) &=& f(x^\m) \; , \\
  \phi(x^\p , x^\m = 0) &=& g(x^\p) \; , \\
  f (x^\m = 0) &=& g(x^\p = 0) \; ,
\eea
where the last identity is a continuity condition. Consistency is
again checked by Taylor expanding, this time around $(x^\p , x^\m) =
0$, 
\bea
  \partial^\p \phi &=& \partial^\p f \equiv f^\prime \; , \nn \\
  \partial^\m \phi &=& \partial^\m g \equiv \dot g \; , \nn \\
  \partial^\p \partial^\m \phi &=& m^2 \phi = m^2 f = m^2 g \; , \nn
  \\
  \partial^\p \partial^\p \phi &=& f^{\prime\prime} \; , \nn \\
  \partial^\m \partial^\m \phi &=& \ddot g \; , \nn \\
  \partial^\p \partial^\p \partial^\m \phi &=& m^2 f^\prime \; , \nn
  \\
  \partial^\m \partial^\m \partial^\p \phi &=& m^2\dot g \; , \nn
  \\ 
  \vdots && \; .
\eea
Whereever a factor of $m^2$ appears we have made use of the
Klein--Gordon equation.  We thus note that the data (together with
their known derivatives) determine \emph{all} partial derivatives of
$\phi$ at the vertex of the cone, $(x^\p , x^\m) = 0$. Intuitively,
this corresponds to the fact that the information spreads from a
source located at origin.

The characteristic initial-value problem amounts to quantization on
{\it two} characteristics, $x^\pm = 0$, i.e., in $d=1+1$, really
\emph{on} the light cone, $x^2 = 0$.  The following two independent
commutators,
\begin{equation}
  [\phi (x), \phi (0)]_{x^\pm = 0} = i \Delta (x)\vert_{x^\pm  = 
  0} = - \sfrac{i}{4} \sgn (x^\mp) \; , 
\end{equation}
are then characteristic data for the Pauli-Jordan function. It turns
out that, in case the field $\phi$ is massless, the above quantization
procedure is the only consistent one (in $d$=1+1), if one wants to use
light--like hypersurfaces \cite{bogolubov:90}.

However, it is important to note that the characteristic initial-value
problem does \emph{not} correspond to light--cone quantization. One
would need two Hamiltonians $P^\m$ \emph{and} $P^\p$, and,
accordingly, two time parameters. This seems somewhat weird, to say
the least, and will not be pursued any further.

\subsubsection{Initial-Boundary-Data: Front-Form.}

In order to find the initial--value problem of the front form with a
single time parameter $x^\p$, let us naively try a straightforward
analog of the Cauchy data and prescribe field and velocity on $\Sigma:
x^\p = 0$, 
\bea
  \phi(x^\p = 0 , x^-) &=& f(x^\m) \; , \\
  \partial^\m \phi (x^\p = 0 , x^-) &=& g(x^\m) \; . 
\eea
It turns out, however, that this \emph{overdetermines} the
system. Namely, from the equation of motion
\be
  \partial^\p \partial^\m \phi = m^2 \phi \; , 
\ee
it is actually possible to obtain the velocity $\partial^\m \phi$ by
inversion of the \emph{spatial} derivative $\partial^\p$, 
\be
\label{VELOCITY}
  \partial^\m \phi = m^2 (\partial^\p)^{-1} \phi = m^2
  (\partial^\p)^{-1} f \; .
\ee
The last identity holds on $\Sigma$ and implies that the data $g$ are
unnecessary (and will even lead to an inconsistency) as the velocity
is already determined by $f$. This is confirmed by the Taylor
expansion on $\Sigma$,
\bea
  \phi &=& f \; , \nn \\
  \partial^\p \phi &=& \partial^\p f \; , \nn \\
  \partial^\m \phi &=& m^2 (\partial^\p)^{-1} f \; , \nn \\
  \partial^\m \partial^\m \phi &=& m^2 (\partial^\p)^{-1} \partial^\m
  \phi =  m^4 (\partial^\p)^{-2} f \; , \nn \\
  \vdots && \; .
\eea
It thus seems that the front form requires only half of the data as
compared to the instant form. This appearance, however, is
deceptive. Note that we have to invert the differential operator $
\partial^\p$. The inverse is nothing but the Green function $G$
defined via 
\be 
\label{GREENFCT_EQ}
  \partial^\p G(x^\m) = \delta (x^\m) \; .
\ee
Clearly, this Green function is determined only up to a homogeneous
solution $h$ satisfying
\be
  \partial^\p h = 0 \; ,
\ee
i.e.~up to a \emph{zero mode} $h = h(x^\p)$ of the operator
$\partial^\p$. Thus, in order to uniquely specify the Green function
(\ref{GREENFCT_EQ}), we have to provide additional information in terms
of \emph{boundary conditions}. The standard choice is to demand
antisymmetry in $x^\m$, whence
\be
\label{GREENFCT}
  G(x^\m) = \frac{1}{4} \sgn (x^\m) \; , 
\ee
which we have already used in (\ref{FUND_COMM}) and (\ref{AS_GREEN}). 
Before we discuss the physical reason for demanding antisymmetry, let
us briefly go to momentum space where we replace $\partial^\p$ by
$i p^\p$.  The equation (\ref{GREENFCT_EQ}) for the Green function
becomes $i p^\p G(p^\p) = 1$, which has the general solution
\be
  G(p^\p) = -i /p^\p + h(p^\m) \delta (p^\p) \; .
\ee
In this identity, $1/p^\p$ has to be viewed as a distribution
corresponding to an arbitrary regularization of the singular function
$1/p^\p$ \cite{gelfand:64}. Any two regularizations differ by terms
proportional to $\delta (p^\p)$, i.e.~a zero mode of $p^\p$. Choosing
an antisymmetric Green function uniquely yields a principal value
prescription, 
\be
  i G(p^\p) = {\cal P} \frac{1}{p^\p} = \frac{1}{2} \left(
  \frac{1}{p^\p + i \epsilon} + \frac{1}{p^\p - i \epsilon} \right) \; ,
\ee
which is the canonical regularization of $1/p^\p$. 

Altogether we have seen that the front form corresponds to prescribing
both initial and boundary conditions, so that one has a `mixed' or
\emph{initial--boundary} value problem. What are the implications for
quantization? We address this question by determining the Poisson
brackets through the requirement that Euler--Lagrange and canonical
equations should be equivalent. To this end we solve the Klein--Gordon
equation (\ref{KG}) for the velocity $\dot \phi \equiv \partial \phi /
\partial x^\p$ as in (\ref{VELOCITY}). This gives
\begin{equation}
  \label{EL_INIT} \dot \phi (x^+, x^-) = -\frac{m^2}{4} \int dy^- G
  (x^-, y^-) \phi (x^+ , y^-) \; ,
\end{equation}
using the Green function $G$ from (\ref{GREENFCT}).  The Hamiltonian
equation of motion is given by the Poisson bracket with $H =
\sfrac{1}{2} m^2 \int dx^\m \phi^2$,
\begin{equation} 
  \label{HAM_INIT}
  \dot  \phi  (x^+ , x^-) =  \frac{m^2}{2}  \int  dy^- \pb{\phi 
  (x^+,   x^-)}{\phi(x^+   ,   y^-)}   \phi   (x^+   ,   y^-)   
\end{equation}
with the bracket of the fields $\phi$ to be determined. Clearly,
Euler--Lagrange and Hamiltonian equation of motion, (\ref{EL_INIT})
and (\ref{HAM_INIT}) become equivalent if one identifies
\begin{equation}
  \pb{\phi  (x^+, x^-)}{\phi  (x^+ , y^-)} \equiv -
  \sfrac{1}{2}G(x^- , y^-) \; . \label{BAS_BRACK}
\end{equation}
We thus see that the fundamental Poisson bracket coincides with the
Green function which, accordingly, justifies the requirement of
antisymmetry. After quantization, (\ref{BAS_BRACK}) of course
coincides with (\ref{FUND_COMM}), the result from Schwinger's action
principle, specialized to $d= 1+1$,
\be
\label{COMM_2D}
  [\phi (x), \phi(0)]_{x^\p = 0} = i \Delta (x)|_{x^\p = 0} =
  - \frac{i}{4} \sgn (x^\m) \; .
\ee
From the momentum space perspective, 
\be
  [\phi (p^\p), \phi(0)] =  \frac{i}{2} G(p^\p) = \frac{1}{2} 
  {\cal P} \frac{1}{p^\p} \; , 
\ee
we conclude that, technically, light--cone quantization is the
inversion of the longitudinal momentum $p^\p$ and as such requires the
specification of initial--\emph{boundary} data. In some sense, this
can also be viewed as an infrared regularization because one provides
a prescription of dealing with a pole at vanishing longitudinal
momentum, $p^\p = 0$. As is well known, a particularly nice way of
regularizing in the infrared is to enclose the system under
consideration in a finite spatial volume. This is the topic of the
next subsection.

\subsection{DLCQ --- Basics}

DLCQ is the acronym for `discretized light--cone quantization',
originally developed by
\cite{maskawa:76,pauli:85a,pauli:85b}\footnote{For a nice overview of
recent developments see \cite{hiller:00}.}. The physical system under
consideration is enclosed in a finite volume with discrete momenta and
prescribed boundary conditions in $x^\m$. Recently, there has been
renewed interest in this method in the context of string theory
\cite{banks:97,susskind:97,lunin:99}.

Our starting point is the Fourier representation for a solution of the
Klein--Gordon equation (still in infinite volume),
\bea
  \phi(x) &=& \int \frac{d^2 p}{2 \pi} \chi (p) \, \delta (p^\p p^\m -
  m^2) \, e^{-i p \cdot x} \nn \\
  &=& \int \frac{dp^\p dp^\m}{4 \pi} \chi(p^\p , p^\m)
  \frac{1}{|p^\p |} \, \delta (p^\m - m^2/p^\p) \, e^{- i p \cdot x} \nn \\
  &=& \int \frac{dp^\p}{4 \pi |p^\p |} \chi (p^\p , \hat p^\m) \, e^{- i
  \hat p \cdot x} \nn \\
  &=& \int_0^\infty \frac{dp^\p}{4 \pi p^\p} \left[\chi(p^\p , \hat
  p^\m) e^{- i \hat p \cdot x} + \chi(-p^\p , - \hat
  p^\m) e^{ i \hat p \cdot x}  \right] \nn \\
  &\equiv& \int \frac{dp^\p}{4 \pi p^\p} \theta (p^\p) \left[a(p^\p)
  e^{- i \hat p \cdot x} + a^* (p^\p) e^{i \hat p \cdot x} \right]
  \label{LC_FOCK} \;
  .
\eea
The following remarks are in order: we have defined the on--shell
energy, $\hat p^\m \equiv m^2 / p^\p$; contrary to the instant form,
the integration over the positive and negative mass hyperboloid is
achieved by a \emph{single} delta function. Again, this is a
consequence of the linearity of the mass--shell constraint in $p^\m$.
The two branches of the mass--shell correspond to positive and
negative values of $p^\p$ (and also $\hat p^\m$), respectively.
Associated with the two signs of the kinematical momentum $p^\p$ are
the positive and negative frequency modes $a$, $a^*$, defined in such a
way that their argument $p^\p$ is always positive (cf.~the step
function $\theta$ in the last line). This can equivalently be viewed
as the reality condition
\be
  a^* (p^\p) = a (- p^\p) \; , 
\ee
as is obvious from the last step in the derivation
(\ref{LC_FOCK}). Upon quantization this implies that annihilation
operators with negative longitudinal momentum $p^\p$ are actually
creation operators for particles with positive $p^\p$. The field
commutator (\ref{COMM_2D}) is reproduced by demanding
\be
  [a (k^\p), a^\dagger (p^\p)] = 4 \pi p^\p \delta (k^\p - p^\p) \; .
\ee
As already indicated, DLCQ amounts to compactifying the spatial
light--cone coordinate, $-L \le x^\m \le L$, and imposing periodic
boundary conditions for the fields,
\be
  \phi(x^\p , x^\m = -L ) = \phi(x^\p , x^\m = L ) \; , 
\ee
which are to hold for all light--cone times $x^\p$. Space--time is
thus endowed with the topology of a cylinder. This implies discrete
longitudinal momenta, $k_n^\p = 2 \pi n /L$, so that the Fock
expansion (\ref{LC_FOCK}) becomes
\begin{equation}
\label{DLCQ_FOCK}
  \phi(x^\p = 0 , x^\m) = a_0 + \sum_{n > 0} \frac{1}{\sqrt{4\pi n}}
  \Big(a_n e^{-i n \pi x^\m / L } + a_n^* e^{i n \pi x^\m / L }\Big)
  \; ,
\end{equation}
Note that we have allowed for a zero momentum mode $a_0$. We will see
in a moment that it actually vanishes in the free theory.  Plugging
(\ref{DLCQ_FOCK}) into the free Lagrangian 
\be
\label{SCAL_LAG_PHI}
  L_0 [\phi] = \sfrac{1}{2} \int dx^\m \left( \sfrac{1}{2} \partial^\p
  \phi \, \partial^\m \phi - \sfrac{1}{2} m^2 \phi^2 \right) \; , 
\ee
we obtain (discarding a total time derivative)
\begin{equation}
  \label{SCAL_LAG} L_0 [a_n ,a_0 ] = -i \sum_{n>0} a_n \dot a_n^* -
  m^2 L a_0^2 -\sum_{n>0} \frac{m^2 L}{4 \pi n}a_n^* a_n \equiv -i
  \sum_{n>0} a_n \dot a_n^* - H \; ,
\end{equation}
with $H$ denoting the Hamiltonian and $\dot a_n^* = \partial a_n^* /
\partial x^\p$. From both representations (\ref{SCAL_LAG_PHI}) and
(\ref{SCAL_LAG}) it is obvious that the light--cone Lagrangian is
linear in the velocity ($\partial^\m \phi$ and $\dot a_n^*$,
respectively). A particularly suited method for quantization in this
case is the one of Faddeev and Jackiw for first order systems
\cite{faddeev:88,jackiw:93}. It avoids many of the technicalities of
the Dirac--Bergmann formalism and is in general more economic. It
reduces phase-space right from the beginning as there are no `primary
constraints' introduced. The method is essentially equivalent to
Schwinger's action principle, especially in the form presented in
\cite{schwinger:53b}. For the case at hand, the method basically boils
down to demanding equivalence of the Euler-Lagrange and Hamiltonian
equations of motion (cf.~last subsection).

The former are given by 
\begin{eqnarray}
  -i \dot a_n + \frac{m^2 L}{4\pi n}a_n &=& 0 \; ,\label{EL_SCAL1} \\ 
  2 m^2 L a_0 &=& 0 \; .  \label{EL_SCAL2}
\end{eqnarray}
The first equation, (\ref{EL_SCAL1}), is just the free Klein--Gordon
equation which can be easily seen upon multiplying by $k_n^\p$. The
second identity, (\ref{EL_SCAL2}), is nondynamical and thus a {\em
constraint} which states the absence of a zero mode for free fields, 
$a_0 = 0$.

The canonical equations are
\begin{equation}
  \label{HAM_SCAL1}
  \dot a_n = \pb{a_n}{H} =  \sum_{k>0} \frac{m^2 L}{4 \pi k}
  \pb{a_n}{a_k^*} a_k  \; , 
\end{equation}
which obviously coincides with (\ref{EL_SCAL1}) if the canonical
bracket is
\begin{eqnarray}
  \label{PB_SCAL}
  \pb{a_k}{a_n^*} = -i \delta_{kn} \; .
\end{eqnarray}
The constraint (\ref{EL_SCAL2}) is obtained by differentiating the
Hamiltonian, 
\begin{equation}
  \label{HAM_SCAL2}
  \pad{H}{a_0} =  2m^2 L a_0 = 0 \; . 
\end{equation}
Let us briefly show that the approach presented above is equivalent to
Schwinger's \cite{schwinger:53b}. From (\ref{SCAL_LAG}) we read off a
generator
\begin{equation}
  \delta G = -i \sum_{n>0} a_n \delta a_n^*
\end{equation}
effecting the transformation
\begin{equation}
  \delta a_n^* = \pb{a_n^*}{\delta G} = -i \sum_{k>0} \pb{a_n^*}{a_k}
  \delta a_k^* \; ,
\end{equation}
which in turn implies the canonical bracket (\ref{PB_SCAL}). 

Quantization is performed as usual by employing the correspondence
principle (\ref{DIRAC_CORR}), so that, from (\ref{PB_SCAL}), the
elementary commutator is given by
\begin{eqnarray}
\label{DLCQ_FOCK_COMM}
  [a_m , a_n^\dagger] = \delta_{mn} \; .
\end{eqnarray}
The Fock space expansion for the (free) scalar field $\phi$ thus becomes
\begin{eqnarray}
  \label{FOCK_EXP2}
  \phi(x^\p = 0 , x^\m) = \sum_{n > 0} \frac{1}{\sqrt{4\pi n}} \Big(a_n
  e^{-i n \pi x^\m / L } + a_n^\dagger e^{i n \pi x^\m / L }\Big)  , 
\end{eqnarray}
Like in the infinite--volume expression (\ref{LC_FOCK}), the Fock
`mea\-sure' $1/\sqrt{4\pi n}$ does not involve any scale like the mass
$m$ or the volume $L$. This is at variance with the analogous
expansion in the instant form which reads
\begin{equation}
  \label{IF_FOCK}
  \phi(x, t=0) = \frac{1}{\sqrt{2L}} \sum_n \frac{1}{\sqrt{2 (k_n^2
  + m^2)}}   \left( a_n e^{i k_n x} + a_n^\dagger e^{-i k_n x} \right) \; ,
\end{equation}
where $-L \le x \le L$, $k_n = \pi n /L$, and $[a_n , a_m^\dagger] =
\delta_{mn}$. Obviously, the `measure' $(k_n^2 + m^2)^{1/2}$
\emph{does} depend on $m$ and $L$. We will discuss some consequences
of this difference in Subsection~3.6. 

We can use the results (\ref{DLCQ_FOCK_COMM}) and (\ref{FOCK_EXP2}) to
calculate the free field commutator at equal light--cone time $x^\p$,
\begin{eqnarray}
  \label{COMM_DLCQ} [\phi(x) , \phi(0)]_{x^\p = 0} = \sum_{n \ne
  0}\frac{1}{4 \pi n} e^{-i n \pi x^\m / L} = -\frac{i}{2} \left[
  \frac{1}{2}\sgn (x^\m) - \frac{x^-}{2L} \right] \; .
\end{eqnarray}
This coincides with (\ref{COMM_2D}) up to a finite size correction
given by the additional term $x^\m / 2L$. The effect of this term is
two--fold. First, it makes the sign function periodic (in the interval
$-L \le x^\m \le L$), and second, it guarantees the absence of a zero
mode which must hold according to (\ref{EL_SCAL2}), (\ref{HAM_SCAL2}),
and
\be
  \intl dx^\m [\phi(x), \phi (0)]_{x^\p = 0} = 0 \; .
\ee
One may equally think of this as the finite--volume analog of the
principal value prescription. 

The commutator (\ref{COMM_DLCQ}) has originally been obtained in
\cite{maskawa:76} using the Dirac-Bergmann algorithm for constrained
systems. The Faddeev--Jackiw method, however, is much more economic and
transparent. In particular, it makes clear that the basic canonical
variables of a light--cone field theory are the Fock operators or their
classical counterparts. The $a_n$ with, say, $-N \le n \le N$ in
(\ref{DLCQ_FOCK}) can be viewed as defining a $(2N + 1)$-dimensional
phase space. A phase space, however, should have even dimension. This
is accomplished by choosing a polarization in terms of positions and
momenta, here $a_n$ and $a_n^\dagger$, with $n > 0$, and by the
vanishing of the zero mode, $a_0 = 0$. It turns out that this
vanishing is a peculiarity of the free theory as is discussed in
\cite{maskawa:76,wittman:89,heinzl:92c}.

At this point one should honestly state that the issue of zero modes
is one of the unsolved problems of light--cone
quantization\footnote{For an overview see e.g. Ch.~7 of
\cite{brodsky:98} and references therein.}. The constraint equations
for the zero modes are in general very hard to solve unless one has
some small parameter like in perturbation theory
\cite{mccartor:92,heinzl:96a} or within a large--$N$ expansion
\cite{borderies:93,borderies:95}. Using a path integral approach, it
has recently been shown \cite{hellerman:99} that integrating out the
zero modes constitutes a strong coupling problem. There are
speculations that this problem might be less severe if one goes beyond
quantum field theory, i.e.~in string or M--theory \cite{banks:99}. 

In the last reference, the author also states that compactification in
a light--like direction ``is close to a space with periodic time'' and
thus ``weird'', in view of possible `grandfather
paradoxes'. Therefore, the natural question arises whether DLCQ is
actually consistent with causality.

\subsection{DLCQ --- Causality}

In this subsection I will address the question under which
circumstances compactification of `space' is compatible with the
requirements of causality. The presented results are based on recent
work with N.~Scheu and H.~Kr\"oger \cite{heinzl:99}. 

In (\ref{PJ}) and (\ref{COMM_CAUSAL}) we have seen that the
(infinite--volume) commutator of two scalar fields vanishes whenever
their space--time arguments are separated by a space--like distance
(cf.~Fig.~2). As already mentioned, this is a manifestation of the
principle of microcausality, which is the general statement that the
commutator of \emph{any} two observables ${\cal O}_1 (x)$ and ${\cal
O}_2 (y)$ must vanish whenever their separation $x-y$ is
space--like. Physically, this implies that measurements of the
observables ${\cal O}_1$ and ${\cal O}_2$ performed at $x$ and $y$, do
not interfere. Some consequences of this principle are the
spin--statistics theorem, analyticity properties of Green functions
leading to dispersion relations etc. \cite{streater:63}.

\begin{figure} 
  \caption{The integrand $I (k^+)$.} 
  \vspace{10pt}
  \begin{center}
  \epsfig{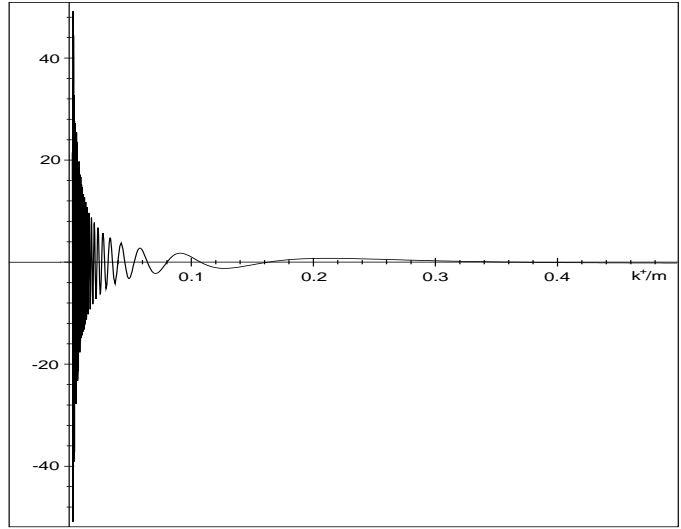}
  \end{center}
  \label{fig:IFF}
\end{figure}

Our starting point are the Fourier representations of the Pauli-Jordan
function, both for the instant and front form (denoted IF and FF,
respectively),
\begin{eqnarray}
  \mbox{IF:} \quad \Delta (x) &=&  - \int \frac{dk^1}{2 \pi \omega_k}
  \sin (k \cdot x)  \equiv \int dk^1 \, I (k^1) 
  \label{DELTA_COV_IF}  \; , \\
  \mbox{FF:} \quad \Delta (x) &=& - \int_0^\infty \frac{dk^+}{2 \pi
  k^+} \sin(k \cdot x) \equiv \int dk^+ \, I(k^+) \; .
  \label{DELTA_COV_FF}
\end{eqnarray}
Both integrals yield the same result (\ref{PJ}) for the Pauli--Jordan
function. Note, however, that the integrand $I(k^+)$ is exploding and
rapidly oscillating for $k^+ \to 0$ (see Fig.~\ref{fig:IFF}) so that
the finite result for the integral is due to sizable cancellations
that occur upon integration.

To obtain the finite--volume representations for the commutator, one
proceeds as in Subsection~3.4 by restricting the spatial coordinates,
$-L \le x^1 , x^- \le L$, and imposing periodic boundary conditions
for the field $\phi$. Momenta become discrete, $k^1_n \equiv \pi n
/L$, and $k_n^+ \equiv 2 \pi n /L$. The finite-volume representations
are \emph{defined} by replacing the integrals (\ref{DELTA_COV_IF}) and
(\ref{DELTA_COV_FF}) by the discrete sums,
\begin{eqnarray}
  \Delta_{IF} (x) &\equiv& - \sum_{n=-N}^N
  \frac{1}{2 \omega_n L} \sin(k_n \cdot x) \; ,
  \label{DELTA_IF} \\ 
  \Delta_{FF} (x) &\equiv&
  - \sum_{n=1}^N \frac{1}{2 \pi n} \sin (k_n \cdot x) \; . 
  \label{DELTA_FF}
\end{eqnarray} 
The on-shell energies for discrete momenta are defined as 
\be
  \omega_n = (n^2 \pi^2/L^2 + m^2)^{1/2}  \quad  \mbox{and}  \quad \hat
  k_n^- = m^2 L/2\pi n \; . 
\ee
For both functions, $\Delta_{IF}$ and $\Delta_{FF}$, the periodicity
in $x^1$ and $x^-$, respectively, with periodicity length $2L$, is
obvious.  The limit $N \to \infty$ is understood unless we perform
numerical calculations where $N$ is kept finite.

\begin{figure} 
  \caption{$\Delta_{IF} (X,T)$ as a function of $X = x^1 /2L$. $T = X^0
  / 2L = 0.2$, $mL = 1$, $N = 50$.}
  \vspace{10pt} 
  \begin{center}
  \epsfig{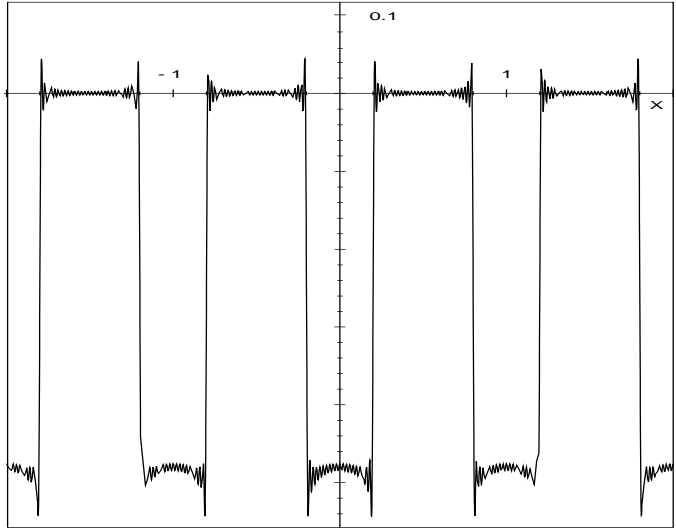}
  \end{center}
  \label{fig:IF}
\end{figure}

The evaluation of the sums (\ref{DELTA_IF}) and (\ref{DELTA_FF}) is
not straightforward. To gain some intuition, we evaluate them
numerically beginning with the IF expression (\ref{DELTA_IF}). The
resulting $\Delta_{IF}$ is plotted in Fig.~\ref{fig:IF}.

Upon inspection, one notes the following: Up to small oscillations
stemming from the (unavoidable) Gibbs phenomenon, $\Delta_{IF}$
vanishes outside the light--cone ($|x^0| < |x^1| < L$), and thus is
causal even in finite volume. If we let the summation cutoff $N$ go to
infinity, $\Delta_{IF}$ approaches the continuum Pauli--Jordan function
$\Delta$ (for $-L < x^0, x^1 < L$). There is a clear physical picture
behind these observations. One can imagine a periodic array of sources
located at the quantization hypersurface $x^0 = 0$ at points $x^1 =
2Ln$. These sources `emit' spherical `waves' into their own future LCs
which start to overlap after time $x^0 > L$. At this point the 'waves'
emanating from the sources begin to interfere. Thus, the influence of
the BC is felt only after a long time (as large as the spatial
extension $L$ of the system). This picture can be confirmed
analytically. An application of the Poisson resummation formula yields
\be
  \Delta_{IF} (x) = \sum_n \Delta (x^0, x^1 + 2Ln) \; , 
\ee
i.e.~a periodic array of $\Delta$'s which are nonoverlapping as long
as $x^0 < L$.

For the front form, the situation turns out to be more
complicated. Using Poisson resummation one can derive the
finite--volume version of the canonical light--cone commutator, at
$x^\p = 0$, which is a periodic sign function,
\be
  \Delta_{FF} (x^\p = 0, x^\m) = - \sfrac{1}{4} \sum_{n} \sgn(x^\m +
  2L n) + x^\m / 4L \; .
\ee
This coincides with (\ref{COMM_DLCQ}) if $x^\m$ is restricted to lie
between $-L$ and $L$.

\begin{figure}[tbp] 
  \caption{$\Delta_{FF} (v,w)$ as a function of $v = x^- / 2L$. $w =
  10000$, $N = 70$. It does not vanish outside the LC, $-1 < v < 0$.}
  \vspace{10pt} \begin{center}
  \epsfig{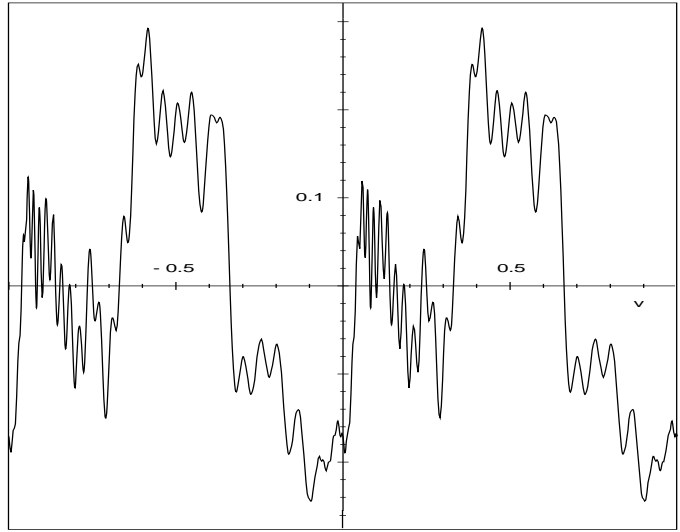} 
  \end{center}
  \label{fig:FF10000}
\end{figure}

For $x^\p \ne 0$, I have evaluated $\Delta_{FF}$ numerically.  The
result is shown in Fig.~\ref{fig:FF10000} as a function of the
dimensionless variables $v \equiv x^\m / 2L$, $w = m^2 L x^\p / 2$. For
large values of $w$, $\Delta_{FF}$ attains a very irregular shape,
though numerically the representation (\ref{DELTA_FF}) converges to a
periodic function. The most important observation, however, is that
$\Delta_{FF}$ does not vanish outside the light--cone, i.e. for $x^\m <
0$, if $x^\p > 0$ as in Fig.~\ref{fig:FF10000}. This a clear
\emph{violation of microcausality} as has first been observed in 
\cite{scheu:98}.

\begin{figure}[tbp] 
  \caption{Comparison of the Fourier representation
  (\protect\ref{DELTA_FF}) with the result of Bernoulli resummation
  (smooth, heavy line).}
  \vspace{10pt}
  \begin{center}
  \epsfig{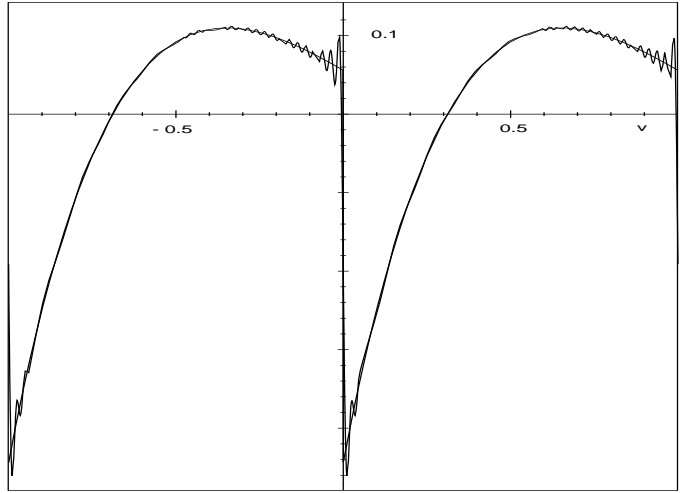}
  \end{center}
  \label{fig:resum}
\end{figure}

As already stated, it is not straightforward to confirm these findings
analytically. Poisson resummation does not work; first, because of the
weak localization properties of $\Delta$ in $x^\m$ (asymptotically, is
goes like $(x^\m)^{-1/4}$); second, and even worse, because the zero
mode $I (k^\p = 0)$ does not exist for $x^\p \ne 0$.  Nevertheless, an
independent confirmation of causality violation {\it can} be obtained
from resumming (\ref{DELTA_FF}) in terms of Bernoulli polynomials,
thereby replacing the Fourier series by a (rapidly converging) power
series in $w$. The result is shown in Fig.~\ref{fig:resum} for
$w=5$. There is nice agreement with the Fourier representation
(\ref{DELTA_FF}) (and no Gibbs phenomenon, as expected). Again,
causality violation is obvious.

From a technical point of view, the violation of causality is not
really astonishing. We are replacing the integral over the severely
oscillating function $I(k^\p)$ by a Riemann sum with equidistant grid
points. In this way, we are sampling the integrand in such a way that
the huge cancellations present in the \emph{integral} do not take
place. Instead, for small $k^\p$, we replace $I(k^\p)$ by a random
`staircase' function which in the end produces the `noise' seen in
Fig.~\ref{fig:FF10000}.

At this point a natural questions arises: is there a remedy for the
causality violation? The answer is positive. We have found two ways
around the problem\footnote{For a third method, see
\cite{chakrabarti:99}.}, both, however, with shortcomings of their
own. The first way is to regularize the integral (\ref{DELTA_COV_FF}),
replacing $\Delta$ by $\Delta_\epsilon$ in such a way that the
associated integrand satisfies $I_\epsilon (k^\p = 0) = 0$. One can
chose e.g.~a principal value regularization
\cite{hornbostel:88} or a more sophisticated prescription
\cite{nakanishi:77}. In this way, one suppresses the
oscillations and the divergence at $k^\p = 0$ at the price of
introducing a small causality violation of order $\epsilon$. But now
$\Delta_\epsilon$ \emph{can} be approximated by a discrete sum if the
momentum grid is sufficiently fine, $\triangle k^\p
\ll \epsilon$. The order of limits, however, becomes important. First, one
has to perform the infinite--volume limit, $\triangle k^\p \to 0$, $L
\to \infty$, and only then the limit $\epsilon \to 0$. For this method,
Poisson resummation should work \cite{salmons:99}. However, it seems
somewhat awkward and not very economic to perform \emph{two}
regularizations (finite $L$ \emph{and} $\epsilon$).

An alternative way of resolving the problem is the following: instead
of an equally spaced grid \`a la DLCQ (i.e.~$\triangle k^\p_n = const$)
one can chose an adapted momentum grid with spacing $\triangle k^\p_n
\sim 1/n$ for small $n$. In this way, one is sampling the
small--$k^\p$-region of $I(k^\p)$ in a more reasonable way. Practically,
the method amounts to viewing $\Delta_{IF}$ as the correct
finite--volume expression and replacing $x^0$ and $x^1$ by $(x^\p \pm
x^\p)/2$, respectively. This is equivalent to introducing \emph{new}
discrete momenta, $k_n^\pm \equiv \omega_n \pm k_n^1$. 

\begin{figure}[tbp] 
  \caption{The causal commutator $\Delta_c$ as a function of $v$. $x^+ /
  2L = 0.2$, $mL = 50$, $N= 50$.}  
  \vspace{10pt}
  \begin{center}
  \epsfig{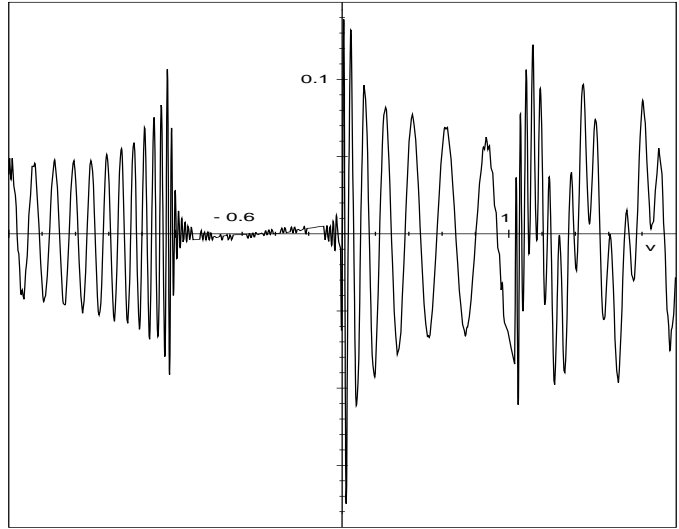}
  \end{center}  
  \label{fig:causal}
\end{figure}

As a result, the point $k^\p = 0$ becomes an accumulation point of the
momentum grid which leads to a causal finite--volume representation
$\Delta_c (x^\p , x^\m)$ of $\Delta$ (see Fig.~\ref{fig:causal}).  This
function, however, is no longer periodic in $x^\m$. We thus find that,
with a light--like direction being compactified, one cannot have both,
periodicity \emph{and} causality. On the other hand, the
regularization method above seems to suggest that the causality
violation is in some sense 'small' and thus may have a minor effect on
the calculation of observables. Whether this statement is true has
still to be worked out in detail. 

\subsection{The Functional Schr\"odinger Picture}

Having discussed some difficulties of light--cone quantization in a
finite `box' we go back to the infinite volume and consider yet
another \emph{method of quantization}, the functional Schr\"odinger
picture. The idea of this method is to mimic quantum mechanics within
quantum field theory. States are described as functionals $\Psi[\phi]$
depending on the field(s) $\phi$, while operators are combinations of
multiplication by functionals of $\phi$ and functional differentiation
with respect to $\phi$.

\subsubsection{Bosons (Instant Form).}

For a free massive scalar field in two dimensions one finds the
following results using instant form dynamics \cite{jackiw:87}. The
canonical momentum (operator) acts via differentiation,
\be
  \hat \pi (x) \Psi [\phi] = -i \fud{}{\phi(x)} \Psi[\phi] \; .
\ee
The ground state $\Psi_0$ of the system, the Fock vacuum, is obtained
by direct analogy with the harmonic oscillator. One rewrites the
vacuum annihilation condition $a(k) | 0 \ket = 0$ as a functional
differential equation,
\be
  \left[ \int dy \Omega (x-y) \phi(y) + \fud{}{\phi(x)} \right] \Psi_0
  [\phi] = 0 \; ,
\ee
which is solved by a Gaussian, 
\be
  \Psi_0 [\phi] \sim \exp \left[ - \sfrac{1}{2} (\phi, \Omega \phi)
  \right] \; .
\ee
Here we have introduced the `quadratic form', 
\be
  (\phi, \Omega \phi) \equiv  \int dx \,  dy \, \phi(x) \Omega (x-y)
  \phi(y) \; ,   
\ee
using the kernel (or \emph{covariance}) $\Omega (x-y)$. The latter is
defined by its Fourier transform,
\be
  \Omega (k) \equiv \sqrt{k^2 + m^2} \equiv \omega_k \; ,
\ee
which is nothing but the on--shell energy of  a free massive
scalar. For later use, it is important to note that the instant form
covariance $\Omega$ is explicitly mass dependent. As a consequence, if
we have two free scalars of masses $m_1$ and $m_2$, respectively,
their Fock vacua are related by a Bogolubov transformation, 
\be
  | \Omega_2 \ket = U_{21} |\Omega_1 \ket \; ,
\ee
with the unitary operator $U_{21}$ explicitly given by
\be
\label{BOGOLUBOV}
  U_{21} = \exp \int \frac{dk}{4\pi} \, \theta_k [a_1 (k) a_1 (-k) -
  a_1^\dagger (k) a_1^\dagger (-k)] \; .
\ee
$\theta_k$ is the Bogolubov angle. We mention in passing that in order
to properly define $U_{21}$ as an operator one should work in finite
volume to avoid infrared singularities \cite{heinzl:98}. Otherwise the
two vacuum states have vanishing overlap. Within the functional
Schr\"odinger picture this has been analysed by \cite{jackiw:87}.

\subsubsection{Fermions (Instant Form).}

For fermions, the situation is slightly more involved. Note that even
within the instant form, the Dirac Lagrangian represents a
first--order system, so that one expects some similarities with
light--cone quantization in this case. This expectation will indeed
turn out to be true. The instant form fermionic field operators are
given by \cite{kiefer:94}
\be
  \hat \psi_\alpha = \frac{1}{\sqrt{2}} \left( u_\alpha +
  \fud{}{u_\alpha^*} \right) \; , \quad \hat \psi_\alpha^\dagger  =
  \frac{1}{\sqrt{2}}  \left( u_\alpha^*  +
  \fud{}{u_\alpha} \right) \; ,
\ee
and thus are linear combinations of multiplication by and
differentiation with respect to the complex--valued Grassmann
functions $u_\alpha(x)$ and $u_\alpha^* (x)$. These functions
characterize the states, for example the ground state (Fock vacuum)
which is again Gaussian,
\be
  \Psi_0 [u , u^*] \sim \exp (u^* , \Omega u) \; .
\ee
For 2$d$ massive fermions, the covariance is found to be
\be
  \Omega (k) = \frac{1}{\sqrt{k^2 + m^2}} (k \sigma^1 - m \sigma^3)
  \; , 
\ee
with $\sigma^1$ and $\sigma^3$ the standard Pauli matrices. Again,
$\Omega$ is explicitly mass dependent. In the massless case, $m=0$, it
becomes particularly simple,
\be
  \Omega (k) = \sgn (k) \sigma^1 \; , 
\ee
or, after Fourier transformation,
\be
  \Omega (x-y) = \frac{i}{\pi} {\cal P} \frac{1}{x-y} \sigma^1 \; .
\ee
Here we have once more made use of the fact that the principle value
${\cal P} (1/x)$ is the Fourier transform of the sign function. 

\subsubsection{Bosons (Front Form).}

As stated above, the latter case is somewhat similar to the generic
situation in light--cone quantization. Let us again consider a
massive free scalar field $\phi$ in 2$d$ with Fock expansion
(\ref{LC_FOCK}). We decompose it into positive and negative frequency
part,
\be
  \phi = \phi^+ [a] + \phi^- [a^*] \equiv u + u^* \; ,
\ee
where $\phi^- = (\phi^+)^*$ as $\phi$ is real. Quantization is
performed by defining field \emph{operators} such that the canonical
light--cone commutator (\ref{FUND_COMM}) is reproduced. The solution
turns out to be somewhat more complicated than for instant form
fields, namely
\bea
  \hat \phi^+ (x^\m) &=& u (x^\m) + \frac{1}{2} \int dy^\m i \Delta_+
  (x^\m - y^\m) \fud{}{u^* (y^\m)} \; , \\
  \hat \phi^- (x^\m) &=& u^* (x^\m) + \frac{1}{2} \int dy^\m i \Delta_-
  (x^\m - y^\m) \fud{}{u (y^\m)} \; .
\eea
$\Delta_+$ and $\Delta_-$ are distributions that sum up to $\Delta$,
$\Delta_+ + \Delta_- = \Delta$, and are explicitly given by
\be
  i \Delta_\pm (x^\m) = \mp \ln (\pm x^\m - i \epsilon) = \mp \ln
  |x^\m| \pm \frac{i}{4} \theta (\mp x^\m) \; .
\ee
The ground state is annihilated by $\hat \phi^+$, $\hat \phi^+ \Psi_0
[u, u^*] = 0$, which yields a functional differential equation, again
with Gaussian solution, 
\be
  \Psi_0 [u, u^*] = \exp [- (u^* , \Omega u)] \; .
\ee
The covariance is given by
\be
\label{FSP_VACUUM}
  \Omega (x^\m) = 2 i \partial^\p \delta (x^\m) \; , \quad \Omega
  (k^\p) = 2 k^\p  \; ,
\ee
and thus is a \emph{local} expression which is a very peculiar
finding. In momentum space, the ubiquitous longitudinal momentum
$k^\p$ appears. One thing that is particularly obvious from
(\ref{FSP_VACUUM}) is the fact that the light--cone Fock vacuum is
mass independent, $\Psi_0 (m_1) = \Psi_0 (m_2)$ which means that the
analog of the Bogolubov transformation (\ref{BOGOLUBOV}) is trivial,
i.e.~$U_{21} = \Eins$. This has been checked explicitly for several
examples, including the Nambu--Jona-Lasinio model
\cite{dietmaier:89} and bosons and fermions coupled to external sources
\cite{heinzl:98}.

I believe that the locality of the covariance has far reaching
consequences which, however, are still to be worked out in the present
framework. Within ordinary Fock space language, some properties of the
light--cone vacuum will be discussed in what follows.

\subsection{The Light--Cone Vacuum}

One of the basic axioms of quantum field theory states that the
spectrum of the four--momentum operator is contained within the closure
of the forward light--cone \cite{streater:63,bogolubov:75}.  The
four--momentum $P^\mu$ of any physical, that is, observable particle
thus obeys
\begin{equation}
  \label{SPECTRUM_COND}
  P^2 \ge 0 \; , \quad P^0 \ge 0 \; ,
\end{equation}
which is, of course, consistent with the mass--shell constraint, $p^2
= m^2$. The tip of the cone, the point $P^2 = P^0 = 0$, corresponds to
the vacuum. From the spectrum condition (\ref{SPECTRUM_COND}) we infer
that
\begin{equation}
  P_0^2 - P_3^2 \ge P_\perp^2 \ge 0 \quad \mbox{or} \quad P^0 \ge |P^3| \; . 
\end{equation}
This implies for the longitudinal light--cone momentum,
\begin{equation}
  P^\p = P^0 + P^3 \ge |P^3| + P^3 \ge 0 \; .
\end{equation}
We thus have the important kinematical constraint that physical states
must have nonnegative longitudinal momentum,
\begin{equation}
  \bra \mbox{phys} | P^\p | \mbox{phys} \ket \ge 0 \; .
\end{equation}
The spectrum of $P^\p$ is thus bounded from below. Due to Lorentz
invariance, the vacuum $| 0 \ket$ must have vanishing four-momentum, and
in particular
\begin{equation}
  P^\p | 0 \ket = 0 \; .
\end{equation}
Therefore, the vacuum is an eigenstate of $P^\p$ with the lowest
possible eigenvalue, namely zero. We will be interested in the
phenomenon of spontaneous symmetry breaking, i.e.~in the question
whether---roughly speaking---the vacuum is degenerate. Let us thus
analyse whether there is another state, $| p^\p = 0 \ket$, having the
same eigenvalue, $p^\p = 0$, as the vacuum. If so, it must be possible
to create this state from the vacuum with some operator $U$,
\begin{equation}
  | p^\p = 0 \ket = U | 0 \ket \; , 
\end{equation}
where $U$ must not produce any longitudinal momentum. Note that within
ordinary quantization such a construction is straightforward and quite
common, for example in BCS theory. A state with vanishing
three-momentum can be obtained via
\begin{equation}
  \label{BCS}
  | \vc{p} = 0 \ket = \int d^3 k f(\vc{k}) a^\dagger (\vc{k}) a^\dagger
    (-\vc{k}) | 0 \ket  \; ,
\end{equation}
where $f$ is an arbitrary wave function. Evidently, the contributions
from modes with positive and negative momenta cancel each other. It is
obvious as well, that within light--cone quantization things must be
different as there cannot be an analogous cancellation for the
longitudinal momenta which are always nonnegative. Instead, one could
imagine something like
\begin{equation}
  \label{LC_BCS}
  |p^\p = \vc{p}_\perp = 0 \ket = \int_0^\infty dk^\p \! \int d^2
   k_\perp \, 
   f(\vc{k}_\perp) \delta(k^\p) a^\dagger (k^\p , \vc{k}_\perp)
   a^\dagger(k^\p, - \vc{k}_\perp) | 0 \ket \; .
\end{equation}
The problem thus boils down to the question whether there are Fock
operators carrying light--cone momentum $k^\p = 0$. As we have seen in
Subsection~3.4, there are no such operators, and a construction like
(\ref{LC_BCS}) is impossible.

The only remaining possibility is that, if $U$ contains a creation
operator $a^\dagger (k^\p > 0)$ carrying longitudinal momentum $k^\p \ne
0$, there must be annihilators that annihilate exactly the same amount
$k^\p$ of momentum. Thus, after Wick ordering, $U$ must have the general
form
\begin{eqnarray}
  \label{U}
  U = \bra 0 | U | 0 \ket &+& \int\limits_{k^\p > 0} dk^\p \, f_2 (k^\p)
  a^\dagger (k^\p) a(k^\p)  \nn \\
  &+& \int\limits_{p^\p > 0} dp^\p \int\limits_{k^\p > 0} dk^\p \, 
  f_3 (k^\p , p^\p )
  a^\dagger (p^\p + k^\p) a(p^\p) a(k^\p) \nn \\
  &+& \int\limits_{p^\p > 0} dp^\p \int\limits_{k^\p > 0} dk^\p \, 
  \tilde f_3 (k^\p , p^\p )
  a^\dagger (p^\p ) a^\dagger (k^\p) a(k^\p + p^\p) \nn \\
  &+&  \ldots \; .  
\end{eqnarray}
It follows that the light--cone vacuum $|0\ket$ is an eigenstate of $U$,
\begin{equation}
\label{UVAC}
  U | 0 \ket = \bra 0 | U | 0 \ket | 0 \ket \; .
\end{equation}
As we only deal with rays in Hilbert space, the action of $U$ on the
vacuum does not create a state distinct from the vacuum. One says that
the vacuum is \emph{trivial}. Put differently, `there is no vacuum but
the Fock vacuum'. Note that this is actually consistent with our
findings in the last subsection: the light--cone vacuum is the same
irrespective of the masses of the particles; the Bogolubov
transformation present in the instant form becomes trivial.

Let us analyse the dynamical implications of the general result
(\ref{UVAC}). Any quantity that is obtained by integrating some
functional of the fields over $x^\m$, i.e.,
\begin{equation}
  F[\phi] = \int dx^\m \mathcal{F}[\phi] \; ,
\end{equation}
is of the form (\ref{U}), because the integration can be viewed as a
projection onto the longitudinal momentum sector $k^\p = 0$. The most
important examples for such quantities are the Poincar\'e generators, as
is obvious from the representations (\ref{FF_P}, \ref{FF_M}).  This
implies in particular that the trivial light--cone vacuum is an
eigenstate of the fully interacting light--cone Hamiltonian $P^\m$,
\begin{equation}
  P^\m | 0 \ket = \bra 0 | P^\m | 0 \ket | 0 \ket \; . 
\end{equation}
This can be seen alternatively by considering
\begin{equation}
  P^\p P^\m |0 \ket = P^\m P^\p | 0 \ket = 0 \; ,
\end{equation}
which says that $P^\m | 0 \ket$ is a state with $k^\p = 0$, so that
$P^\m$ must have a Fock representation like $U$ in (\ref{U}).

The actual value of $\bra 0 | P^\m | 0 \ket$ is not important at this
point as it only defines the zero of light--cone energy.  Note that,
within the instant form, the Fock or trivial vacuum is {\em not} an
eigenstate of the full Hamiltonian as the latter usually contains
terms with only creation operators where positive and negative
three-momenta compensate to zero as in (\ref{BCS}). The instant--form
vacuum thus is unstable under time evolution.  Such a vacuum, a
typical example of which is provided by (\ref{BCS}), is called
`nontrivial'.

This concludes the general discussion of light--cone quantum field
theory. Having understood the foundations of the approach we are now
heading for the applications. We begin with a survey of the
light--cone Schr\"odinger equation.

\vfill\eject

\section{Light--Cone Wave Functions}

In this section we collect some basic facts about the eigenvalue
problem of the light--cone Hamiltonian, or, in other words, about the
light--cone Schr\"odinger equation and its solutions, the light--cone
wave functions.  Throughout this section, I will use the conventions
of  \cite{brodsky:89}.

\subsection{Kinematics}

To set the stage for the definition of light--cone wave functions let
me first introduce some relevant kinematical variables. Consider a
system of many particles which, for the time being, will be assumed as
noninteracting. Let the $i^{th}$ particle have mass $m_i$ and
light--cone four--momentum
\begin{equation}
  p_i = (p_i^\p , \vc{p}_{\perp i} , p_i^\m) \; .
\end{equation}
As the particles are free, the total four--momentum is conserved and
thus given by the sum of the individual momenta,
\begin{equation}
  \label{P_SUM}
  P = \sum_i p_i \; .
\end{equation}
The individual particles  are  on--shell, so their four--momentum
squared is
\be
  p_i^2 = p_i^\p p_i^\m - p_{\perp i}^2 = m_i^2 \; .
\ee  
The square of the total four--momentum, on the other hand, defines the
\emph{free invariant mass squared},
\be
\label{FIMS1}
  P^2 = P^\p P^\m - P_\perp^2 \equiv M_0^2 \; ,
\ee
a quantity that will become important later on. We introduce {\em
relative} momentum coordinates $x_i$ and $\vc{k}_{\perp i}$ via
\begin{eqnarray}
  p_i^\p &\equiv& x_i P^\p \; , \\
  \vc{p}_{\perp i} &\equiv& x_i \vc{P}_\perp + \vc{k}_{\perp i} \; .
\end{eqnarray}
Thus, $x_i$ and $\vc{k}_{\perp i}$ denote the longitudinal momentum
fraction and the relative transverse momentum of the $i^{th}$ particle,
respectively. Comparing with (\ref{P_SUM}) we note that these variables
have to obey the constraints
\begin{equation}
  \label{RELMOM_SUM}
  \sum_i x_i = 1 \; , \quad   \sum_i \vc{k}_{\perp i} = 0 \; .
\end{equation}
A particularly important property of  the relative momenta is their boost
invariance. To show this we calculate, using (\ref{P_BOOST}),
\begin{equation}
  \label{X_INV}
  x_i^\prime = e^\omega p_i^\p / e^\omega P^\p = x_i \; .
\end{equation}
From this and (\ref{PERP_BOOST}) we find in addition
\begin{equation}
  \vc{k}_{\perp i}^\prime = \vc{p}_{\perp i}^\prime - x_i
  \vc{P}_{\perp}^\prime 
  = \vc{p}_{\perp i}+ \vc{v}_\perp p_i^\p - x_i (\vc{P}_\perp
  + \vc{v}_\perp P^\p ) = \vc{k}_{\perp i} \; ,
\end{equation}
which indeed proves the frame independence of $x_i$ and $\vc{k}_{\perp
  i}$.

Let us calculate the total light--cone energy of the system in terms of
the relative coordinates. Making use of the constraints
(\ref{RELMOM_SUM}), we obtain
\begin{eqnarray}
  P^\m &=& \sum_i p_i^\m = \sum_i \frac{p_{\perp i}^2 +
  m_i^2}{p_i^\p} 
  = \sum_i \frac{(x_i \vc{P}_\perp + \vc{k}_{\perp i})^2 + m_i^2}{x_i
  P^\p} \nn \\
  &=& \frac{1}{P^\p}\left( P_\perp^2 + \sum_i \frac{k_{\perp i}^2 +
  m_i^2}{x_i} \right) \equiv P_{\mathrm{CM}}^\m  + P_{\mathrm{r}}^\m 
  \label{SEP_HAM}\; .
\end{eqnarray}
This is another important result: the light--cone Hamiltonian $P^\m$
separates into a center--of--mass term,
\begin{equation}
  \label{P_CM}
  P_{\mathrm{CM}}^\m = P_\perp^2 / P^\p \; ,
\end{equation}
and a term containing only the relative coordinates,
\begin{equation}
  \label{P_REL}
  P_{\mathrm{r}}^\m = \frac{1}{P^\p}\left( \sum_i \frac{k_{\perp i}^2 +
  m_i^2}{x_i} \right) = \frac{M_0^2}{P^\p} \; .
\end{equation}
The second identity, which states that $P_r^\m$ is essentially the free
invariant mass squared, follows upon multiplying (\ref{SEP_HAM}) by
$P^\p$,
\begin{equation}
  \label{FIMS2}
  P^\p P_r^\m = P^\p P^\m - P_\perp^2 = M_0^2 = \sum_i \frac{k_{\perp i}^2 +
  m_i^2}{x_i} \; . 
\end{equation}
To simplify things even more, one often goes to the `transverse rest
frame'  where $\vc{P}_\perp$ and therefore the center--of--mass
Hamiltonian $P_{\mathrm{CM}}^\m$ from (\ref{P_CM}) vanish. 

In the interacting case, the dynamical Poincar\'e generators acquire
`potential' terms as I have shown in Section~3.1. The light--cone
Hamiltonian, e.g.~becomes $P^\m = P_0^\m + V$, leading to a
four--momentum squared 
\be
  P^2 = P^\p (P_0^\m + V) - P_\perp^2 \equiv M^2 \; .
\ee
Subtracting (\ref{FIMS1}) we obtain the useful relation,
\be
\label{LC_SEQ}
  M^2 - M_0^2 = P^\p V \equiv W \; .
\ee
In the quantum theory, this operator identity, when applied to physical
states, is nothing but the light--cone Schr\"odinger equation. 
   
Summarizing we note that the special behavior under boosts together
with the transverse Galilei invariance leads to frame independent
relative coordinates and a separation of the center--of--mass motion,
reminiscent of ordinary nonrelati\-vis\-tic physics. This is at
variance with the instant form, where the appearance of the notorious
square root in the energy, $P^0 = (\vc{P}^2 + M_0^2)^{1/2}$, prohibits
a similar separation of variables.

\subsection{Definition of Light--Cone Wave Functions}

Let us first stick to a discrete notation and, for the time being, stay
in 1+1 dimensions. We thus have a Fock basis of states
\begin{eqnarray}
  && | n \ket = a_n^\dagger | 0 \ket \; , \nn \\
  && | m , n \ket = a_m^\dagger a_n^\dagger |0 \ket \; , \nn \\
  && \vdots \nn \\
  && | n_1 , \ldots , n_N \ket = a_{n_1}^\dagger \ldots a_{n_N}^\dagger | 0 \ket
  \; . \label{FOCK_BASIS}
\end{eqnarray}
This leads to a completeness relation defining the unit operator in Fock
space, 
\begin{eqnarray}
  \Eins &=& | 0 \ket \bra 0 | + \sum_{n > 0} | n \ket \bra n | +
  \sfrac{1}{2} \sum_{m,n > 0} | m,n \ket \bra m,n | + \ldots \nn \\ 
  &=&  | 0 \ket \bra 0 | + \sum_{N = 1}^\infty \frac{1}{N!} \sum_{n_1,
  \ldots , n_N >0} |n_1 , \ldots , n_N \ket \bra n_1 , \ldots , n_N | \; .
\end{eqnarray}
An arbitrary state $|\psi \ket$ can thus be expanded as
\begin{equation}
  |\psi  \ket = \sum_{n>0} \bra n | \psi \ket | n \ket +
   \sfrac{1}{2} \sum_{m,n > 0} \bra m,n | \psi \ket | m,n \ket + \ldots
   \; .
\end{equation}
The sums are such that the longitudinal momenta in each Fock sector add
up to the total longitudinal momentum of $|\psi \ket$.  Note that the
vacuum state does not contribute as it is orthogonal to any particle
state, $\bra 0 | \psi \ket = 0$. The normalization of this state is
obtained as
\begin{eqnarray}
  \bra \psi | \psi \ket &=& \sum_{n>0} |\bra n | \psi \ket|^2 +
  \sfrac{1}{2} \sum_{m ,n >0} |\bra m ,n | \psi \ket|^2 + \ldots \nn \\
  &=& \sum_{N=1}^\infty \frac{1}{N!} \sum_{n_1 , \ldots , n_N > 0} |\bra
  n_1 , \ldots , n_N | \psi \ket|^2 \; .
\end{eqnarray}
Let us assume that the state $|\psi \ket$ corresponds to a bound state
obeying the light--cone Schr\"odinger equation derived from
(\ref{LC_SEQ}), 
\begin{equation}
  (M^2 - \hat M_0^2) | \psi \ket = \hat W | \psi \ket \; .
\end{equation}
We want to project this equation onto the different Fock sectors. For
this we need the eigenvalues of the free invariant mass squared when
applied to an $N$-particle state
\begin{equation}
  | N \ket \equiv | n_1 , \ldots n_N \ket \; .
\end{equation}
We find that $|N \ket$ is an eigenstate of $\hat M_0^2$,
\begin{equation}
  \hat M_0^2 |N \ket =  \sum_{i=1}^N
  \frac{m_i^2}{x_i} | N \ket \equiv M_{N}^2  | N \ket \; .
\end{equation}
The light--cone Schr\"odinger equation thus becomes a system of coupled
eigenvalue equations,
\begin{equation}
  \label{GEN_LCBSE} 
  \left \lceil  \begin{array}{cl}
                  (M^2 - M_1^2) & \bra l | \psi \ket \\
                  (M^2 - M_2^2) & \bra k l | \psi \ket \\
                  \vdots & 
               \end{array} \right \rceil = 
  \left \lceil \begin{array}{r@{}c@{}l r@{}c@{}l c}
                \bra l | & W & | m \ket & \bra l | & W & | m n \ket &
                \ldots  \\
                \bra k l |& W & | m \ket & \bra k l |& W & | m 
                n \ket & \ldots  \\
                & \vdots & & & \vdots & & \ddots 
              \end{array} \right \rceil 
  \left \lceil \begin{array}{c}
                \bra l | \psi \ket \\
                \bra k l | \psi \ket \\
                \vdots    
             \end{array} \right \rceil \; . 
\end{equation}
Clearly, this represents an infinite number of equations which in
general will prove impossible to solve unless the matrix is very
sparse and/or the matrix elements are small. The former condition is
usually fulfilled as the interaction $W$ in the light--cone
Hamiltonian typically changes particle number at most by
two\footnote{Note that terms with only creation operators are
forbidden by $k^\p$--conservation.  Still, in 1+1 dimensions, things
can become messy as interactions with polynomials of arbitrary powers
in $\phi$ are allowed without spoiling renormalizability
\cite{zinn-justin:96}}. Assuming the matrix elements to be small
amounts to dealing with a perturbative situation. This will be true
for nonrelativistic bound states of heavy constituents, but not for
light hadrons which we are mainly interested in. There are, however,
situations where the magnitude of the amplitudes decreases enormously
with the particle number $N$, so that it is a good approximation to
restrict to the lowest Fock sectors. In instant form field theory this
has long been known as the Tamm--Dancoff method
\cite{tamm:45,dancoff:50}.

Let us turn to the more realistic case of 3+1 dimensions in a
continuum formulation.  The invariant normalization of a momentum
eigenstate $|P^\p, \vc{P}_\perp \ket \equiv | \vcg{P}
\ket $ is given by
\begin{equation}
  \label{MOM_NORM}
  \bra \vcg{P} | \vcg{K} \ket = 16 \pi^3 P^\p
  \delta^3 (\vcg{P} - \vcg{K})  \; .
\end{equation}
We already know that the bare Fock vacuum is an eigenstate of the
interacting Hamiltonian. It thus serves as an appropriate ground state
on top of which we can build a reasonable Fock expansion. If we
specialize immediately to the case of QCD, we are left with the Fock
basis states
\begin{eqnarray}
  && |0 \ket \; , \nn \\
  && | q \bar q : \vcg{k}_i , \alpha_i \ket = b^\dagger
  (\vcg{k}_1 , \alpha_1) d^\dagger (\vcg{k}_2 , \alpha_2)
  |0 \ket \; , \\
  && | q \bar q g : \vcg{k}_i , \alpha_i \ket = b^\dagger
  (\vcg{k}_1 , \alpha_1) d^\dagger (\vcg{k}_2 , \alpha_2)
  a^\dagger (\vcg{k}_3 , \alpha_3) |0 \ket \; , \nn \\
  && \vdots
\end{eqnarray}
In these expressions, $b^\dagger$, $d^\dagger$ and $a^\dagger$ create
quarks $q$, antiquarks $\bar q$ and gluons $g$ with momenta
$\vcg{k}_i$ from the trivial vacuum $|0 \ket$. The $\alpha_i$ denote
all other relevant quantum numbers, like helicity, polarization,
flavor and color.

In a more condensed notation we can thus describe, say, a pion with
momentum $\vcg{P} = (P^\p, \vc{P}_\perp )$, as
\begin{equation}
  \label{GEN_PI_WF}
  | \pi (\vcg{P}) \ket = \sum_{n , \lambda_i} \int \overline{\prod_i}
    dx_i \frac{d^2 k_{\perp i}}{16\pi^3} \psi_{n/\pi} 
    (x_i, \vc{k}_{\perp i}, \lambda_i ) \Big| n: x_i
    P^\p , x_i \vc{P}_\perp + \vc{k}_{\perp i} ,
    \lambda_i \Big \rangle  \; ,
\end{equation}
where we have suppressed all discrete quantum numbers apart from the
helicities $\lambda_i$. The integration measure takes care of the
constraints (\ref{RELMOM_SUM}) which the relative momenta in each
Fock state (labeled by $n$) have to obey,
\begin{eqnarray}
  \overline{\prod_i} dx_i &\equiv& \prod_i dx_i \, \delta \bigg(1 - \sum_j
  x_j \bigg)  \; , \\ \relax
  \overline{\prod_i} d^2 k_{\perp i} &\equiv& 16 \pi^3 \prod_i   
  d^2k_{\perp i} \, \delta^2 \bigg(\sum_j \vc{k}_{\perp j} 
  \bigg) \; . 
\end{eqnarray}
As a mnemonic rule, we note that any measure factor $d^2 k_{\perp i}$
in (\ref{GEN_PI_WF}) is always accompanied by $1 / 16\pi^3$.

The most important quantities in (\ref{GEN_PI_WF}) are the
\emph{light--cone wave functions}
\begin{equation}
  \psi_{n/\pi} (x_i, \vc{k}_{\perp i}, \lambda_i) \equiv \bra n: x_i
  P^\p , x_i \vc{P}_\perp + \vc{k}_{\perp i} , \lambda_i |\pi
  (\vcg{P}) \ket \; ,
\end{equation}
which are the amplitudes to find $n$ constituents with relative momenta
$p_i^\p = x_i P^\p$, $\vc{p}_{\perp i} = x_i \vc{P}_\perp +
\vc{k}_{\perp i}$ and helicities $\lambda_i$ in the pion. Due to the
separation properties of the light--cone Hamiltonian the wave functions
do not depend on the total momentum $\vcg{P}$ of the pion. Applying
(\ref{MOM_NORM}) to the pion state (\ref{GEN_PI_WF}), we obtain the
normalization condition
\begin{equation}
  \label{LCWF_NORM}
  \sum_{n , \lambda_i} \int \overline{\prod_i}  dx_i  \frac{d^2 k_{\perp
  i}}{16\pi^3}   |\psi_{n/\pi} (x_i,
  \vc{k}_{\perp i} , \lambda_i) |^2 = 1 \; .
\end{equation}
The light--cone bound--state equation for the pion is a straightforward
generalization of (\ref{GEN_LCBSE}),  
\begin{equation}
  \label{PI_LCBSE} 
  \left \lceil  \begin{array}{cl}
                  (M^2 - M_{q \bar q}^2) & \bra q \bar q| \pi \ket  \\
                  (M^2 - M_{q \bar q g}^2) & \bra q \bar q g | \pi \ket \\
                  \vdots & 
               \end{array} \right \rceil = 
  \left \lceil \begin{array}{r@{}c@{}l r@{}c@{}l c}
                \bra q \bar q | & W & | q \bar q \ket & \bra q \bar q |
                & W & | q \bar q g \ket &
                \ldots  \\
                \bra q \bar q g |& W & | q \bar q \ket & \bra q \bar q g
                | & W & | q \bar q g \ket & \ldots  \\
                & \vdots & & & \vdots & & \ddots 
              \end{array} \right \rceil 
  \left \lceil \begin{array}{c}
                \bra q \bar q | \pi \ket \\
                \bra q \bar q g | \pi \ket \\
                \vdots    
             \end{array} \right \rceil \; . 
\end{equation}
If a constituent picture for the pion were true, the valence state
would dominate,
\be
  |\psi_{2/\pi}|^2 \gg |\psi_{n/\pi}|^2 \; , \quad n > 2 \; , 
\ee
and, in the extreme case, the pion wave function would be entirely
given by the projection $\bra q \bar q| \pi \ket$ onto the valence
state. All the higher Fock contributions would vanish and the
unitarity sum (\ref{LCWF_NORM}) would simply reduce to
\begin{equation}
  \label{VAL_NORM}
  \sum_{\lambda \lambda^\prime}  \int_0^1 dx \int \frac{d^2
  k_\perp}{16\pi^3}   |\psi_{q \bar q/\pi} (x,
  \vc{k}_{\perp }, \lambda , \lambda^\prime ) |^2 = 1 \; .
\end{equation}
We will later discuss a model where this is indeed a good approximation
to reality.

\subsection{Properties of Light--Cone Wave Functions}

Let us rewrite the light--cone bound--state equation (\ref{GEN_LCBSE}) by
collecting all light--cone wave functions $\psi_n = \bra n | \psi \ket$
into a  vector $\Psi$, 
\begin{equation}
  \label{PSI}
  \Psi  = \frac{W \Psi}{M^2 - M_0^2} \; .
\end{equation}
From this expression it is obvious that all light--cone wave functions
tend to vanish whenever the denominator
\begin{equation}
  \epsilon \equiv M^2 - M_0^2 = M^2 - \Big( \sum_i p_i \Big)^2 =  
  M^2 - \sum_i \frac{k_{\perp i}^2 +  m_i^2}{x_i} 
\end{equation}
becomes very large. This quantity measures how far off energy shell
the total system, i.e.~the bound state is,
\begin{equation}
  P^\m - \sum_{i}p_i^\m = \epsilon/P^\p \; .
\end{equation}
For this reason, $\epsilon$ is sometimes called the `off--shellness'
\cite{namyslowski:85,lavelle:87}.  We thus learn from (\ref{PSI}) that
there is only a small overlap of the bound state with Fock states that
are far off shell. This implies the limiting behavior
\begin{equation}
  \label{WF_BC}
  \psi(x_i , \vc{k}_{\perp i} , \lambda_i) \to 0 \quad \mbox{for} \;
  x_i \to 0 \; , \; k_{\perp i}^2 \to \infty \; .
\end{equation}
These boundary conditions are related to the self--adjointness of the
light--cone Hamiltonian and to the finiteness of its matrix elements.
Analogous criteria have been used recently to relate wave functions of
different Fock states $n$ \cite{antonuccio:97a} and to analyse the
divergence structure of light--cone perturbation theory
\cite{burkardt:98}.

Omitting spin, flavor and color degrees of freedom, a light--cone
wave function will be a scalar function $\phi(x_i , \vc{k}_{\perp i})$ of
the parameter $\epsilon$. This is used for building models, the most
common one being to assume a Gaussian behavior, originally suggested
by \cite{terentev:76}, 
\begin{equation}
  \phi(x_i , \vc{k}_{\perp i}) = N \exp(-|\epsilon|/\beta^2) \; ,
\end{equation}
where $\beta$ measures the size of the wave function in momentum space.
Note, however, that a Gaussian ansatz is in conflict with perturbation
theory which is the appropriate tool to study the high--$\vc{k}_\perp$
behavior and indicates a power decay of the {\em renormalized} wave
function --- up to possible logarithms \cite{brodsky:89}. For the
unrenormalized wave functions the boundary conditions (\ref{WF_BC}) are
violated unless one uses a cutoff as a regulator (see next section).

As the off--shellness $\epsilon$ is the most important quantity
characterizing a light--cone wave function let us have a closer look by
specializing to the simplest possible system, namely two bound particles
of equal mass. One can think of this, for instance,  as the valence
wave function of the pion or positronium. The off--shellness becomes
\begin{equation}
  \label{2P_OFFS}
  \epsilon = M^2 - \frac{k_\perp^2 + m^2}{x (1-x)} 
  = - \frac{1}{x(1-x)} \Big[ M^2 \big(x - \sfrac{1}{2}\big)^2 
  + \underbrace{\frac{4m^2 -
  M^2}{4}}_{\ge 0} + k_\perp^2 \Big] \; .
\end{equation}
The second term in square brackets is positive because, for a bound
state, the binding energy,
\begin{equation}
  \label{EB}
  E = M - 2m \; , 
\end{equation}
is negative so that $2m > M$. As a result, the off--shellness is always
negative. Only for free particles it is zero, because {\em all} momentum
components (including the energy) sum up to the total momentum. In this
case, each individual term in (\ref{2P_OFFS}) vanishes,
\begin{equation}
  M = 2m \; , \quad x = \sfrac{1}{2} \; , \quad \vc{k}_\perp = 0 \; .
\end{equation}
It follows that the light--cone wave function of a two--particle system
(composed of equal--mass constituents) with the binding energy
switched off `adiabatically', is of the form
\begin{equation}
  \phi (x , \vc{k}_\perp ) \sim \delta (x - 1/2) \, \delta^2 (\vc{k}_\perp)
  \; .
\end{equation}

\subsection{Examples of Light--Cone Wave Functions}

From the discussion above, one expects that for weak binding, in
particular for nonrelativistic systems, the wave functions will be
highly peaked around $ x= 1/2$ (in the equal mass case) and
$\vc{k}_\perp = 0$. Let us check this explicitly for hydrogen--like
systems which constitute our first example \cite{lepage:81}.

\subsubsection*{Example 1: Hydrogen--like Systems.}

Let me recall the ordinary Schr\"odinger equation of the Coulomb
problem written in momentum space,
\be
  \left( E - \frac{\vc{p}^2}{2m} \right) \psi (\vc{p}) = \int
  \frac{d^3 k}{(2 \pi)^3} V(\vc{p} - \vc{k}) \psi (\vc{k}) = \int
  \frac{d^3 k}{(2 \pi)^3} \frac{4 \pi \alpha}{(\vc{p} - \vc{k})^2}
  \psi(\vc{k}) \; .
\ee
This integral equation looks very similar to a light--cone
Schr\"odinger equation within a two--particle truncation. The Coulomb
kernel is due to the exchange of an instantaneous photon having a
propagator proportional to $\delta (x^0)$. One can actually solve the
Coulomb problem directly in momentum space \cite{fock:35,bethe:57} but
for our purposes it is simpler just to Fourier transform the ground
state wave function $\psi_0 (r) = N \exp (-m \alpha r)$, yielding
\be
\label{HYDRO_GROUND}
  \psi_0 (\vc{p}) = 8 \pi N  \frac{m \alpha}{(\vc{p}^2
  + m^2 \alpha^2)^2} \; ,
\ee
with $\alpha = e^2/4\pi = 1/137$ being the fine structure constant,
$m$ the reduced mass and $\vc{p}$ the relative momentum.

How does this translate into the light--cone language? To answer this
question, we go to the particle rest frame with $\vc{P} = 0$ or $P^\p
= P^\m = M$ and $\vc{P}_\perp = 0$, implying $\vc{p}_{\perp i} =
\vc{k}_{\perp i}$. In this frame, the nonrelativistic limit is
defined by the following inequalities for the constituent masses and
momenta (in ordinary instant--form coordinates),
\begin{equation}
  \label{HIERARCHY}
  p_i^0 - m_i \simeq \frac{\vc{p}_i^2}{2 m_i}   \ll |\vc{p}_i| 
  \ll m_i \; .
\end{equation}
The prototype systems in this class are of hydrogen type where we have
for binding energy and r.m.s.~momentum,
\begin{eqnarray}
  |E| &=&  \frac{\bra \vc{p}^2 \ket}{2m} = \frac{m\alpha^2}{2} \; , \\
  \bra p \ket &=& m \alpha  \; .
\end{eqnarray}
In this case, the hierarchy (\ref{HIERARCHY}) becomes
\begin{equation}
  \frac{\alpha^2}{2} \ll \alpha \ll 1 \; ,
\end{equation}
which is fulfilled to a very good extent in view of the smallness of
$\alpha$. 

Consider now the longitudinal momentum of the $i^{th}$ constituent,
\begin{equation}
  p_i^\p = p_i^0 + p_i^3 \simeq m_i + \frac{\vc{p}_i^2}{2m_i} + p_i^3 
  = x_i P^\p =   x_i M \; .
\end{equation}
We thus find that we should replace $p_i^3$ in instant--form
nonrelativistic wave functions by
\begin{equation}
  \label{REPLACE}
  p_i^3 = x_i M - m_i \; ,
\end{equation}
where we neglect terms of order $\vc{p}_i^2 / m_i$.  Let us analyze the
consequences for the off--shellness. The latter is in ordinary
coordinates
\begin{equation}
  \label{IF_OFFS}
  \epsilon = M^2 - M_0^2 = (M + \sum_i p_i^0)(M - \sum_i p_i^0) \; .
\end{equation}
We thus need 
\begin{equation}
  \sum_i p_i^0 \simeq \sum_i m_i + \sum_i \frac{\vc{p}_i^2}{2m_i} = M -
  E + \sum_i \frac{\vc{p}_i^2}{2m_i} \; ,
\end{equation}
with $E = M - \sum m_i$ denoting the mass--defect, which is a small
quantity, $E \ll M$. The off--shellness (\ref{IF_OFFS}), therefore,
becomes
\begin{equation}
  \epsilon \simeq 2M \bigg(E - \sum_i \vc{p}_i^2 / 2m_i \bigg) \simeq
  - 2M  \sum_i \frac{k_{\perp i}^2 + (Mx_i - m_i)^2}{2m_i}  \; ,
\end{equation}
where we have performed the replacement (\ref{REPLACE}) in the second
identity. The light--cone wave functions will be peaked where the
off--shellness is small, that is, for
\begin{equation}
\label{NR_XK}
  x_i = m_i / M \, , \quad \mathrm{and} \quad \vc{k}_{\perp i} = 0 \; ,
\end{equation}
as expected from the noninteracting case. 

To be explicit, we consider the ground state wave function of
positronium, given by (\ref{HYDRO_GROUND}) with the reduced
mass $m$ being half the electron mass $m_e$.  Using the replacement
prescription (\ref{REPLACE}) once more, we obtain
\begin{equation}
  \label{PSI_POS}
  \psi (x , \vc{k}_\perp) = 8 \pi N  \frac{m \alpha}{\Big[k_\perp^2  +
  (xM - m_e)^2 + m^2 \alpha^2 \Big]^2} \; , 
\end{equation}
where $M \simeq 2 m_e$ is the bound state mass. This result is valid for
small momenta, i.e.~when $k_\perp^2, (xM - m_e)^2 \ll m_e^2$. It is
obvious from (\ref{PSI_POS}) that the positronium wave function is
sharply peaked around $x = m_e /M \simeq 1/2$ and $k_\perp^2 = 0$.

\subsubsection*{Example 2: 't~Hooft Model.}

The `t~Hooft model \cite{thooft:74,thooft:75} is QCD in two
space--time dimensions with the number $N_C$ of colors being
infinite. The Lagrangian is
\be 
  {\cal L} = \bar \psi (i \partial \!\!\!/ - m ) \psi -
  \frac{1}{4} F_{\mu\nu} F^{\mu\nu} \; .
\ee
The limit of large $N_C$ is taken in such a way that the expression
$g^2 N_C$, $g$ denoting the Yang--Mills coupling, stays finite. In two
dimensions, $g$ has mass dimension one, which renders the theory
superrenormalizable and provides a basic unit of mass, namely,
\be
\label{BASIC_SCALE}
  \mu_0 \equiv \sqrt{g^2 N_C / 2\pi} \; .
\ee
The model is interesting because it contains physics analogous or
similar to what one finds in `real' QCD\footnote{For a recent review on the
`t~Hooft model, see \cite{abdalla:96}.}. First of all, the model is
(trivially) confining due to the linear rise of the Coulomb potential
in 2$d$. This is most easily exhibited by working in light--cone
gauge, $A^\p = 0$, and eliminating $A^\m$ via Gauss's law. In this
way it becomes manifest that there are no dynamical gluons in
2$d$. Within covariant perturbation theory, the Coulomb potential can
be understood as resulting from the exchange of an instantaneous gluon.

In the next section, we will discuss the spontaneous breakdown of
chiral symmetry in QCD. It turns out that in the `t~Hooft model a
similar phenomenon occurs: chiral symmetry is `almost' spontaneously
broken \cite{witten:78,zhitnitsky:86}. As a consequence, there arises
a {\em massless} bound state in the chiral limit of vanishing quark
mass \cite{thooft:74,thooft:75} which we will call the `pion' for
brevity\footnote{As explained in \cite{witten:78,zhitnitsky:86}, this
is not in contradiction with Coleman's theorem \cite{coleman:73c} as
the `pion' is not a Goldstone boson.}. Furthermore, there is a
nonvanishing quark condensate in the model which has first been
calculated by \cite{zhitnitsky:86},
\be
\label{COND_THOOFT}
  \cond / N_C = -0.28868 \; .
\ee
Note that the condensate is proportional to $N_C$ as it involves a
color trace. 

It turns out that the `t~Hooft model has one big advantage for the
application of light--cone techniques which is due to the large--$N_C$
limit. The matrix elements entering the light--cone Schr\"odinger
equation in the two--particle sector have the following
$N_C$--dependence, 
\be
  \bra 2 | W | 2n \ket \sim \left( \frac{g^2 N_C}{N_C} \right)^n \sim
  N_C^{-n} \; .
\ee
As a result, those diagrams which correspond to a change in particle
number (like $2 \to 4$, $2 \to 6, \ldots$) are suppressed by powers of
$1/N_C$. The truncation to the two-particle sector therefore becomes
exact: the `pion' is a pure quark--antiquark state; there are no
admixtures of higher Fock states. A constituent picture is thus
realized, and we are left with the light--cone Schr\"odinger equation,
\be
\label{BSEQ1}
  \left[ M^2 - \frac{m^2}{x (1-x)} \right] \phi (x) = {\cal P}
 \int_0^1 dy \, \frac{\phi(x) - \phi(y)}{(x-y)^2} \; .
\ee
This expression defines the `Coulomb problem' of the `t~Hooft model.
It corresponds to the first line of (\ref{PI_LCBSE}) where, as a
result of the truncation, only the matrix element $W_2
\equiv \bra q \bar q | W | q \bar q \ket$ has been retained.  We will
refer to (\ref{BSEQ1}) as the 't~Hooft equation in what
follows. $\phi(x)$ denotes the valence part of the `pion' wave
function, $x$ and $y$ are the momentum fractions of the two quarks
(with equal mass $m$) in the meson.  The symbol $ {\cal P}$ indicates
that the integral is defined as a principal value
\cite{thooft:74,gelfand:64}. It regularizes\footnote{\cite{wu:77} has
suggested a theoretical alternative to the principal value which
nowadays is called `Leibbrandt--Mandelstam prescription
\cite{mandelstam:83,leibbrandt:84}. It leads to completely different
physics. This apparent contradiction has only recently been clarified
\cite{bassetto:00}.} the Coulomb singularity $1/(x-y)^2$ in the matrix
element $W_2$. All masses are expressed as multiples of the basic
scale $\mu_0$ defined in (\ref{BASIC_SCALE}).  The eigenvalue $M$
denotes the mass of the lowest lying bound state (the `pion').  Our
objective is to calculate $M$ and $\phi$.

In his original work on the subject, \cite{thooft:74,thooft:75},
't~Hooft used the following ansatz for the wave function
\begin{equation}
  \phi (x) = x^\beta (1-x)^\beta \, .
  \label{THO_ANS}
\end{equation}
This  ansatz  is  symmetric  in $x \leftrightarrow  1-x$  (charge
conjugation odd), and $\beta$ is supposed to lie between zero and
one  so  that  the  endpoint  behavior  is  nonanalytic.  As  a
nontrivial  boundary condition, one has the exact solution of the
massless case, $m = 0$ (the `chiral limit'),
\begin{equation}
  M^2  = 0  \; , \quad  \mbox{and}  \quad  \phi  (x)  = 1 \; ,
  \quad \mbox{{\it i.e.}} \quad \beta = 0 \; . 
  \label{MASSLESS} 
\end{equation}
In this limit, the `pion' wave function is constant, i.e.~the pion has
no structure and is point-like.  The main effect of a nonzero quark
mass is the vanishing of the wave functions at the endpoints implying
a nonzero $\beta$.  This suggests the following series expansion for
$\beta$,
\begin{equation}
  \beta (m) = \beta_1 m + \beta_2 m^2 + \beta_3 m^3 + O(m^4) \; ,
  \label{BETA_EXP}
\end{equation}
and for the `pion' mass squared,
\begin{equation}
  M^2 = M_1 m + M_2 m^2 + M_3 m^3 + O(m^4) \; .  
  \label{M2_EXP}
\end{equation}
As is obvious from the last two expressions, we are working in the
limit of small quark mass $m$. The expansion (\ref{BETA_EXP}) shows
that also $\beta$ is small in this case so that the wave function will
be rather flat (for intermediate values of $x$). On the other hand we
know from Example~1 that light--cone wave functions are highly peaked
near $x=1/2$ in case the binding is weak. This suggests that, for
small quark mass $m$, the `pion' is rather strongly bound.

The exponent $\beta$ in (\ref{THO_ANS}) can actually be determined
exactly by studying the small-$x$ behavior of the bound state equation
(\ref{BSEQ1}).  To this end we evaluate the principal value integral
for $x \to 0$ and plug it into (\ref{BSEQ1}). This yields the
transcendental equation \cite{thooft:74},
\begin{equation}
  m^2 - 1 + \pi \beta \cot \pi \beta = 0 \; .
  \label{COT}
\end{equation}
Using this expression we can determine $\beta$ either numerically for
arbitrary $m$ or analytically for small $m$, which yields the
coefficients of (\ref{BETA_EXP}),
\begin{equation}
  \beta = \frac{\sqrt{3}}{\pi} m  + O(m^3) \; . 
  \label{BETA_EXC} 
\end{equation}
The `pion' mass is determined by calculating the expectation value of
the light--cone Hamiltonian (\ref{BSEQ1}) in the state given by
`t~Hooft's ansatz (\ref{THO_ANS})\footnote{The relevant integrals can
be found in \cite{bardeen:80} and \cite{harada:94}.}. This yields
$M^2$ as a function of $\beta$ and $m$,
\be
\label{EIGENVALUE1}  
  M^2 = \frac{2}{\beta_1} m + O(m^2)  \; . 
\ee
Upon comparing with (\ref{BETA_EXP}) and (\ref{BETA_EXC}) the lowest
order coefficient $M_1$, which is the slope of $M^2 (m)$ at $m=0$, is
found to be
\be
\label{M_1}
  M_1 = 2 \pi / \sqrt{3} \; .
\ee
Note that $M^2$ indeed vanishes in the chiral limit. 

Having obtained an approximate solution for the mass and wave function
of the `pion' we are in the position to calculate `observables'. It
turns out that, for small $m$, all of them can be expressed in terms
of the lowest order coefficient, $M_1$. We begin with the 'pion decay
constant' \cite{callan:76b,zhitnitsky:86}, which is given by the `wave
function at the origin', i.e.~the integral over the (momentum space)
wave function
\be
\label{F_PI_THOOFT}
  f_\pi \equiv \bra 0 | \bar \psi i \gamma_5 \psi | \pi \ket =
  \sqrt{\frac{N_C}{\pi}} \frac{M^2}{2m} \int dx \, \phi(x) =
  \sqrt{\frac{N_C}{4 \pi}} M_1 \; .
\ee
The quark condensate is obtained via a sum rule using the chiral Ward
identity \cite{zhitnitsky:86,heinzl:96b},
\be
\label{COND_M_1}
  \cond = -m \frac{f_\pi^2}{M^2} = - \frac{N_C}{4\pi} M_1 \; .
\ee
Inserting the value (\ref{M_1}) for $M_1$ this coincides with
(\ref{COND_THOOFT}). The last identity can actually be viewed as the
`Gell-Mann--Oakes--Renner relation' \cite{gell-mann:68} of the `t~Hooft
model,
\be
   M^2 = - 4 \pi \,  m \, \cond / N_C + O (m^2) \; ,
\ee
which provides a relation between the particle spectrum (the `pion'
mass) and a \emph{ground state} property (the condensate). This is
conceptually important because it implies that we can circumvent the
explicit construction of a nontrivial vacuum state by calculating the
spectrum of excited states, i.e.~by solving the light--cone
Schr\"odinger equation. The eigenvalues and wave functions actually
contain information about the structure of the vacuum!  This point of
view has been adopted long ago in the context of chiral symmetry
breaking within the (light--cone) parton model
\cite{casher:71}: ``In this framework the spontaneous symmetry
breakdown must be attributed to the properties of the hadron's wave
function and not to the vacuum'' \cite{casher:74}. Related ideas have
been put forward more recently in \cite{lenz:91}.

In the above, we have been using the value for $\beta$ given in
(\ref{BETA_EXC}). One can equally well use $\beta$ as a variational
parameter and minimize the expectation value of the mass operator with
respect to it. This yields the same result for $M_1$, namely
(\ref{M_1}). The variational method, however, is better suited if one
wants to go beyond the leading order in expansion (\ref{M2_EXP}). This
has been done in \cite{harada:98}. The results are shown in
Table~\ref{II} where we list the expansion coefficients $M_i$ of the
`pion' mass squared, $M^2$, as they change with increasing number of
variational parameters, ($a, b, c, d$).

\begin{figure}[!ht]
  \caption{\protect\label{fig-83} \textsl{The light--cone wave function 
    of the 't~Hooft model `pion' for $m$ = 0.1.  The solid curve
    represents the result from 't~Hooft's original ansatz, the dashed
    curve our best result (with maximum number of variational
    parameters).  At the given resolution, however, the curves of {\em
      all} extensions of 't~Hooft's ansatz ($a, b, c, d$) lie on top of
    each other.}}  
\begin{center}
\epsfig{file=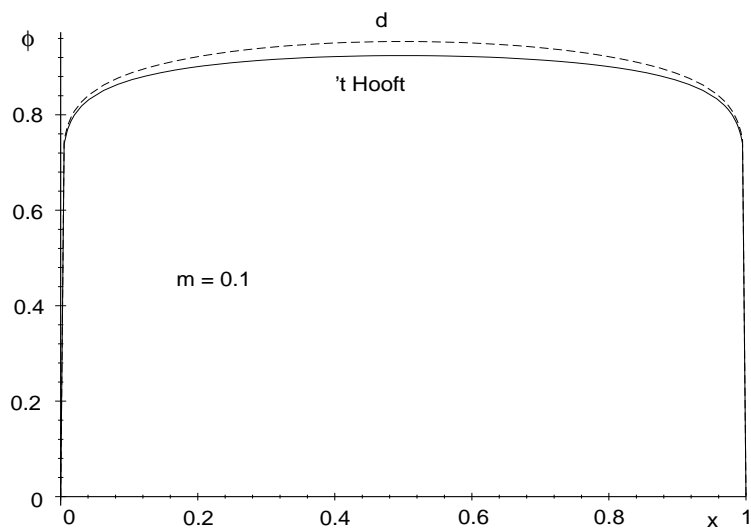,height=7cm,width=10cm}
\end{center}
\end{figure}

\begin{figure}[!tbp]
  \caption{\label{fig:excited-wavefunc} 
    Wave functions of the first eight excited states in the `t~Hooft
    model as obtained via variational methods
    \protect\cite{stern:99}. Note that all wave functions vanish at the
    end points, $x=0$ and $x=1$.}
\vspace{0.5cm} 
  \begin{minipage}{\halfwidth}
    \centering
    \includegraphics[width=\fullwidth]{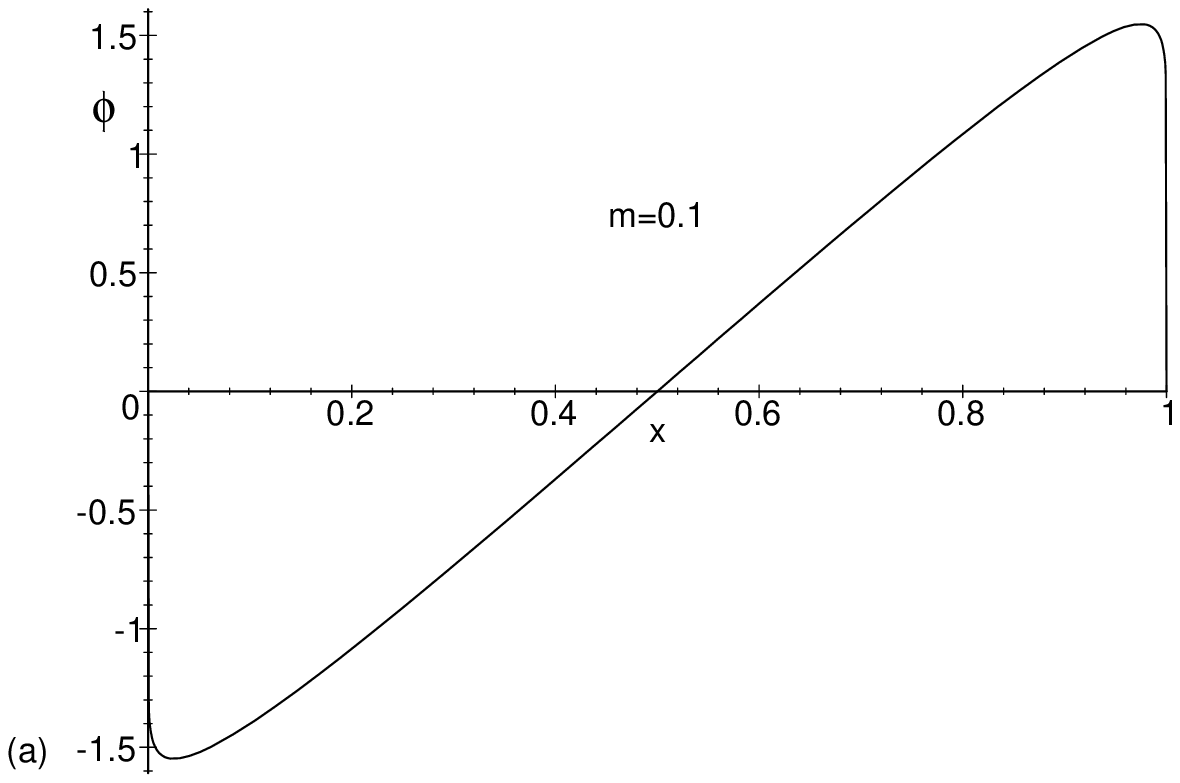}
  \end{minipage}
  \begin{minipage}{\halfwidth}
    \centering
    \includegraphics[width=\fullwidth]{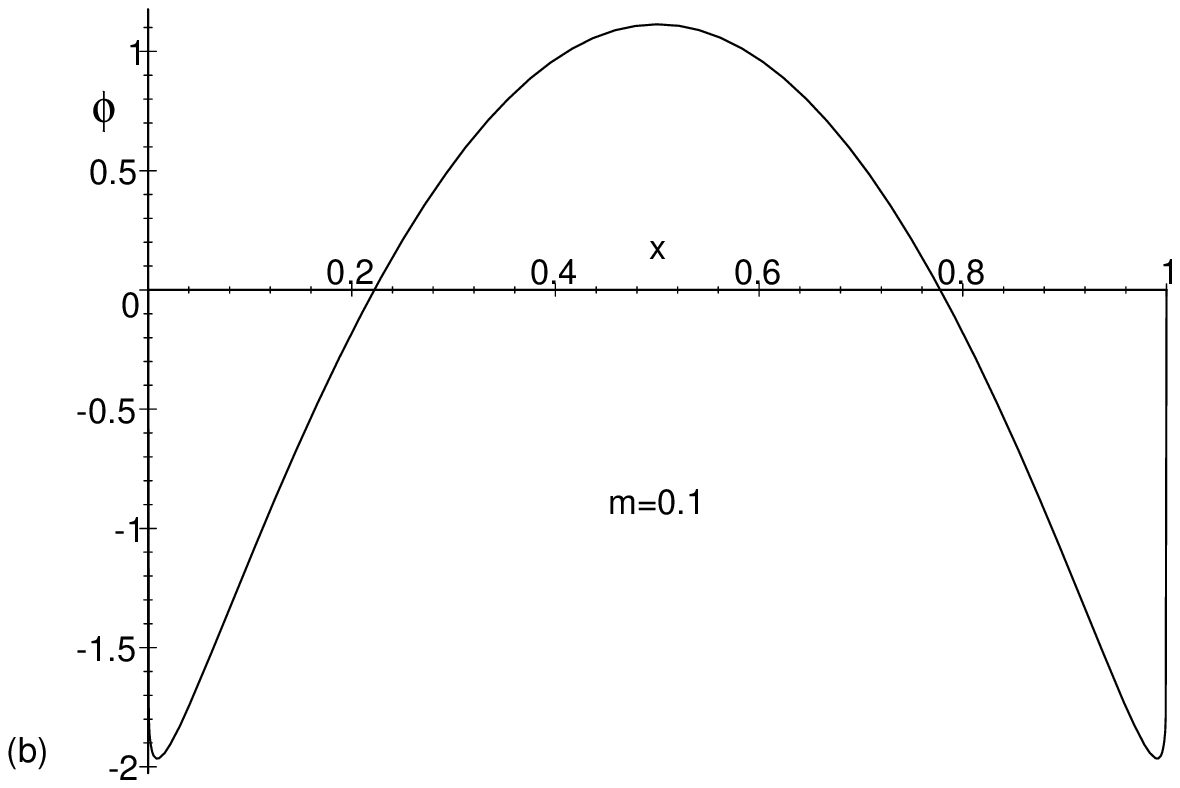}
  \end{minipage}
  \vspace{0.2cm}
  \begin{minipage}{\halfwidth}
    \centering
    \includegraphics[width=\fullwidth]{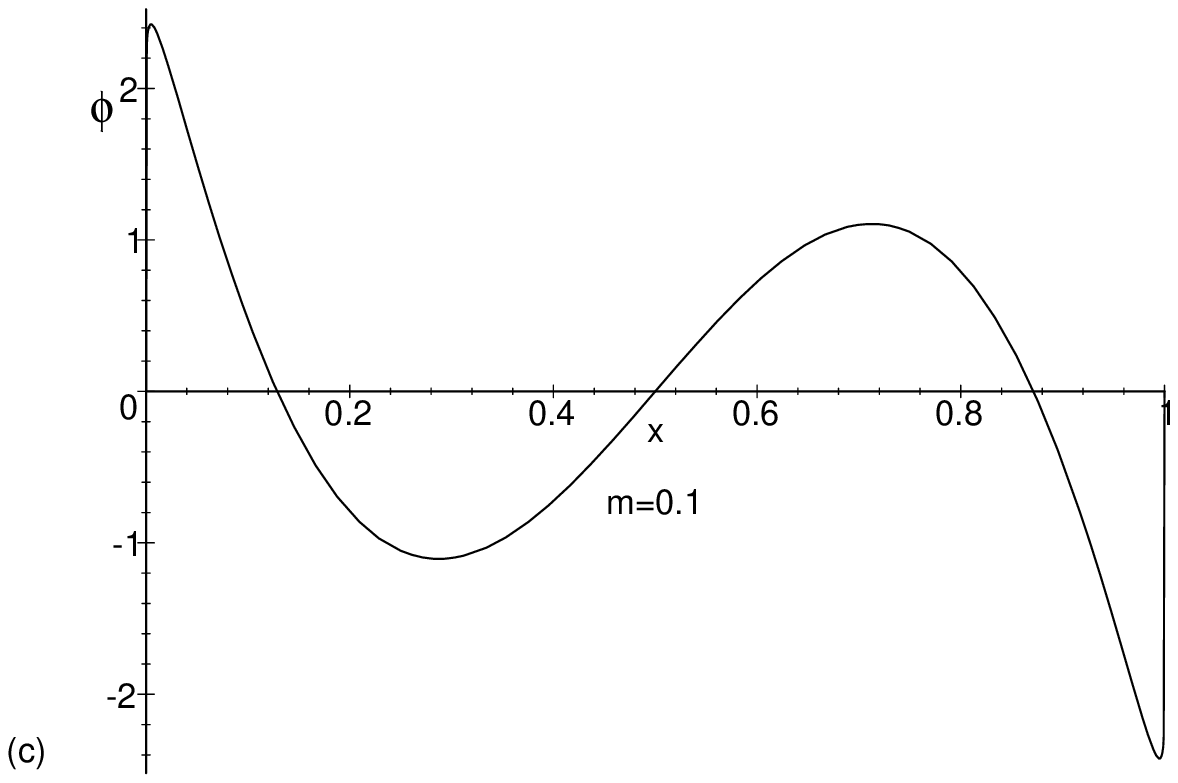}
  \end{minipage}
  \begin{minipage}{\halfwidth}
    \centering
    \includegraphics[width=\fullwidth]{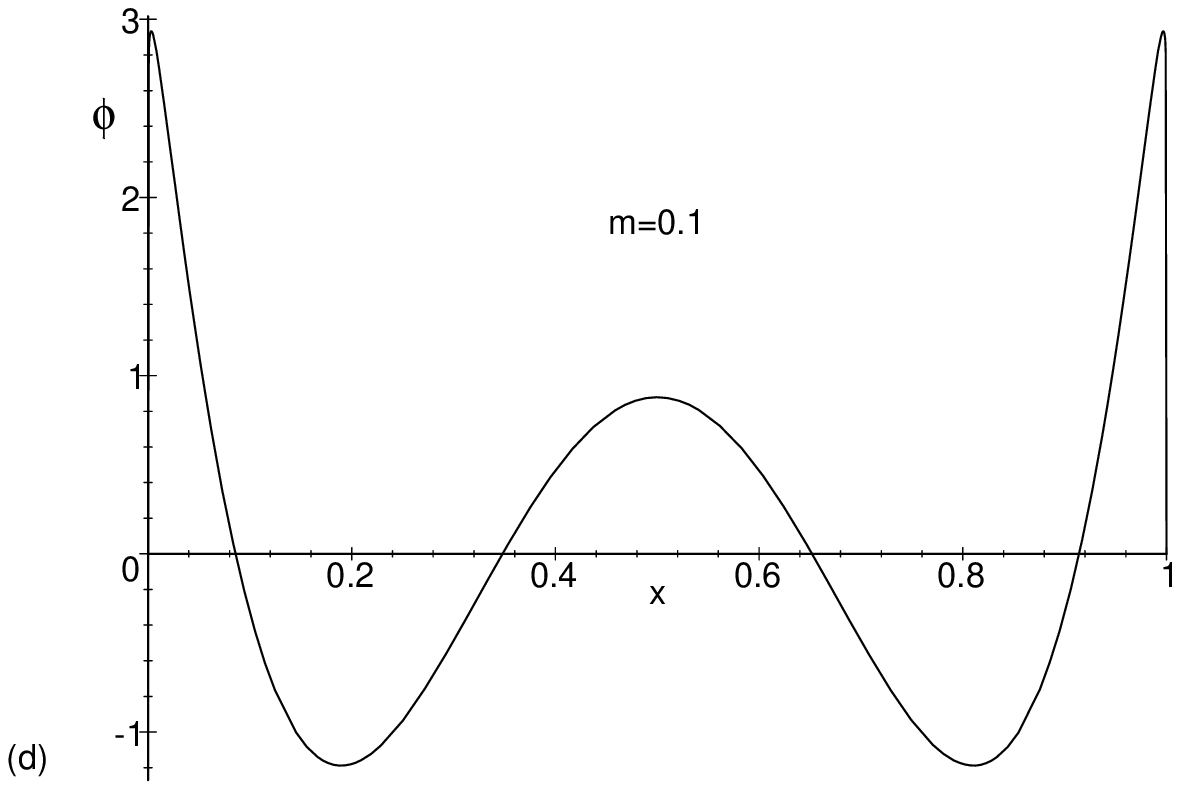}
  \end{minipage}
  \vspace{0.2cm}
  \begin{minipage}{\halfwidth}
    \centering
    \includegraphics[width=\fullwidth]{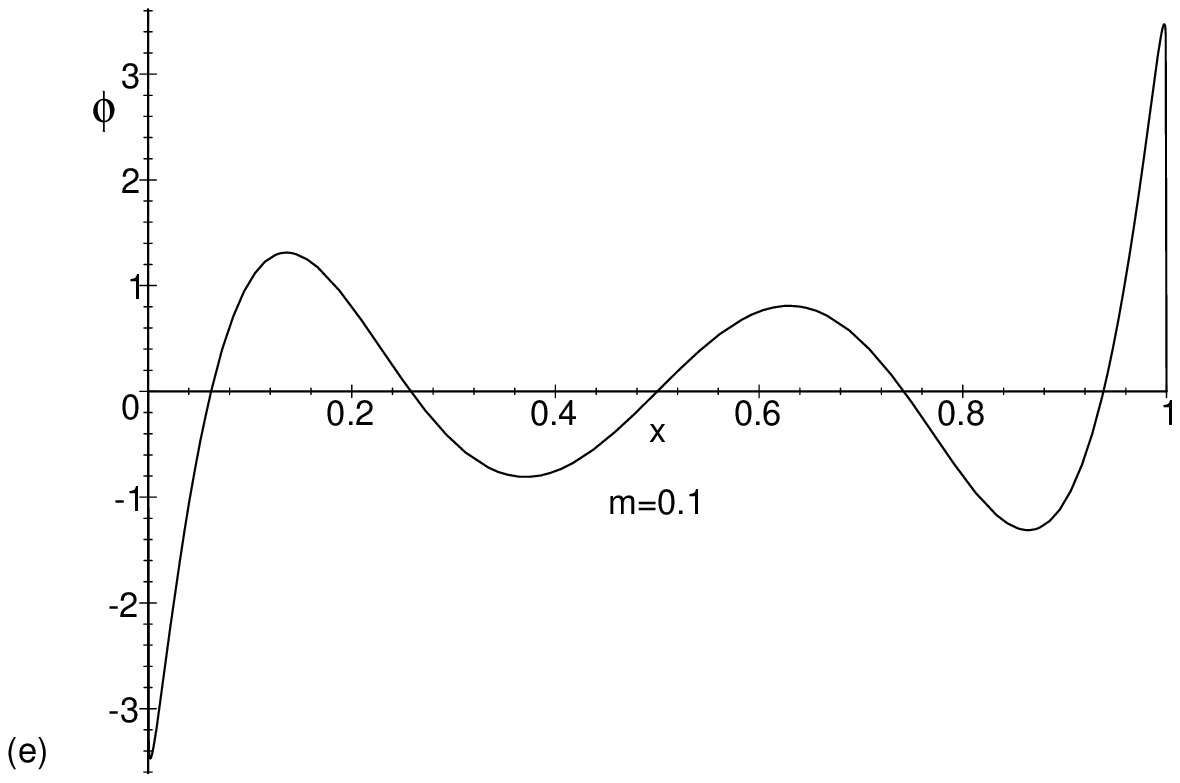}
  \end{minipage}
  \begin{minipage}{\halfwidth}
    \centering
    \includegraphics[width=\fullwidth]{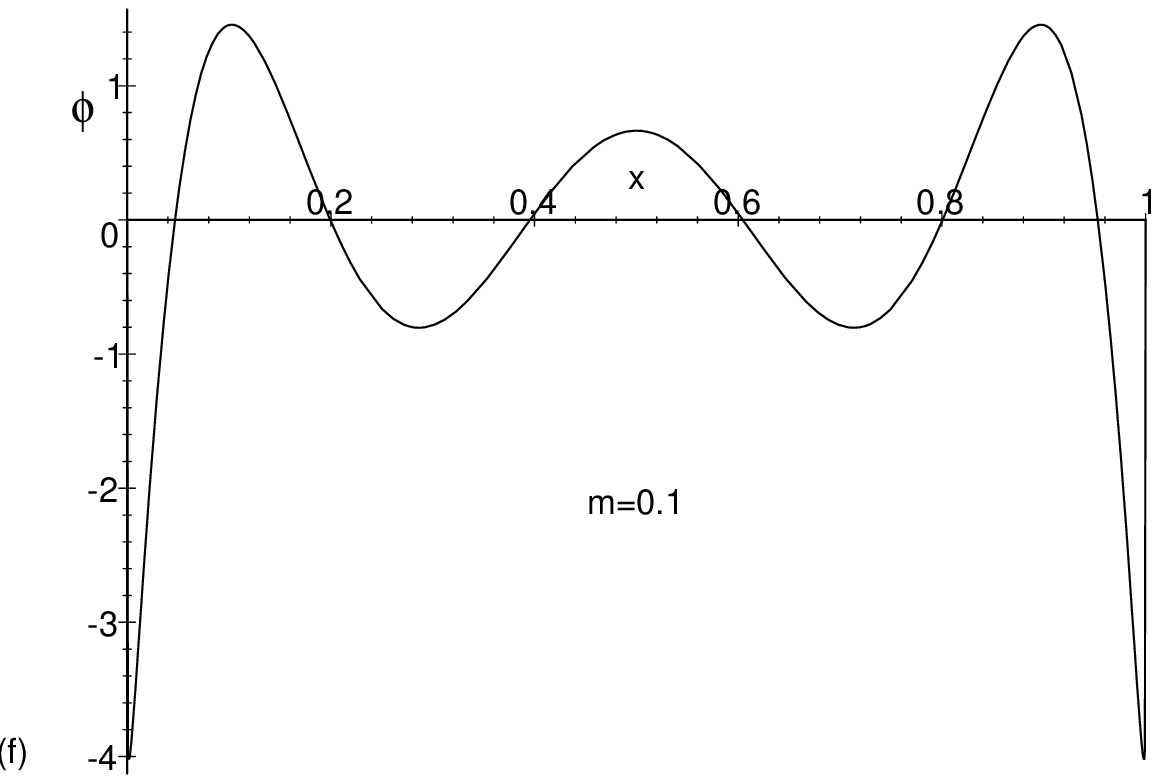}
  \end{minipage}
  \vspace{0.2cm}
  \begin{minipage}{\halfwidth}
    \centering
    \includegraphics[width=\fullwidth]{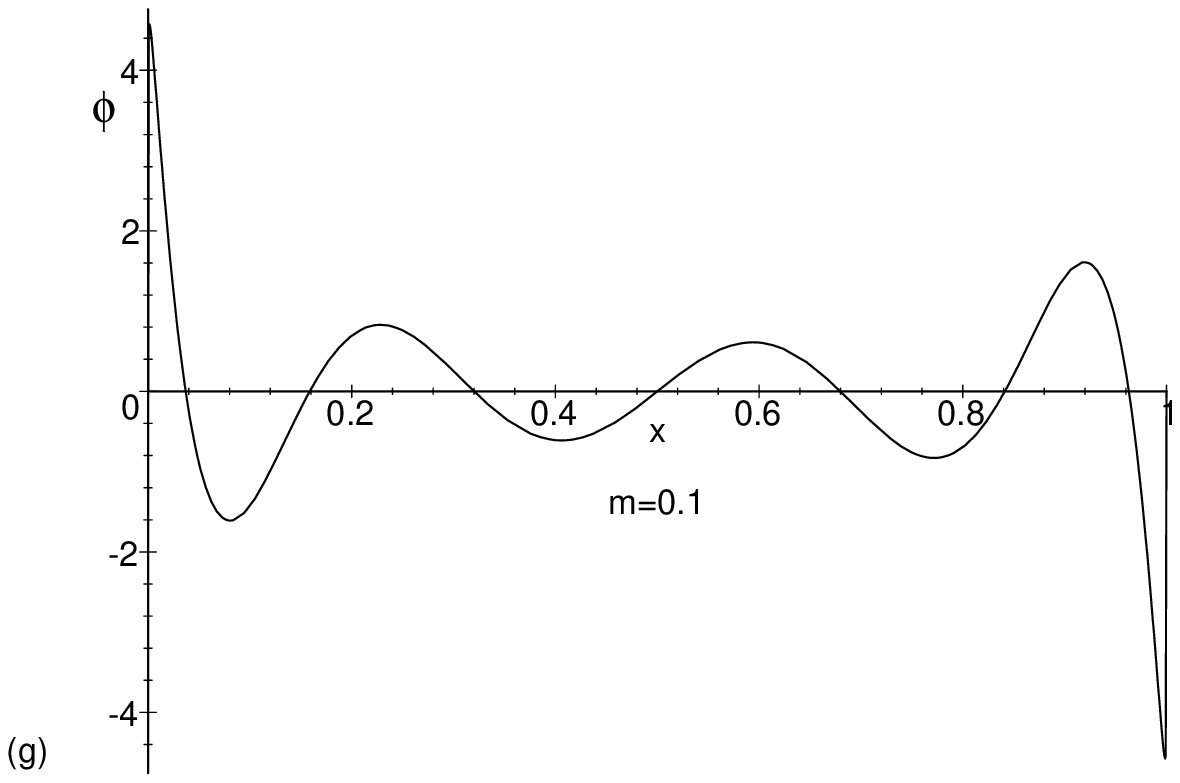}
  \end{minipage}
  \begin{minipage}{\halfwidth}
    \centering
    \includegraphics[width=\fullwidth]{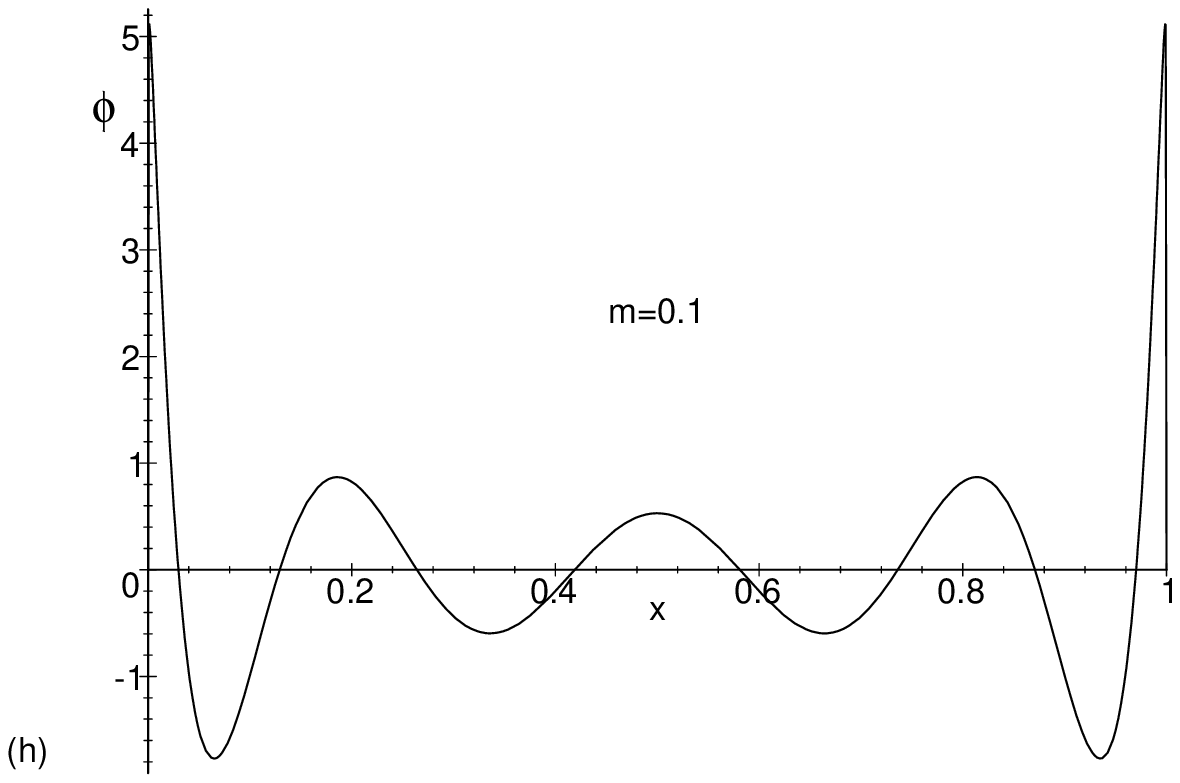}
  \end{minipage}
\end{figure}

Interestingly, the value of $M_1$ does not change at all by enlarging
the space of trial functions. $M_2$ and $M_3$, on the other hand, do
change and show rather good convergence.  For $M_2$ we finally have a
relative accuracy of $8 \cdot 10^{-7}$, and for $M_3$ of $4 \cdot
10^{-5}$.  Furthermore, the coefficients are getting \emph{smaller} if
one adds more basis functions, in accordance with the variational
principle. The associated light--cone wave functions are shown in
Fig.~\ref{fig-83}. There are only minor changes upon including more
variational parameters. In Fig.~\ref{fig:excited-wavefunc} we display
the eight lowest excited states. They habe been obtained using the
position of the nodes as additional variational parameters
\cite{stern:99}.
\begin{table}[!ht]
\renewcommand{\arraystretch}{1.2}
\caption{\label{II} \textsl{Expansion   coefficients   of  $M^2$  for  the
't~Hooft model obtained by successively enlarging the space of
variational parameters.  $M_1$ is the standard 't~Hooft result
(\protect\ref{M_1}). Note the good convergence towards the bottom of
the table.}}
\vspace{.3cm}
\begin{tabular*}{\textwidth}[t]{l @{\extracolsep\fill} c c c }
  \hline\hline
  \it Ansatz & $M_1 = 2\pi/\sqrt{3}$ & $M_2$ & $M_3$ \\
  \hline
  't~Hooft & 3.62759873 & 3.61542218 & 0.043597197  \\ 
  $a$      & 3.62759873 & 3.58136872 & 0.061736701  \\ 
  $b$      & 3.62759873 & 3.58107780 & 0.061805257  \\ 
  $c$      & 3.62759873 & 3.58105821 & 0.061795547  \\ 
  $d$      & 3.62759873 & 3.58105532 & 0.061793082  \\ 
  \hline\hline
\end{tabular*}
\end{table}

\subsubsection*{Example 3: Gaussian Model.}

A very simple and intuitive example of a light--cone wave function is
provided by the Gaussian model in the form presented by \cite{lepage:81},
\be
\label{GAUSSIAN}
  \phi (x , \vc{k}_\perp) = N \exp \left[ - \frac{1}{\Lambda^2}
  \frac{k_\perp^2 + m^2}{x (1-x)} \right] \; .
\ee
This is a scalar two--particle wave function that drops exponentially
with the free invariant mass squared of two constituents with equal
mass $m$. The shortcoming of the model is that the wave function is
not derived dynamically, i.e.~as a solution of a light--cone
Schr\"odinger equation. However, it satisfies all the requirements
derived in Subsection~4.3. The benefit of the model is its
simplicity. The latter is even enhanced if one neglects the
constituent mass $m$ by assuming $m \ll \Lambda$. Then one is left
with only two parameters, the normalization $N$ and the transverse
size $\Lambda$.

One enforces a constituent picture by normalizing to unity,
\be
\label{GAUSSIAN_NORM}
  || \phi (x, \vc{k}_\perp ) ||^2 \equiv \int_0^1 dx \int \frac{d^2
  k_\perp}{16 \pi^3} \phi^2 (x, \vc{k}_\perp) = \frac{N^2
  \Lambda^2}{192 \pi^2} \stackrel{!}{=} 1 \; . 
\ee
This obviously relates $N$ and $\Lambda$. If we now calculate the
r.m.s.~transverse momentum in the bound state described by
(\ref{GAUSSIAN}), we find using (\ref{GAUSSIAN_NORM}), 
\be
\label{K_PERP_GAUSSIAN}
  \bra k_\perp^2 \ket \equiv \int_0^1 dx \int \frac{d^2
  k_\perp}{16 \pi^3} \, k_\perp^2 \, \phi^2 (x, \vc{k}_\perp) = \frac{N^2
  \Lambda^4}{1920 \pi^2} = \Lambda^2 / 10 \; .
\ee
If we view the Gaussian wave function as a crude model for, say, the
pion we can actually estimate the width parameter $\Lambda$.  The pion
is highly relativistic, so we expect its r.m.s.~transverse momentum to
be of the order of the constituent quark  mass, $\bra k_\perp^2
\ket^{1/2} \simeq m_Q \simeq 330$  MeV. This leads to a typical width of
$\Lambda \simeq 1$ GeV.

Having thus determined the two parameters of the model we could go on
and calculate observables \cite{lepage:81}. This will actually be done
in the next section using a more realistic model for the (determination
of) the pion wave function.

\section{The Pion Wave Function in the NJL Model}

The ultimate goal of light--cone field quantization is to derive and
solve light--cone Schr\"odinger equations, in particular the one of
QCD. This would yield hadron  masses and light--cone wave functions and
thus detailed information on the internal structure of mesons and
baryons. In order to successfully pursue this program, a number of
problems has to be overcome. 

In dealing with gauge theories in a Hamiltonian framework one has to
solve Gauss's law together with the light--cone specific
constraints. The only known solution so far is in the light--cone
gauge $A^\p = 0$ \cite{tomboulis:73,casher:76,lepage:81} which,
however, is beset by infrared problems of its own---see
e.g.~\cite{bassetto:91a}\footnote{Recently, some progress has also
been made in the covariant gauge \cite{srivastava:99}.}. A Hamiltonian
formulation analogous to the Weyl gauge, where one fixes the residual
gauge freedom and solves Gauss's law \emph{after} quantization, seems
particularly difficult \cite{heinzl:95b}.

Even after successful light--cone quantization of QCD one encounters a
severe problem: the theory has to be renormalized. Otherwise,
light--cone wave function will not be normalizable and physically
meaningful. Renormalization in a Hamiltonian framework presents
enormous difficulties as there is no explicit covariance. In addition,
rotational invariance is not manifest within the front form (see
Section~2). Due to the lack of these important symmetries, there is an
abundance of possible counterterms which even can be nonlocal,
e.g.~behave like $ \sim 1/k^\p$. As a result, to the best of my
knowledge, the renormalization program has not been extended beyond
one loop --- with one notable exception in QED \cite{brodsky:73b}.

The issue of renormalization is, of course, delicately intertwined
with solving the light--cone bound state equation. The latter attempt
will in general only be feasible if a Tamm--Dancoff truncation in
particle number is performed \cite{harindranath:90}. This again
violates important symmetries (even gauge symmetry). One hopes,
however, that a Wilsonian renormalization group explicitly taylored
for this case will restore these symmetries \cite{wilson:94}. The
present status of this program is nicely reviewed in \cite{perry:99}.

A conceptual problem has already been mentioned. The light--cone
vacuum is trivial as argued in Section~3. On the other hand, it is
well known that the instant form vacuum is populated by all sorts of
nonperturbative quantum fluctuations leading to nonvanishing vacuum
expectation values or condensates. The whole business of QCD sum rules
\cite{shifman:79} is based upon this picture. How do we
reconcile this with the triviality of the light--cone vacuum? A
possible resolution to this `triviality problem' has been given in the
last section. The spectrum of excited states actually carries implicit
information on the structure of the vacuum. The task then is to make
this information \emph{ex}plicit.

The list of problems just given is yet another manifestation of the
`principle of conservation of difficulties'. In a first attempt to
tackle these problems I will simply side--step most of them by
considering an (instructive) model instead of QCD. This model,
however, is designed to capture some important physical features of
`real' QCD. The idea is originally due to Nambu and Jona-Lasinio
(NJL), who, back in 1961, invented a ``Dynamical Model of Elementary
Particles Based on an Analogy with Superconductivity''
\cite{nambu:61a}. It was meant to provide a microscopic mechanism for
the generation of \emph{nucleon} masses, with the mass gap being the
analog of the BCS energy gap in a superconductor. Nowadays, with the
nucleons replaced by quarks, the model serves as a low--energy
effective theory of QCD explaining the spontaneous breakdown of chiral
symmetry. Let me thus give a brief introduction to the latter
phenomenon before I come to the detailed explanation of the model.

\subsection{A Primer on Spontaneous Chiral Symmetry Breaking}

If we have a look at Table \ref{T2}, which provides a list of all
quark flavors, we realize that there are large differences in the
quark masses as they appear in the QCD Hamiltonian. In particular,
there is a hierarchy,
\begin{equation}
  \underbrace{m_u , m_d \ll m_s}_{\mbox{light quarks}} \ll
  \underbrace{m_c, m_b, m_t}_{\mbox{heavy quarks}} \; .
\end{equation}
As the masses of heavy and light quarks are separated by the very same
scale ($\simeq$ 1 GeV) as the perturbative and nonperturbative regime,
one expects different physics associated with those two kinds of
quarks. This expectation turns out to be true. The physics of heavy
quarks is governed by a symmetry called `heavy quark symmetry' leading
to a very successful `heavy quark effective theory'
\cite{neubert:94}. The physics of light quarks, on the other hand, is
governed by chiral symmetry which we are now going to explain.
\begin{table}
\begin{center}
\renewcommand{\arraystretch}{1.2}
\caption{\label{T2} \textsl{The presently observed quark flavors. $Q/e$ is the
  electric charge in units of the electron charge. The (scale
  dependent!) quark masses are given for a scale of 1 GeV.}}
\vspace{.5cm}

\begin{tabular*}{\textwidth}[h]{ @{\extracolsep\fill} l c c r c }
\hline\hline
flavor & & $Q/e$ & mass & \\
\hline
down    & $d$ & $-1/3$ & 10  & MeV \\
up      & $u$ & $+2/3$ & 5   & MeV \\
strange & $s$ & $-1/3$ & 150 & MeV \\
\hline
charm   & $c$ & $+2/3$ & 1.5 & GeV \\
bottom  & $b$ & $-1/3$ & 5.1 & GeV \\
top     & $t$ & $+2/3$ & 180 & GeV \\
\hline\hline
\end{tabular*}
\end{center}
\end{table}

Let us write the QCD Hamiltonian in the following way,
\begin{equation}
  \label{QCD_HAM}
  H_{\mathrm{QCD}} = H_\chi + \bar \psi \mathcal{M} \psi \; , 
\end{equation} 
${\cal M} = diag(m_u , m_d, m_s)$ being the mass matrix for the light
flavors. To a good approximation, one can set ${\cal M} = 0$. In
this case, the QCD Hamiltonian $H_\chi$ is invariant under the
symmetry group $SU(3)_R \otimes SU(3)_L$, the chiral flavor
group. Under the action of this group, the left and right handed
quarks independently undergo a chiral rotation. Due to Noether's
theorem, there are sixteen conserved quantities, eight vector charges
and, more important for us, eight pseudo-scalars, the chiral charges
$Q_5^a$ satisfying
\begin{equation}
  [Q_5^a , H_\chi ] = 0 \; .
\end{equation}
This states both that the chiral charges are conserved, and that
$H_\chi$ is chirally invariant. Under parity, $Q_5^a \to -
Q_5^a$. Now, if $|A \ket$ is an eigenstate of $H_\chi$, so is $Q_5^a
|A \ket$ with the same eigenvalue. Thus, one expects (nearly)
degenerate parity doublets in nature, which, however, do not exist
empirically. The only explanation for this phenomenon is that chiral
symmetry is spontaneously broken. In contradistinction to the
Hamiltonian, the QCD ground state (the vacuum) is {\em not} chirally
invariant,
\begin{equation}
  Q_5^a | 0 \ket \ne 0 \; . 
\end{equation}
For this reason, there must exist a nonvanishing vacuum expectation
value, the quark condensate,
\begin{equation}
  \bra \bar \psi \psi \ket = \bra \bar \psi_R \psi_L + \bar \psi_L
  \psi_R \ket  \; .
\end{equation}
This condensate is not invariant (it mixes left and right) and
therefore serves as an order parameter of the symmetry breaking. Note
that in QCD the quark condensate is a renormalization scale dependent
quantity. A recent estimate can be found in \cite{dosch:98}, with a
numerical value,
\begin{equation}
  \cond (1 \; \mathrm{GeV}) \simeq (-229 \; \mathrm{MeV})^3 \; .
\end{equation}
The spontaneous breakdown of chiral symmetry thus implies that the
(QCD) vacuum is nontrivial: it must contain quark--antiquark pairs
with spins and momenta aligned in a way consistent with vacuum quantum
numbers. A possible analog is the BCS ground state given in
(\ref{BCS}).

In terms of the full quark propagator,
\begin{equation}
  S(p) =  \frac{p \!\!\!/ + M(p)}{p^2 - M^2(p)} \; ,
\end{equation}
where we have allowed for a momentum dependent (or `running') mass
$M(p)$, the quark condensate is given by
\begin{equation}
\label{COND_PROP}
  \cond = - i \int \frac{d^4 p}{(2\pi)^4} \tr S(p) = - 4i N_C \int \frac{d^4
    p}{(2\pi)^4} \frac{M(p)}{p^2   - M^2 (p)} \; .
\end{equation}
We thus see that the involved Dirac trace yields a nonvanishing
condensate only if the effective quark mass $M(p)$ is nonzero.  This
links the existence of a quark condensate to the mechanism of
dynamical mass generation. In this way we have found another argument
that the bare quarks appearing in the QCD Hamiltonian (\ref{QCD_HAM})
indeed acquire constituent masses. Of course we are still lacking a
microscopic mechanism for that.

Goldstone's theorem \cite{goldstone:61} now states that for any
symmetry generator which does not leave the vacuum invariant, there
must exist a massless boson with the quantum numbers of this
generator. This results in the prediction that in massless QCD one
should have an octet of massless pseudoscalar mesons. In reality one
finds what is listed in Table~\ref{T3}.

\begin{table}
\begin{center}
\renewcommand{\arraystretch}{1.2}
\caption{\protect\label{T3} \textsl{Masses of the  pseudoscalar octet
    mesons (in MeV).}} 
\vspace{.5cm}

\begin{tabular*}{\textwidth}[h]{ @{\extracolsep\fill} l  c c c c c }
\hline\hline
meson & $\pi^0$ & $\pi^\pm$ & $K^0$, $\bar K^0$ & $K^\pm$ & $\eta$ \\
\hline
mass  & 135 & 140 & 500 & 494 & 549 \\
\hline\hline
\end{tabular*}
\end{center}
\end{table}

The nonvanishing masses of these mesons are interpreted as stemming
from the nonvanishing quark masses in the QCD Hamiltonian which break
chiral symmetry explicitly. Being small, they can be treated as
perturbations. 

It should be stressed that chiral symmetry has nothing to say about
the mechanism of confinement which presumably is a totally different
story.  This is also reflected within a QCD based derivation of chiral
symmetry breaking in terms of the instanton model, as presented
e.g.~in \cite{diakonov:95}. This model explains many facts of
low--energy hadronic physics but is known not to yield confinement. It
is therefore possible that confinement is not particularly relevant
for the understanding of hadron structure
\cite{diakonov:95}.

The instanton vacuum actually leads to an effective theory very close
to the NJL model, the basics of which are our next topic.

\subsection{NJL Folklore}

Before I consider the light--cone formulation of the model, let me
briefly recall its main physical features\footnote{For recent reviews,
see \cite{vogl:91,klevansky:92,hatsuda:94}.}.  In its
standard form, the NJL model has a chirally invariant four--fermion
interaction, which can be imagined as the result of `integrating out'
the gluons in the QCD Lagrangian.  For simplicity I concentrate on the
case of one flavor. Extension to several flavors is
straightforward. In the chiral limit (quark mass $m_0 = 0$), the
Lagrangian is
\be
  {\cal L} = \bar \psi i \partial \!\!\!/ \psi - G (\bar \psi
  \gamma_\mu \psi)^2 \equiv {\cal L}_0 + {\cal L}_{\mathrm{int}}
  \; .  
\ee
Its four--fermion interaction is chirally symmetric under $U(1)_L
\times U(1)_R$. We shall see in a moment that this symmetry is 
spontaneously broken.

It is important to observe that the coupling $G$ of the model has
negative mass dimension, $[G] = -2$, hence, it is not
renormalizable. Accordingly, it requires a cutoff which is viewed as a
parameter of the model that is to be fixed by phenomenology. We thus
follow the general spirit of effective field theory
\cite{holstein:95,kaplan:95,manohar:96}. From the point of view of the
light--cone formulation to be developed later, the
nonrenormalizability is an advantage: it enables us to `circumvent'
the difficulties of the light--cone renormalization program. There
simply is no `need' to renormalize.

Following the standard approach \cite{nambu:61a,vogl:91} we treat the
model in mean--field approximation (which actually coincides with the
large--$N$ limit). We begin with a Fierz transformation by
schematically rewriting the interaction Lagrangian in Fierz symmetric
form
\be
\label{FIERZ}
  {\cal L}_{\mathrm{int}} = G \sum_i c_i (\bar \psi \Gamma_i
  \psi)^2 = G (\bar \psi \psi)^2 + \ldots \; ,
\ee
where $i = S,P,V,A$ enumerates the different Dirac bilinears (scalar,
pseudoscalar, vector and axial vector, respectively). In
(\ref{FIERZ}), I have only displayed the scalar part of the
interaction because only the scalar density $S \equiv \bar \psi \psi$
can have a vacuum expectation value, the quark condensate,
\be
  \bra S \ket = \cond \; .
\ee
Let us determine this quantity in mean--field approximation. To define
the latter we calculate, 
\be
  S^2 = (S - \bra S \ket + \bra S \ket)^2 = (S - \bra S \ket)^2 + 2 S
  \bra S \ket - \bra S \ket^2 \simeq 2 S \bra S \ket + const \; .
\ee
Thus, by neglecting quadratic fluctuations of $S$ around its
expectation value, we linearize the interaction and obtain the
mean--field Lagrangian,
\be
\label{MF_LAG}
  {\cal L}_{\mathrm{MFA}} = \bar \psi (i \partial\!\!\!/ + 2G \bra S
  \ket ) \psi \; .
\ee
The mean--field solution has a very intuitive explanation. One
essentially argues that the main effect of the interaction is to
generate the mass of the quarks which become quasi--particles that
interact only weakly. Neglecting this interaction entirely, one can
view the process of mass generation as the transition of quarks with
mass $m_0 = 0$ to mass $m$ resulting in a mass term $m
\bar \psi \psi$ in the Lagrangian (\ref{MF_LAG}). We thus find that
the dynamically generated mass is determined by the \emph{gap
equation}
\be
\label{GAP}
  m = -2 G \cond \; .
\ee
How do we actually calculate the condensate $\cond$? To this end we go
back to (\ref{COND_PROP}) and express the condensate in terms of the
full propagator `at the origin', i.e.~at space-time point $x = 0$,
\be
\label{COND_PROP2}
  \cond_m = -i \, \tr S_F(x = 0) = -i \, \tr \int \frac{d^4 p}{(2 \pi)^4}
  \frac{1}{p \!\!\!/ - m + i \epsilon} \; .
\ee
The following remarks are in order. First we note that within
mean--field approximation the dynamically generated mass $m$ is
constant, i.e.~independent of the momentum $p$. Furthermore, the full
propagator $S_F$ is defined in terms of the constituent mass $m$, so
that the gap equation (\ref{GAP}) becomes an implicit
self--consistency condition where the mass $m$ to be determined
appears on both sides. This equation will be solved in a moment. The
integral appearing on the r.h.s.~of (\ref{COND_PROP2}) is
quadratically divergent\footnote{As the condensate involves the
product of field operators at coinciding space-time points, this
clearly is a short--distance singularity.} and has to be regulated. If
we use a cutoff $\Lambda$ we have, for dimensional reasons,
\be
\label{COND_DIM}
  \cond \sim \Lambda^2 m \; .
\ee
A particularly intuitive way to calculate the condensate (which will
also be used in the light--cone case) is based on the
Hellmann--Feynman theorem. This states that `the derivative of an
expectation value is the expectation value of the derivative', if the
expectation is taken between normalizable states. If applied to the
vacuum expectation value of our mean--field Hamiltonian,
\be
  {\cal H} (m) = {\cal H}_0 + m \bar \psi \psi \; , 
\ee
which is the vacuum energy density ${\cal E} (m) \equiv \bra 0 | {\cal H}
(m) | 0 \ket $, the theorem yields,
\be
  \cond_m = \pad{}{m} {\cal E} (m) = - \pad{}{m} \int \frac{d^3
  p}{(2 \pi)^3} 2 \sqrt{\vc{p}^2 + m^2} \; .
\ee
The factor of two in the integrand is due to the fermion spin
degeneracy. If we choose a noncovariant three--vector cutoff,
$|\vc{p}| < \Lambda_3$, the result for the condensate is of the form
(\ref{COND_DIM}),
\be
\label{COND_IF_EVAL}
  \cond_m = - \frac{m \Lambda_3^2}{2 \pi^2} \left[ 1 + \frac{\delta^2}{2} \ln
   \delta^2 + O (\delta^2) \right ] \; , 
\ee
with $\delta \equiv m/\Lambda_3$. The dynamical mass is found by
inserting this result into the gap equation (\ref{GAP}). A nontrivial
solution $m = m(G) \ne 0$ arises above a critical coupling $G_c$ which
is determined by the identity $m(G_c) = 0$. It is known from the
theory of critical phenomena that the mean--field approximation leads
to a square root behaviour of the mass around $G_c = 4 \pi^2 /
\Lambda^2$,
\be
\label{G_CRIT}
  m(G) \sim (G - G_c)^{1/2} \; , \quad G \ge G_c \; .
\ee
We thus have seen that the NJL model describes the transformation of
bare quarks of mass $m_0 (=0)$ into dressed (or constituent) quarks
($Q$) of mass $m \ne 0$. By considering the bound state equation in
the pseudoscalar channel one can also verify Goldstone's mechanism:
there is a massless $Q\bar Q$ bound state, which is identified with
the pion, exactly if the gap equation holds
\cite{nambu:61a,vogl:91}. This concludes the presentation of
spontaneous chiral symmetry breaking within the model. One should keep
in mind that it follows the same pattern as in QCD.

Let me conclude the NJL `crash course' by reemphasizing that the model
does not confine. This means in particular, that there is a
nonvanishing probability for mesons to decay into constituent
quarks. We thus cannot expect to obtain reliable estimates for strong
decay widths of mesons or other quantities that are not dominated by
their chiral properties.

It is getting time to discuss the light--cone formulation of the
model. For a `light--cone physicist' the model is interesting for
several reasons.  I have already pointed out that the lack of
renormalizability is welcome because we only have to worry about a
proper regularization. Furthermore, the model addresses the
conceptually important questions of spontaneous symmetry breaking and
condensates. Are these in conflict with a trivial vacuum? Finally, a
constituent picture seems to be realized which should make a
truncation of the light--cone Schr\"odinger equation feasible. Now, we
again encounter the standard rule of physics that there is nothing
like free lunch. Here, this is mainly due to the appearance of
complicated constraints for part of the fermionic degrees of
freedom. Let me thus make a small aside on the special features of
light--cone fermions.

\subsubsection{Light--Cone Fermions.}

The solution of the Dirac equation (for free fermions of mass $m_0$)
has the following light--cone Fock expansion at $x^\p = 0$,
\begin{equation}
  \label{FOCKEXP_FERM} \psi(\vcg{x},0) = \sum_\lambda
  \int_0^{\infty}\frac{dk^\p}{ k^\p}\int \frac{d^2k_{\perp}}{16 \pi^3}
  \, \left[b(\vcg{k},\lambda) u(\vcg{k}, \lambda) e^{- i \svcg{k}
  \cdot \svcg{x}} + d^{\dagger} (\vcg{k},\lambda) v(\vcg{k}, \lambda)
  e^{i \svcg{k} \cdot \svcg{x}} \right] \; ,
\end{equation}
where we recall the notations
\begin{equation}
  \vcg{k} \equiv (k^\p,\vc{k}_{\perp}) \; , \quad
  \vcg{x} \equiv  (x^\m , \vc{x}_{\perp}) \; , \quad
  \vcg{k} \cdot \vcg{x} \equiv \sfrac{1}{2} k^\p x^\m -
  \vc{k}_{\perp} \cdot \vc{x}_{\perp} \; .  
\end{equation}
Like for scalars, the Fock measure is independent of the mass
$m_0$. The Fock operators satisfy the canonical anti--commutation
relations,
\be  
  \pb{b(\vcg{k},\lambda)}{b^\dagger (\vcg{p},\lambda^\prime)} =
  \pb{d(\vcg{k},\lambda)}{d^\dagger (\vcg{p},\lambda^\prime)} = 16
  \pi^3 k^\p \delta^3 (\vcg{k} - \vcg{p}) \; , 
\ee
The basis spinors $u$ and $v$ obey the Dirac equations,
\begin{eqnarray}
  (k\!\!\!/ - m_0) \, u(\vcg{k}, \lambda) &=& 0  \; , \\
  (k\!\!\!/ + m_0) \, v(\vcg{k}, \lambda) &=& 0  \; ,  
\end{eqnarray}
and are explicitly given by
\bea
  u(\vcg{k}, \lambda) = \frac{1}{\sqrt{k^\p}} (k^\p + \beta m +
  \alpha_i k_i) X_\lambda \; , \label{SPINOR_U} \\
  v(\vcg{k}, \lambda) = \frac{1}{\sqrt{k^\p}} (k^\p - \beta m +
  \alpha_i k_i) X_{-\lambda} \label{SPINOR_V} \; . 
\eea
The four--spinor $X$ will be defined in a moment; $\alpha_i$ and
$\beta$ are the standard hermitean Dirac matrices.  The crucial point
is the decomposition of the fermion field, $\psi = \psi_+ + \psi_-$
into `good' ($+$) and `bad' ($-$) components, $\psi_\pm \equiv
\Lambda_\pm \psi$, by means of the projection matrices,
\begin{equation} 
  \Lambda_{\pm} \equiv   \frac{1}{4} \gamma^\mp
  \gamma^{\pm}   = \frac{1}{2}
    \left( \begin{array}{c c} 
               \Eins & \pm  \sigma_3  \\
               \pm \sigma_3 & \Eins 
           \end{array} 
    \right) \; . 
\end{equation}
The spinor $X$ appearing in (\ref{SPINOR_U}) and (\ref{SPINOR_V}) is
an eigenspinor of $\Lambda_+$, $\Lambda_+ X_\lambda = X_\lambda$. The
Dirac equation decomposes accordingly into \emph{two} equations. The
one for $\psi_-$ reads
\be
  2i \partial^\p \psi_- = (-i \gamma^i \partial^i + m_0) \gamma^\p
  \psi_+ \; , 
\ee
and does not contain a light--cone time derivative. Therefore, the bad
component $\psi_-$ is constrained; it can be expressed in terms of the
good component, i.e.~$\psi_- = \psi_- [\psi_+]$. Again, this requires
the inversion of the notorious spatial derivative $\partial^\p$. As a
result, the free Dirac Hamiltonian (density) only depends on $\psi_+$,
\be
  {\cal H} = \psi_+^\dagger \, \frac{- \partial_\perp^2 + m_0^2}{i
  \partial^\p} \, \psi_+ \equiv {\cal H} [\psi_+] \; , 
\ee
where we easily recognize the light--cone energy, $(k_\perp^2 +
m_0^2)/k^\p$.

It turns out that in case of a four--fermion interaction like in the
NJL model the constraint becomes rather awkward to solve. In
particular, its solution has to be consistent with the mean field
approximation employed \cite{dietmaier:89}. This can be achieved most
elegantly by using the large--$N$ expansion
\cite{bentz:99,itakura:99}. Nevertheless, the light--cone Hamiltonian
of the model is a rather complicated expression . We will therefore
follow an alternative road which is the topic of the next subsection.

\subsection{Schwinger--Dyson Approach}

The first derivation, analysis and solution of a light--cone
bound--state equation appeared in 't~Hooft's original paper on what is
now called the 't~Hooft model \cite{thooft:74}. We have discussed this
model in the last section where we also rederived 't~Hooft's solution.
Interestingly, 't~Hooft did not use the light--cone formalism in the
manner we presented it and which nowadays might be called
standard. This amounts to deriving the canonical light--cone
Hamiltonian and setting up the associated system of bound--state
equations by projecting on the different sectors of Fock space
(cf.~Section~4).  Instead, he started from covariant equations, namely
the Schwinger--Dyson equations for the quark propagator (or
self--energy), and the Bethe--Salpeter equation for the bound--state
amplitude, which needs the quark self--energy as an input. The
light--cone Schr\"odinger equation was then obtained by projecting the
Bethe--Salpeter equation onto hypersurfaces of equal light--cone
time. In this way, one avoids to explicitly derive the light--cone
Hamiltonian, which, as explained above, can be a tedious enterprise in
view of complicated constraints one has to solve. Let us therefore
have a closer look at this way of proceeding.

\subsubsection{The Schwinger--Dyson Equation for the Propagator.}

The first step in the program\footnote{For recent literature on the
Schwinger--Dyson approach, see
e.g.~\cite{roberts:94,roberts:00,maris:00}.} is to solve the
Schwinger--Dyson equation for the propagator, or, equivalently, for
the quark self--energy. As this cannot be done exactly, one resorts to
mean--field (or large--$N$) approximation. This is essentially what
has been done in the last subsection. Let us rewrite this in terms of
Schwinger--Dyson equations. The one for the full propagator $S$ reads
\begin{equation}
  S = S_0 + S_0 \Sigma S \; , 
\end{equation}
and is formally solved by
\begin{equation}
  S (p)  =  \frac{1}{S_0^{-1} (p) - \Sigma (p)} \; ,
\end{equation}
where $S_0$ is the free propagator,
\begin{equation}
  S_0 (p)  =  \frac{1}{p \!\!\!/ - m_0 } \; ,
\end{equation}
and $\Sigma$ the quark self--energy. In mean--field approximation, it
is momentum independent and defines the constituent mass through the
gap equation, $\Sigma = const = m$, see (\ref{GAP}) and
(\ref{COND_PROP2}).

To solve the latter in the light--cone framework, we basically just
have to calculate the condensate.  As in the standard approach, this
can be obtained via the Feynman-Hellman theorem by differentiating the
energy density of the quasi--particle Dirac sea,
\begin{eqnarray}
  \cond_m    &=&    \pad{}{m}    {\cal E}    (m)   =   \pad{}{m}
  \int\limits_{-\infty}^0    dk^\p    \int   \frac{d^2
  k_\perp}{16\pi^3}  2 \frac{m^2  + k_\perp^2}{k^\p}   \nn \\
  &=&   -\frac{m}{4\pi^3}    \int\limits_0^{\infty}
  \frac{dk^\p}{k^\p}  \int  d^2  k_\perp  \; , 
  \label{COND_LC} 
\end{eqnarray}
Again, as it stands, the integral is divergent and requires
regularization.  In the most straightforward manner one might choose
$m^2/\Lambda \le k^\p \le \Lambda$ and $|\vc{k}_\perp| \le \Lambda$, so
that the condensate becomes
\begin{equation}
  \cond_m  =  - \frac{m}{4\pi^2}  \int\limits_{m^2/\Lambda}^\Lambda
  \frac{dk^\p}{k^\p}  \int\limits_0^{\Lambda^2}  d(k_\perp^2)  =  -
  \frac{m}{4\pi^2} \Lambda^2 \ln \frac{\Lambda^2}{m^2} \; .
  \label{COND_WRONG}
\end{equation}
Plugging this result into the gap equation (\ref{GAP}) one finds for the
dynamical mass squared,
\begin{equation}
  m^2  (G)  =  \Lambda^2  \exp  \left(-\frac{2\pi^2}{G  \Lambda^2}
  \right) \; .
  \label{DYN-MASS1}
\end{equation}
The  critical  coupling  is determined  by the vanishing  of this
mass,  $m (G_c)  = 0$,  and from  (\ref{DYN-MASS1})  we find  the
surprising result
\begin{equation}
  G_c = 0 \; .
  \label{GC1}
\end{equation}
This result, however, is wrong since one knows from the conventional
treatment of the model that the critical coupling is finite of the order
$\pi^2/\Lambda^2$, both for covariant and noncovariant cutoff
\cite{nambu:61a}.  In addition, it is quite generally clear that in the
{\em free} theory ($G=0$) chiral symmetry is not broken (as $m_0 = 0$)
and, therefore, this should not happen for arbitrarily small coupling,
either, cf.~(\ref{G_CRIT}).  The remedy is to use the \emph{invariant--mass
cutoff} \cite{lepage:80},  
\begin{equation}
\label{IMC1}  
  M_0^2  \equiv \frac{k_\perp^2 + m^2}{x (1-x)} \le \Lambda^2 \; ,
\end{equation}
where we have defined the longitudinal momentum fraction, $x \equiv
k^\p / \Lambda$. This provides a cutoff both in $x$ (or $k^\p$)
\emph{and} $k_\perp$, 
\bea
  \label{IMC2}
  0 &\le& k_\perp^2 \le \Lambda^2 x (1-x) - m^2 \; , \\
  x_0 &\le& x \le x_1 \; , \\
  x_{0,1} &\equiv& \sfrac{1}{2} (1 \mp \sqrt{1 - 4 \epsilon^2}) \; ,
\eea 
with $\epsilon^2 \equiv m^2 / \Lambda^2$. Note that the transverse
cutoff becomes a polynomial in $x$. The $k_\perp$-integration thus has
to be performed before the $x$-integration.

For the condensate (\ref{COND_LC}) the invariant--mass cutoff
(\ref{IMC1}) yields an analytic structure different from
(\ref{COND_WRONG}),
\begin{equation}
  \cond_m  =  -  \frac{m \Lambda^2}{8\pi^2}   \left( 1 +  2 \epsilon^2
\ln \epsilon^2 + O(\epsilon^2) \right) \; ,
  \label{COND_LC_EVAL}
\end{equation}
where we have neglected sub-leading terms in $\epsilon^2$. The result
(\ref{COND_LC_EVAL}) coincides with the standard one,
(\ref{COND_IF_EVAL}), if one identifies the noncovariant cutoffs
according to
\be
  \Lambda^2 \equiv 4 (\Lambda_3^2 + m^2) \; .
\ee
This has independently been observed by \cite{bentz:99}. From
(\ref{COND_LC_EVAL}), one infers the correct cutoff dependence of the
critical coupling,
\begin{equation}
  G_c = \frac{4\pi^2}{\Lambda^2} \; .
  \label{GC2}
\end{equation}
The moral of this calculation is that even in a nonrenormalizable
theory like the NJL model, the light--cone regularization prescription is
a subtle issue.  

In the NJL model with its second--order phase transition of mean-field
type, the usual analogy with magnetic systems can be made.  Chiral
symmetry corresponds to rotational symmetry, the vacuum energy density
to the Gibbs free energy, and the mass $m$ to an external magnetic
field.  The order parameter measuring the rotational symmetry breaking
is the magnetization.  It is obtained by differentiating the free energy
with respect to the external field.  This is the analogue of expression
(\ref{COND_LC}) derived from the Feynman--Hellmann theorem.

\subsubsection{The Bound--State Equation.}

Once the physical fermion mass $m$ is known by solving the gap
equation, it can be plugged into the Bethe--Salpeter equation for
quark--antiquark bound states (mesons), given by
\begin{equation}
  \label{BSE1}
  \bsa =  S_1 S_2 K \bsa \; .
\end{equation}
$S_1$, $S_2$ denote the full propagators of quark and anti--quark, $K$
the Bethe--Salpeter kernel, and $\bsa$ the Bethe--Salpeter amplitude.
From the latter, one obtains the light--cone wave function via
integration over the energy variable $k^\m$
\cite{michael:82,brodsky:85,chakrabarty:89,liu:93},
\begin{equation}
\label{BSE2}
  \lca  (\vcg{k})  = \int  \frac{dk^\m}{2\pi}  \bsa  (k) \; ,
  \quad \vcg{k} = (k^\p, \vc{k}_\perp) \; . 
\end{equation}

In ladder approximation (again equivalent to the large--$N$ limit),
(\ref{BSE1}) and (\ref{BSE2}) become
\begin{equation}
  \lca  (\vcg{k})  = \int \frac{dk^\m}{2\pi}   S(k)  S(k-P)
  \int \frac{d^3 p}{(2\pi)^3} \int \frac{d p^\m}{2\pi}
  K (k, p ) \bsa (p) \; , 
\end{equation}
with $P$ denoting the bound--state four--momentum.  On the
left--hand--side, the projection onto $x^\p = 0$ (i.e.  the
$k^\m$--integration) has already been carried out.  On the
right--hand--side, the two integrations over $k^\m$ and $ p^\m$ still
have to be performed. Whether this can easily be done depends of
course crucially on the kernel $K$, which in principle is a function
of both energy variables.  For the NJL model, however, $K$ assumes the
very simple form,
\begin{equation}
  K(k  ,  p)  = 2 \gamma_5  \otimes  \gamma_5  - \gamma_\mu
  \gamma_5 \otimes \gamma^\mu \gamma_5 \; , 
\end{equation}
i.e.~it  is momentum  independent  due to the four-point
contact   interaction, 
\be
  W \sim \int d^4 x \int d^4 y \, (\bar \psi \Gamma \psi)(x) \, \delta^4
  (x-y) \, (\bar \psi \Gamma \psi)(y) \; .
\ee
Thus, the $p^\m$--integration immediately yields $\lca$, and the
$k^\m$--integration can be performed via residue techniques and is
completely determined by the poles of the propagators, $S(k)$ and
$S(k-P)$.  As a result, one finds a nonvanishing result only if $0
\le k^\p \le P^\p$, and {\em one} of the two particles is put
on-shell, e.g.~$k^2 = m^2$, as already observed by \cite{gross:88}.

The upshot of all this is nothing but the light--cone bound--state
equation, which explicitly reads
\begin{eqnarray}  
  \label{NJL_LFBSE}
  \lca  (x,  \vc{k}_\perp)  &=&  - \frac{2G}{x(1-x)} \frac{(\hat  k
  \!\!\!/  + m ) \gamma_5  (\hat k \!\!\!/ - P \!\!\!\!/  + m)}{M^2 -
  M_0^2 }  \nn \\
  &\times& \int_0^1 dy \int \frac{d^2 p_\perp}{8\pi^3} \tr
  \left[   \gamma_5   \lca  (y  ,   {\vc{p}}_\perp)   \right]
  \theta_\Lambda (y , {\vc{p}}_\perp) \nn \\
  &+&  \frac{G}{x(1-x)} \frac{(\hat  k  \!\!\!/  +  m  ) \gamma_\mu
  \gamma_5  (\hat  k \!\!\!/  - P \!\!\!\!/  + m)}{M^2  - M_0^2  }
  \nn \\
  &\times& \int_0^1  dy \int \frac{d^2   p_\perp}{8\pi^3}  \tr \left[
  \gamma^\mu  \gamma_5  \lca  (y  , {\vc{p}}_\perp)  \right]
  \theta_\Lambda (y , {\vc{p}}_\perp) \; . \nn \\ 
\end{eqnarray}
Here we have defined the longitudinal momentum fractions $x = k^\p /
P^\p$, $y =  p^\p / P^\p$, the on-shell momentum $\hat k = (\hat
k^\m , \vc{k}_\perp , k^\p)$ with $\hat k^\m = (k_\perp^2 + m^2)/k^\p$,
and the bound state mass squared, $M^2 = P^2$, which is the eigenvalue
to be solved for.  $\theta_\Lambda (x , \vc{p}_\perp )$ denotes the
invariant mass cutoff (\ref{IMC1}).

Now, while (\ref{NJL_LFBSE}) may appear somewhat complicated it is
actually very simple; indeed, it is basically already the solution of
the problem.  The crucial observation is to note that the two integral
expressions are mere normalization constants,
\begin{eqnarray}
  C_\Lambda &\equiv& \int_0^1 dy \int \frac{d^2  p_\perp}{8\pi^3}
  \tr  \left[  \gamma_5  \lca  (y , {\vc{p}}_\perp)  \right]
  \theta_\Lambda (y , {\vc{p}}_\perp) \; , \\
  D_\Lambda   P^\mu   &\equiv&  \int_0^1   dy  \int  \frac{d^2   
  p_\perp}{8\pi^3}  \tr \left[ \gamma^\mu \gamma_5 \lca (y , 
  {\vc{p}}_\perp) \right] \theta_\Lambda (y , {\vc{p}}_\perp)
  \; .
\end{eqnarray}
Thus, the solution of the light--cone bound--state equation
(\ref{NJL_LFBSE}) is
\begin{eqnarray}
  \lca   (x,   \vc{k}_\perp)   &=&   -   \frac{2G C_\Lambda}{x(1-x)}
  \frac{(\hat k \!\!\!/ + m ) \gamma_5 (\hat k \!\!\!/ - P \!\!\!\!/ +
  m)}{M^2 - M_0^2 } \nn \\
  &+&  \frac{G  D_\Lambda}{x(1-x)} \frac{(\hat  k  \!\!\!/  + m ) P
  \!\!\!\!/  \gamma_5  (\hat k \!\!\!/ - P \!\!\!\!/  + m)}{M^2 - M_0^2
  } \; ,
  \label{NJL_LCA}
\end{eqnarray}
with yet undetermined normalization constants $C_\Lambda$ and
$D_\Lambda$.  As a first check of our bound--state wave function
(\ref{NJL_LCA}) we look for a massless pion in the chiral limit.  To
this end we decompose the light--cone wave function into Dirac
components according to \cite{lucha:91},
\begin{equation}
  \lca = \phi_{\mbox{\tiny  S}} + \phi_{\mbox{\tiny  P}} \gamma_5 +
  \phi_{\mbox{\tiny A}}^\mu \gamma_\mu \gamma_5 + \phi_{\mbox{\tiny
  V}}^\mu    \gamma_\mu     +    \phi_{\mbox{\tiny     T}}^{\mu\nu}
  \sigma_{\mu\nu} \; .
\end{equation}
Multiplying (\ref{NJL_LCA}) with $\gamma_5$, taking the trace and
integrating over $\vcg{k}$ we find
\begin{eqnarray}
  C_\Lambda  &=& - \frac{G C_\Lambda}{2  \pi^3} \int_0^1 dx \int d^2
  k_\perp  \frac{M^2  x + M_0^2  (1-x)}{x  (1-x)  (M^2  - M_0^2)  }
  \theta_\Lambda (x , \vc{k}_\perp) \nn \\
  &+& \frac{G D_\Lambda}{2\pi^3}  M^2 \int_0^1  dx \int  d^2 k_\perp
  \theta_\Lambda (x , \vc{k}_\perp) \; .
\end{eqnarray}
In the  chiral  limit  one  expects  a solution  for  $M=0$,  the
Goldstone pion. In this case one obtains
\begin{equation}
  \label{GAP_INT}
  1 = \frac{G}{2\pi^3}  \int_0^1 dx \int d^2 k_\perp \theta_\Lambda
  (x , \vc{k}_\perp) \; .
\end{equation}
This is exactly the gap equation (\ref{GAP}) using the definition
(\ref{COND_LC}) of the condensate (with the invariant--mass cutoff
(\ref{IMC1}) understood in both identities). Note once more the
light--cone peculiarity that the (Fock) measure in (\ref{GAP_INT}) is
entirely mass independent. All the mass dependence, therefore, has to
come from the (invariant--mass) cutoff. Otherwise one will get a wrong
behavior of the dynamical mass $m$ as a function of the coupling $G$,
as was the case in (\ref{DYN-MASS1}).
 
With this in mind, we see that the Goldstone pion is a solution of the
light--cone bound--state equation exactly if the gap equation holds.
This provides additional evidence for the self--consistency of the
procedure.  The deeper reason for the fact that the quark self--energy
and the bound--state amplitude satisfy essentially the same equation,
is the chiral Ward identity relating the quark propagator and the
pseudoscalar vertex \cite{savkli:98}.

Our next task is to actually evaluate the solution (\ref{NJL_LCA}) of
the bound--state equation. $\lca$ is a Dirac matrix and therefore is not
yet a light--cone wave function as defined in Section~4. The relation
between the two quantities has been given by \cite{liu:93},
\begin{equation}
  2 P^\p \psi (x , \vc{k}_\perp, \lambda , \lambda^\prime) = \bar u (x
  P^\p , \vc{k}_\perp , \lambda) \gamma^\p \lca (\vcg{k}) \gamma^\p v
  (\bar x P^\p , - \vc{k}_\perp , \lambda^\prime) \; ,
\end{equation}
where we have denoted $\bar x \equiv 1-x$ to save space. A somewhat
lengthy calculation yields the result \cite{heinzl:98},
\begin{equation}
  \label{PSI_PROV} \psi (x , \vc{k}_\perp, \lambda , \lambda^\prime) =
  \frac{2 G P^\p / \sqrt{x \bar x}}{M^2 - M_0^2} \left( \frac{2
  C_\Lambda}{M} \bar u_\lambda M \gamma_5 v_{\lambda^\prime} -
  D_\Lambda \bar u_\lambda P \!\!\!\!/ \, \gamma_5 v_{\lambda^\prime}
  \right) \; ,
\end{equation}
with the arguments of the spinors $\bar u$ and $v$ suppressed. At this
point we have to invoke another symmetry principle. \cite{ji:92} have
pointed out that the spin structure ($\bar u \Gamma v$) should be
consistent with the one obtained form the instant form spinors via a
subsequent application of a Melosh transformation
\cite{melosh:74} and a boost. Using this recipe, one obtains the
following relation between the constants $C_\Lambda$ and $D_\Lambda$,
\begin{equation}
  \label{CDN}
  2 C_\Lambda / M = - D_\Lambda \equiv N  / 2 G  \; .
\end{equation}
As a result, the spin structure in (\ref{PSI_PROV}) coincides with the
standard one used e.g.~in
\cite{dziembowski:88,chung:88b,jaus:90,ji:90,ji:92}. The NJL wave
function of the pion thus becomes
\begin{equation}
  \psi(x, \vc{k}_\perp , \lambda, \lambda^\prime ) = \frac{N P^\p /
  \sqrt{x \bar x}}{M^2 - M_0^2} \; \bar u_\lambda (M + P\!\!\!\!/) \,
  \gamma_5 \, v_{\lambda^\prime} \, \theta(\Lambda^2 - M_0^2) \; .
\end{equation}
Not surprisingly, the off--shellness $M^2 - M_0^2$ appears in the
denominator. $N$ is the normalization parameter defined in
(\ref{CDN}), and the spin (or helicity) structure is given by
\begin{eqnarray}
  &&\bar u (x P^\p , \vc{k}_\perp , \lambda ) \, (M + P\!\!\!\!/) \,
  \gamma_5 \, v (\bar x P^\p , - \vc{k}_\perp, \lambda^\prime) = \nn
  \\ && = \frac{1}{\sqrt{x \bar x} P^\p} \left[ \lambda \Big( mM + m^2
  - \vc{k}_\perp^2 + M^2 x \bar x \Big) \delta_{\lambda,
  -\lambda^\prime} - k_{-\lambda} (M + 2m) \delta_{\lambda
  \lambda^\prime} \right] \; \;
\end{eqnarray}
where we have used (\ref{SPINOR_U}), (\ref{SPINOR_V}) and denoted
$k_\lambda \equiv k^1 + i \lambda k^2$. The first term with spins
anti--parallel corresponds to $L_z =0$, the second one (with spins
parallel) to $L_z = \pm 1$. It has already been pointed out by
Leutwyler that both spin alignments should contribute to the pion wave
function \cite{leutwyler:74a,leutwyler:74b}. Note that the latter is a
cutoff dependent quantity. This is necessary in order to render the
wave function normalizable. A single power of the off--shellness in
the denominator is not sufficient for that. Only the cutoff guarantees
the boundary conditions (\ref{WF_BC}) so that the wave function drops
off sufficiently fast in $x$ and $k_\perp$.

As we are interested in analyzing the quality of a constituent picture,
we approximate the pion by its valence state, denoting $\psi \equiv
\psi_2$, 
\begin{equation}
\label{PION_STATE}
  | \pi: \vcg{P} \ket = \sum_{\lambda, \lambda^\prime} \int_0^1
  dx \int \frac{d^2 k_\perp}{16\pi^3} \, \psi_2 (x,
  \vc{k}_\perp, \lambda, \lambda^\prime ) \, | q \bar q : x, \vc{k}_\perp, 
  \lambda, \lambda^\prime \ket \, 
\end{equation}
which should be compared with the general expression (\ref{GEN_PI_WF}). 
The normalization of this state is given by (\ref{MOM_NORM}),
\begin{equation}
  \bra \pi: \vcg{P}^\prime | \pi: \vcg{P} \ket = 16 \pi^3
  P^\p \delta^3 (\vcg{P} - \vcg{P}^\prime) \; .
  \label{PI_NORM}
\end{equation}
As usual we work in a frame in which the total transverse momentum
vanishes, i.e.~$\vcg{P} = (P^\p , \vc{P}_\perp = 0)$. Expression
(\ref{PI_NORM}) yields the normalization (\ref{VAL_NORM}) of the wave
function,
\begin{equation}
  \sum_{\lambda \lambda^\prime} \int_0^1 dx \int \frac{d^2 k_\perp}{16
  \pi^3} | \psi_2 (x, \vc{k}_\perp , \lambda , \lambda^\prime ) |^2
  \equiv \| \psi_2 \|^2  = 1 \;
  .
  \label{PSI_NORM}
\end{equation}
It is of course a critical assumption that the probability to find the
pion in its valence state is one. In this way we enforce a constituent
picture {\it by fiat}, and it is clear that such an assumption has to
be checked explicitly by comparing with phenomenology.

\subsection{Observables}

With the light--cone wave function at hand, we are in the position to
calculate observables. To proceed we will employ the following two
simplifications. First of all, we will always work in the chiral limit
of vanishing quark mass, $m_0 = 0$, which, as we have seen, leads to a
massless Goldstone pion, $M=0$. We write the pion wave function as a
matrix in helicity space,
\bea
  \label{PI_WF_MATRIX}
  \psi_2 (x , \vc{k}_\perp) &=& 
  \left( \begin{array}{cc} 
              \psi_{2 \uparrow \uparrow} & \psi_{2 \uparrow\downarrow} \\
              \psi_{2 \downarrow\uparrow}  & \psi_{2 \downarrow\downarrow} 
              \end{array}
       \right) \nn \\ 
  &=& - \frac{N}{k_\perp^2 +
  m^2} \left( \begin{array}{cc} 
              -2m (k^1 - ik^2) & m^2 - k_\perp^2 \\
              k_\perp^2 - m^2  & - 2m (k^1 + ik^2 )
              \end{array}
       \right) \theta(\Lambda^2 - M_0^2) \, . \nn \\
\eea
Note that, in the chiral limit,  the wave function becomes independent
of $x$ (apart from cutoff effects). This actually agrees with our
findings in the two--dimensional `t~Hooft model,
cf.~(\ref{MASSLESS}). 

The diagonal terms in (\ref{PI_WF_MATRIX}) correspond to parallel
spins, the off--diagonal ones to anti--parallel spins. The different
components are related by the symmetry properties,
\be
  \psi_{2 \downarrow\downarrow} = \psi_{2 \uparrow\uparrow}^* \; ,
  \quad \psi_{2 \downarrow\uparrow} = - \psi_{2 \uparrow\downarrow} \;
  .
\ee
Second, we will go to the large--cutoff limit, that is, we will keep
only the leading order in $\epsilon^2 = m^2 / \Lambda^2$. We thus
assume that the cutoff is large compared to the constituent mass. From
the standard values, $\Lambda \simeq 1$ GeV, $m \simeq 300$ MeV, we
expect that this assumption should induce an error of the order of
10\%. The technical advantage of the large--cutoff limit is a simple
analytic evaluation of all the integrals we will
encounter. Furthermore, the leading order will be independent of the
actual value of the constituent mass. It should be mentioned that the
same procedure has been used in calculations based on the instanton
model of the QCD vacuum
\cite{petrov:98}. There, the ratio $\epsilon^2$ can be related to
parameters of the instanton vacuum, namely
\be
  \epsilon^2 = (m \rho)^2 \sim (\rho/R)^4 \; , 
\ee
where $\rho \simeq 1/3$ fm is the instanton size and $R$ the mean
distance between instantons. Thus, $\epsilon^2$ can be identified with
the `diluteness parameter' or `packing fraction' of the instanton
vacuum and hence is parametrically small, $\epsilon^2 \simeq 1/4$.

The upshot of all this is that we work with the extremely simple model
wave function \cite{radyushkin:95}
\be
\label{STEP}
  \psi_{2 \uparrow\downarrow} \simeq N \, \theta(\Lambda^2 - M_0^2) \;
  , \quad \psi_{2 \uparrow\uparrow} = 0 \;,
\ee
which is entirely determined by two parameters, the normalization
constant $N$ and the cutoff $\Lambda$. We thus need two constraints on
the wave function to fix our two parameters.

\subsubsection{Normalization.}

As announced, we enforce a constituent picture by demanding
(\ref{PSI_NORM}) which decomposes into
\be
  1 = || \psi_2 ||^2 = || \psi_{2 \uparrow\downarrow} ||^2 + ||
  \psi_{2 \uparrow\uparrow} ||^2 + || \psi_{2 \downarrow\uparrow} ||^2
  + || \psi_{2 \downarrow\downarrow} ||^2 \; .
\ee
Explicitly, one finds 
\bea
\label{NJL_NORM}
  || \psi_{2 \uparrow\downarrow} ||^2 &=& || \psi_{2
  \downarrow\uparrow} ||^2 = N^2 \int_0^1 dx \int \frac{d^2
  k_\perp}{16 \pi^3} \, \theta (\Lambda^2 - M_0^2) \nn \\
  &=& \frac{N^2}{16 \pi^2} \int_0^1 dx \int_0^{\Lambda^2 x (1-x)} dk_\perp^2
  = \frac{N^2 \Lambda^2}{96 \pi^2} \stackrel{!}{=} 1/2 \; , 
\eea
while the components with parallel spins have vanishing norm in the
large--cutoff limit, $|| \psi_{2 \uparrow\uparrow} ||^2 = ||
\psi_{2 \downarrow\downarrow} ||^2 = 0$, cf.~(\ref{STEP}).

\subsubsection{Pion Decay Constant.}

A second constraint on the wave function is provided by the pion decay
constant $f_\pi$, which appears in the semi--leptonic process $\pi \to
\mu \nu$. The relevant matrix element is 
\be
\label{V-A}
  \bra 0 | \bar \psi_{\bar d} (0) \gamma^\p \gamma_5 \psi_u (0) |
  \pi^\p (P^\p) \ket = i \sqrt{2} P^\p f_\pi \; ,
\ee
$\bar \psi_{\bar d}$ and $\psi_u$ denote the field operators of the
$\bar d$ and $u$ quark in the pion. If we insert all quantum numbers,
the pion state to the right of the matrix element is given by
\be
  | \pi^\p \ket = \psi_{\bar d u} \otimes \frac{1}{\sqrt{6}} \Big( |
  \bar d_{c \uparrow} u_{c \downarrow} \ket - | \bar d_{c \downarrow}
  u_{c \uparrow} \ket \Big) \; .
\ee
The spatial (or internal) structure of the state is encoded in the
light--cone wave function $\psi_2 \equiv \psi_{\bar d u}$. If we
insert the Fock expansions (\ref{FOCKEXP_FERM}) for $\bar \psi_{\bar
d}$ and $\psi_u$ as well as the pion state (\ref{PION_STATE}), we
obtain the following constraint on the pion wave function,
\be
\label{F_PI_CONSTRAINT}
  \int_0^1 dx \int \frac{d^2 k_\perp}{16 \pi^3} \, \psi_{2 \uparrow
  \downarrow} (x, \vc{k}_\perp) = \frac{f_\pi}{2 \sqrt{3}} \; .
\ee
The left--hand--side is basically the (position space) `wave function
at the origin'. Quark ($u$) and antiquark ($\bar d$) thus have to sit
on top of each other in order to have sizable probability for 
decay. Note that only the $L_z = 0$ component contributes. Concerning
the effect of higher Fock states, it can be shown
\cite{lepage:81,brodsky:89} that indeed only the valence wave function
contributes to (\ref{F_PI_CONSTRAINT}). This constraint is therefore
\emph{exact} and holds beyond a constituent picture. Empirically, the
pion decay constant is $f_\pi$ = 92.4 MeV \cite{holstein:95}.

As already stated, this is our second source of phenomenological
information to fix cutoff and normalization. Using the explicit form
(\ref{STEP}) of the wave function, the constraint
(\ref{F_PI_CONSTRAINT}) becomes
\be
\label{F_PI_NJL}
  \int_0^1 dx \int \frac{d^2 k_\perp}{16 \pi^3} \, \psi_{2 \uparrow
  \downarrow} (x, \vc{k}_\perp) = \frac{N \Lambda^2}{96 \pi^2}
  \stackrel{!}{=} \frac{f_\pi}{2 \sqrt{3}} \; .
\ee
With (\ref{F_PI_NJL}) and (\ref{NJL_NORM}) we now have two equations
for our two parameters which accordingly are determined as
\bea
  N &=& \sqrt{3} / f_\pi \; , \label{PARAMETERS1} \\
  \Lambda &=& 4 \pi f_\pi \simeq 1.16 \; \mbox{GeV} \label{PARAMETERS2} \; .
\eea
The value (\ref{PARAMETERS2}) for the cutoff $\Lambda$ is the standard
scale below which chiral effective Lagrangians are believed to make
sense \cite{manohar:84}. It is reassuring that within our
approximations we get exactly this value. This means that we are not
doing something entirely stupid. A more severe test of consistency is
provided by the next constraint to be satisfied.

\subsubsection{Constraint from $\pi^0 \to 2 \gamma$.}

This constraint has also been derived by Brodsky and Lepage
\cite{lepage:81} within an analysis of the $\pi \gamma$
transition form factor. It assumes the very simple form,
\be
\label{PI_GAMMA_CONSTRAINT}
  \int_0^1 dx \, \psi_{2 \uparrow \downarrow} (x, \vc{0}_\perp) =
  \frac{\sqrt{3}}{f_\pi} \; . 
\ee
Inserting the light--cone wave function (\ref{STEP}), the
right--hand--side simply becomes the normalization $N$ which is indeed
consistent with our findings (\ref{PARAMETERS1}) and
(\ref{PARAMETERS2}). We mention in passing that the constraint
(\ref{PI_GAMMA_CONSTRAINT}) usually is the simplest way to fix the
normalization $N$. Its derivation, however, is more complicated than
that of (\ref{F_PI_CONSTRAINT}).

\subsubsection{Pion Form Factor.}

We proceed by calculating the pion electromagnetic formfactor. It is
defined by the matrix element of the electromagnetic current
$J_{\mathrm{em}}^\mu$ between pion states,
\begin{equation}
  \bra \pi : \vcg{P} | J_{\mathrm{em}}^\mu | \pi : \vcg{P}^\prime \ket = 2 (P
  + P^\prime)^\mu F(Q^2) \; , \quad Q^2 \equiv - (P - P^\prime)^2 \; .
\end{equation}
Considering $\mu = +$ in a frame where $\vcg{P} = (P^\p , \vc{0})$ and
$\vcg{P}^\prime = (P^\p , \vc{q}_\perp)$ one is led to the the Drell--Yan
formula \cite{drell:70c,brodsky:89}, 
\begin{equation}
  \label{DY}
  F(\vc{q}_\perp^2 ) = \sum_{\lambda \lambda^\prime} \int_0^1 dx \int
  \frac{d^2 k_\perp}{16\pi^3} \, \psi^* (x , \vc{k}_\perp^\prime , \lambda ,
  \lambda^\prime ) \, \psi (x , \vc{k}_\perp , \lambda , \lambda^\prime) \; .
\end{equation}
The transverse momentum of the struck quark is $\vc{k}_\perp^\prime =
\vc{k}_\perp + (1 - x) \vc{q}_\perp$. The formula (\ref{DY}) with
its overlap of two wave functions on the right--hand--side is rather
similar to the nonrelativistic result as will be shown in what
follows.

The form factor of a nonrelativistic system is given by the Fourier
transform of the charge distribution (normalized to one), that is, 
\be
\label{FF_NR}
  F(\vc{p}) = \int d^3 r \, \psi^* (\vc{r}) \, \psi(\vc{r}) \, e^{i \svc{p}
  \cdot \svc{x}} = \int \frac{d^3 k}{(2 \pi)^3} \, \psi^* (\vc{k} + \vc{p})
  \, \psi(\vc{k}) \; . 
\ee
It is important to note that $\vc{k}$ and $\vc{k}^\prime \equiv \vc{k}
+ \vc{p}$ are \emph{relative} momenta,
\be
  \vc{k} \equiv \frac{1}{M} (m_2 \vc{k}_1 - m_1 \vc{k}_2) \; , 
\ee
(and analogously for $\vc{k}^\prime$), so that the $\vc{k}_{1,2}$ are
the actual particle momenta. Accordingly, $\vc{p}$ is the
\emph{relative} momentum transfer,
\be
  \vc{p} = \vc{k}^\prime - \vc{k} = \frac{m_2}{M} (\vc{k}_1^\prime -
  \vc{k}_1) \equiv \frac{m_2}{M} \vc{q} = x_2 \vc{q} \; ,
\ee
where, in the last step, we have used (\ref{NR_XK}). Plugging this
into the form factor (\ref{FF_NR}) we obtain the formula,
\be
  F(\vc{q}) = \int \frac{d^3 k}{(2 \pi)^3} \, \psi^* (\vc{k} + x_2
  \vc{q}) \, \psi(\vc{k}) \; , 
\ee
which, as promised, is quite similar to (\ref{DY}). 

If one sets the momentum transfer $\vc{q}_\perp$ in (\ref{DY}) equal
to zero, one is left with the normalization integral (\ref{PSI_NORM}),
so that the form factor is automatically normalized to one (the same
is true in the nonrelativistic case). 

\begin{figure}
  \caption{\protect\label{fig-ff} \textsl{The pion form factor squared
  vs.~momentum transfer $q^2 \equiv q_\perp^2$. The full line is the
  monopole fit of \protect\cite{amendolia:86}, $|F|^2 = n/(1 +
  q_\perp^2 r_\pi^2 /6)^2$ with $n = 0.991$, $r_\pi^2 = 0.431$ fm$^2$;
  the dashed line is the same fit with our values, $n=1$, $  r_\pi^2
  = 0.36$ fm$^2$. The agreement is consistent with the
  expected accuracy of 10\%.}}  
  \vspace{0.5cm} 
  \begin{center}
  \includegraphics[height=10cm,width=10cm]{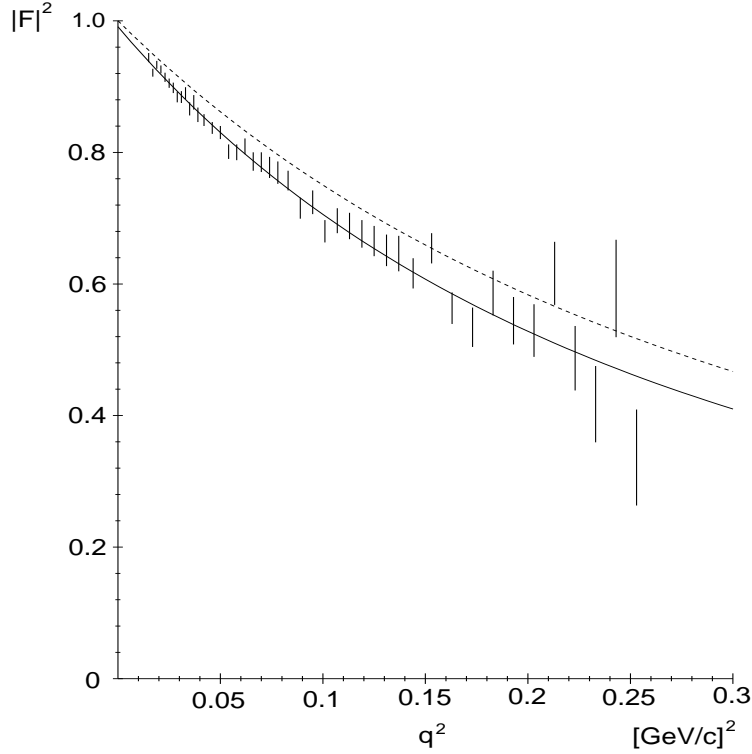}
  \end{center}
\end{figure}

We will use the Drell--Yan formula for the form factor to determine the
pion charge radius $r_\pi$, which is given by the slope of the form
factor at vanishing momentum transfer,
\be
  F(q_\perp^2) \equiv 1 - \frac{r_\pi^2}{6} \, q_\perp^2 +
  O(q_\perp^4) \; .
\ee
Using (\ref{DY}) this results in the nice explicit formula,
\be
\label{R_PI}
  r_\pi^2 = - \frac{3}{2} \int_0^1 dx \int \frac{d^2 k_\perp}{16 \pi^3}
  \left. \frac{\partial^2}{\partial q_i \partial q_i} \psi_2^*
  (\vc{k}_\perp + \bar x \vc{q}_\perp) \right|_{q_\perp = 0} \psi
  (\vc{k}_\perp) \; .
\ee
Upon inserting the wave function (\ref{STEP}), however, one encounters
a problem. The sharp cutoff (corresponding to a step function) is too
singular to lead to a reasonable result. The derivatives in
(\ref{R_PI}) are concentrated at the boundary of the support of the
wave function which in the end leads to artificial infinities. Thus,
for the time being, we resort to a smooth cutoff, 
\be
\label{NJL_GAUSSIAN}
  \theta_\Lambda^s (x, \vc{k}_\perp) \equiv \exp \left[ -
  \frac{k_\perp^2}{\Lambda^2 x (1 - x)} \right] \; , 
\ee
which basically transforms the sharp--cutoff wave function
(\ref{STEP}) to the Gaussian (\ref{GAUSSIAN}) (with $m$ set to
zero). Plugging this into (\ref{R_PI}) yields the pion charge radius
\be
\label{R_PI_EVAL}
  r_\pi^2 = \frac{12}{\Lambda^2} = \frac{3}{4 \pi^2 f_\pi^2} = (0.60
  \; \mbox{fm})^2  \; .
\ee
This is the standard result for the NJL model \cite{blin:88} and has
also been obtained within the instanton model \cite{diakonov:95}. It
is slightly smaller than the experimental value, $r_\pi = 0.66$
\cite{amendolia:86}, a discrepancy which is usually attributed to the
use of the large--cutoff limit. A pole fit using our value of the pion
charge radius is displayed in Fig.~\ref{fig-ff}.

\subsubsection{Transverse Size.}

As in Section 4, Example~3, we can use the light--cone wave function
(\ref{STEP}) to calculate the r.m.s.~transverse momentum which leads
to
\be
  \bra k_\perp^2 \ket \equiv \int_0^1 dx \int \frac{d^2 k_\perp}{16
  \pi^3} \, k_\perp^2 \, || \psi_{2 \uparrow \downarrow} (x, \vc{k}_\perp)
  ||^2 = \frac{\Lambda^2}{10} \simeq (370 \; \mbox{MeV})^2 \; .
\ee
This actually coincides with the result (\ref{K_PERP_GAUSSIAN}) for
the smooth--cutoff Gaussian wave function (\ref{GAUSSIAN}) or
(\ref{NJL_GAUSSIAN}). Therefore, unlike the charge radius $r_\pi$, the
r.m.s.~trans\-verse momentum is insensitive to the details of the cutoff
procedure. We thus have $\bra k_\perp^2 \ket^{1/2} \simeq m > M_\pi$,
which confirms that the pion is highly relativistic.

The r.m.s.~transverse momentum can easily be translated into a
transverse size scale $R_\perp$,
\be
\label{R_PI_CORE}
  R_\perp^2 \equiv 1/ \bra k_\perp^2 \ket \simeq (0.54 \; \mbox{fm})^2
  \; .
\ee
This is slightly smaller than the charge radius which we attribute to
the fact that the charge distribution measured by the charge radius
does not coincide with the distribution of baryon density. The `core
radius' $R_\perp$ is sometimes related to the decay constant $f_\pi$ via
the dimensionless quantity $C = f_\pi R_\perp$ \cite{weise:84}. In
constituent quark models one typically gets $C \simeq 0.4$. This
implies the fairly large value $R_\perp \simeq $ 0.8 fm. Using standard
many--body techniques, Bernard et al.~have calculated this quantity in
a model treating the pion as a collective excitation of the QCD
vacuum, and find $C \simeq 0.2$ \cite{bernard:85}. This result is
close to what we get from (\ref{R_PI_CORE}),
\begin{equation}
  C = f_\pi R_\perp \simeq 0.25 \; .
\end{equation}
Thus, though we work within a constituent picture, we do get a
reasonable value. We believe that this is due to the intrinsic
consistency of the light--cone framework with the requirements of
relativity, a feature that is lacking in the constituent quark model.

\subsubsection{(Valence) Structure Function.}

The pion structure function arises in the description of deep
inelastic scattering off charged pions. In terms of light--cone wave
functions it is defined as the momentum fraction $x$ times the sum of
`quark distributions' $f_q$ weighted by the quark charges $e_q$. The
quark distributions are given by the squares of light--cone wave
functions integrated over $k_\perp$. For the valence structure
function of the pion we thus have the formula
\cite{lepage:81,brodsky:89},
\be
  F_2^v (x) = x \Big[ e_u^2 \, f_u^v (x) + e_d^2 \, f_{\bar d}^v (x)
  \Big] = \frac{5}{9} \, x  f^v (x) 
  = \frac{5}{9} \, x \sum_{\lambda \lambda^\prime} \int \frac{d^2
  k_\perp}{16 \pi^3} \left| \psi_2  (x, \vc{k}_\perp , \lambda , 
  \lambda^\prime)  \right|^2 ,
\ee
where the $f^v$ denote the valence quark distributions.  For the model
wave function (\ref{STEP}) the structure function becomes
\be
  F_2^v (x) = \frac{10}{3} x^2 (1-x) \; ,
\ee
which in turn leads to the (valence) quark distribution
\be
\label{QUARK_DISTR}
  f^v (x) \equiv f_u^v (x) = f_{\bar d}^v  (x) = 6 \, x (1-x) \; .
\ee
The following consistency checks can be made. The probability to find a
valence quark in the pion,
\be
  \int_0^1 dx \, f^v (x) = 1 \; , 
\ee
is unity, as it should. For the mean value of the momentum fraction
$x$ carried by one of the quarks one finds
\be
\label{X_AV}
  \bra x \ket \equiv \int_0^1 dx \, x f^v (x) = 1/2 \; .
\ee
Thus, on average, the quarks share an equal amount of longitudinal
momentum, which again, of course, is the correct result. 

If one compares with other NJL calculations of the structure function
\cite{shigetani:93,bentz:99} and with the empirical parton
distributions in the literature \cite{glueck:99}, one finds reasonable
qualitative agreement.

Let me finally point out that it is not entirely obvious to which
actual momentum scale our results correspond. The transverse--momentum
cutoff is $x$--dependent, $\Lambda^2 (x) \simeq
\Lambda^2 \, x (1-x)$, so, if we use the average $x$ of (\ref{X_AV}),
$\bra x \ket \simeq 1/2$, a natural scale seems to be\footnote{I thank
W.~Schweiger for discussions on this point.}
\begin{equation}
\label{SCALE}
  Q \equiv \Big[\bra x \ket (1 -\bra x \ket )\Big]^{1/2} \Lambda \simeq
  \Lambda /2 \simeq 600 \; \mathrm{MeV} \; .
\end{equation}

\subsubsection{Pion Distribution Amplitude.}

The pion distribution amplitude was originally introduced to describe
hard exclusive processes involving pions
\cite{lepage:79a,lepage:79b,lepage:80}. The pion formfactor at
\emph{large} $Q^2$, for example, is given by the following convolution
formula, 
\be
\label{CONVOLUTION}
  F(Q^2) = \int_0^1 dx \int_0^1 dy \, \phi^* (x,Q) \, T_H (x,y; Q) \,
  \phi (y, Q) \, \Big[1 + O(1/Q) \Big] \; ,
\ee
where $Q$ denotes the \emph{large} momentum transfer, $T_H$ a `hard
scattering amplitude' and $\phi$ the pion distribution
amplitude. While the amplitude $T_H$ is the sum of all perturbative
contributions to the scattering process, $\phi$ is nonperturbative in
nature. The convolution formula is thus a prominent example where we
see the principle of \emph{factorization} into `soft' and `hard'
physics at work. 

The pion distribution amplitude is rather straightforwardly related to
the light--cone wave function of the pion,
\bea
  \phi (x, Q) &\sim& \int \frac{dz^\m}{4 \pi} \, e^{i x P^\p z^\m / 2}
  \bra 0 | \bar \psi (0) \gamma^\p \gamma_5 \psi (z^- , \vcg{0}_\perp)
  | \pi (P^\p) \ket \nn \\ &\sim& \int \frac{d^2 k_\perp}{16 \pi^3} \,
  \psi_{q \bar q}^{(Q)} (x, \vc{k}_\perp) \; .
\eea
The normalization is fixed by demanding that $\phi$ integrates to
unity. 

Brodsky and Lepage have shown that $\phi$ obeys an evolution equation
of the form \cite{lepage:79a,lepage:80},
\be
\label{BL_EVOLUTION}
  Q \pad{}{Q} \, \phi (x, Q) = \int_0^1 dy \, V(x,y; Q) \, \phi(y, Q) \; .
\ee
where the evolution kernel $V$ is determined by perturbative QCD. For
$Q \to \infty$, (\ref{BL_EVOLUTION}) has the asymptotic solution
\be
  \phi_{\mathrm{as}} (x) = 6 \, x (1-x) \; ,
\ee
(which is normalized to 1).  The (nonasymptotic) pion distribution
amplitude has been a rather controversial object. For a while people
have tended to believe in a `double--humped' shape of the amplitude
(due to a factor $(1 - 2x)^2$), which was originally suggested by
Chernyak and Zhitnitsky using QCD sum rules \cite{chernyak:84}. In
1995, however, the CLEO collaboration has published data
\cite{savinov:95} that seemed to support an amplitude that is not too
different from the asymptotic one \cite{kroll:96}. Theoretical
evidence for this fact has recently been reported in
\cite{belyaev:98a,belyaev:98b,petrov:97,petrov:98}. Belyaev and
Johnson, for instance, have found two constraints which should be
satisfied by the distribution amplitude \cite{belyaev:98a},
\begin{eqnarray}
  \phi (x = 0.3) &=& 1 \pm 0.2 \; , \nn \\
  \phi (x = 0.5) &=& 1.25 \pm 0.25 \; , \label{BJ}
\end{eqnarray}
which are consistent with an amplitude being close--to--asymptotic. 

Last year, the experimental developments have culminated in a direct
measurement of the distribution amplitude at Fermilab
\cite{ashery:99}. At a fairly low (i.e.~nonasymptotic) momentum scale
of $Q^2 \simeq 10$ GeV$^2$, one finds a distribution amplitude that is
very close to the asymptotic one.

\begin{figure}
\caption{\label{DISAMP_FIG}\textsl{The (asymptotic) pion distribution
  amplitude. The vertical lines mark the constraints
  (\protect\ref{BJ}) of Belyaev and Johnson.}}
\vspace{0.5cm}
  \begin{center}
    \includegraphics[height=9cm,width=8cm]{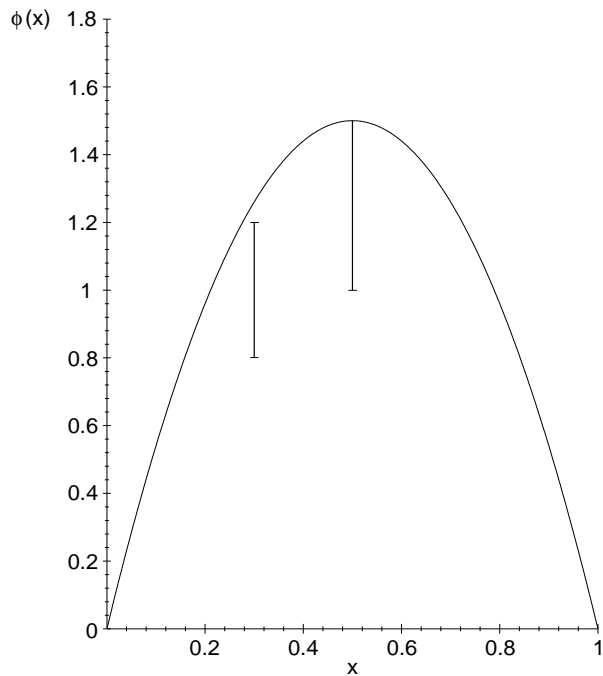}
  \end{center}
\end{figure}

Let us see what we get in the NJL model. The distribution amplitude
is given by
\be
  \phi_{\mathrm{NJL}} (x) = \frac{2 \sqrt{3}}{f_\pi} \int \frac{d^2
  k_\perp}{16 \pi^3} \, \psi_{2 \uparrow\downarrow} (x , \vc{k}_\perp) =
  6 \, x (1-x) = \phi_{\mathrm{as}} (x) \; .
\ee
Thus, in the large cutoff limit, the NJL distribution amplitude is
exactly given by the asymptotic one! Upon comparing with
(\ref{QUARK_DISTR}), we see that $\phi_{\mathrm{NJL}}$ coincides with
the quark distribution $f^v$. This is accidental and stems from the
fact that, due to the use of a step function in (\ref{STEP}), $\psi_2
\sim |\psi_2|^2$. 

In Figure \ref{DISAMP_FIG} we have displayed our distribution
amplitude together with the constraints (\ref{BJ}), represented by the
vertical lines.

Our findings are thus consistent with the recent Fermilab experiment
\cite{ashery:99}. One should, however, be aware of the fact that our
energy scale of $Q \simeq 0.6$ GeV from (\ref{SCALE}) is below the
experimental one of $Q \simeq 3$ GeV. It is also somewhat lower than
$Q \simeq $ 1 GeV, which has been assumed by Belyaev and Johnson in
their analysis of the distribution amplitude in terms of light--cone
quark models \cite{belyaev:98b}.

\section{Conclusions}

In this lecture I have discussed an alternative approach to
relativistic (quantum) physics based on Dirac's front form of
dynamics. It makes use of the fact that for relativistic systems the
choice of the time parameter is not unique. Our particular choice uses
null--planes tangent to the light--cone as hypersurfaces of equal
time. This apparently trivial change of coordinates has far--reaching
consequences:

\begin{itemize}  
  
\item The number of kinematical (i.e.~interaction independent)
  Poincar\'e generators becomes maximal; there are seven of them instead
  of the usual six, among them the boosts.

\item Lorentz boosts in $z$-direction become diagonal; the light--cone
  time and space coordinates, $x^\p$ and $x^\m$, respectively, do not
  get mixed but rather get rescaled.

\item As a consequence, for many--particle systems one can introduce
  frame--indep\-end\-ent relative coordinates, the longitudinal momentum
  fractions, $x_i$, and the relative transverse momenta, $\vc{k}_{\perp
    i}$.    

\item Because of a two--dimensional Galilei invariance, relative and
  center--of--mass motion separate. As a result, many formulae are
  reminiscent of nonrelativistic physics and thus very intuitive.
  
\item This is particularly true for light--cone wave functions which,
  due to the last two properties, are boost invariant and do not
  depend on the total momentum of the bound state.  They are therefore
  ideal tools to study relativistic particle systems.
  
\item The last statement even holds for relativistic quantum field
  theory where one combines the unique properties of light--cone
  quantization with a Fock space picture. The central feature making
  this a reasonable idea is the triviality of the light--cone vacuum
  which accordingly is an eigenstate of the fully interacting
  Hamiltonian. This implies that the Fock operators create the
  \emph{physical} particles from the ground state.

\end{itemize}

As expected, however, the principle of conservation of difficulties is
at work so that there are problems to overcome. Of particular concern
to us was the 'vacuum problem'. In instant--form quantum field theory,
especially in QCD, many nonperturbative phenomena are attributed to
the nontriviality of the vacuum which shows up via the appearance of
condensates. These suggest that the instant--form vacuum is a
complicated many--body state (like, e.g.~the BCS ground state). In
addition, many of these condensates signal the spontaneous (or
anomalous) breakdown of a symmetry. The conceptual problem which
arises at this point is to reconcile the existence of condensates with
the triviality of the light--cone vacuum.

The idea put forward in these lectures is to reconstruct ground state
properties from the particle spectrum. The latter is obtained by
solving the light--cone Schr\"odinger equation for masses and wave
functions of the associated bound states.  For a relativistic quantum
field theory, this, in principle, amounts to solving an infinite
system of coupled integral equations for the amplitudes to find an
ever increasing number of constituents in the bound state. Experience,
however, shows that the \emph{light--cone} amplitudes to find more
than the valence quanta in the bound state tend to become rather
small. Note that the same is \emph{not} true within ordinary, that
is, instant--form quantization. 

A first explorative step towards solving a realistic light--cone
Schr\"odinger equation was performed using an effective field theory,
the NJL model. We have seen that, though we made a number of
approximations, in particular by \emph{enforcing} a constituent
picture, a number of pionic observables are predicted with reasonable
accuracy, among them the pion electromagnetic form factor, the pion
charge and core radius and the pion valence structure function at low
normalization scale. The pion distribution amplitude (in the chiral
and large--cutoff limit) turns out to be asymptotic.

The results presented in these lectures provide some confidence that,
also for real QCD, light--cone quantization may provide a road towards
a reasonable constituent picture, in which hadrons are consistently
described as bound states of a minimal number of constituents. How
this hope can be turned into fact remains to be seen.

\section*{Acknowledgments}

It is a pleasure to thank the organizers of the 39th Schladming winter
school, in particular W.~Schweiger and H.~Latal, for inviting me to
lecture and providing such a stimulating environment, both
scientifically and socially.



\newpage

\begin{thebibliography}


\bibitem{}{abdalla:96}{\protect\citeauthoryear{Abdalla and Abdalla}{Abdalla and
  Abdalla}{1996}}
Abdalla, E. and M.~C.~B. Abdalla (1996).
\newblock Updating QCD in two dimensions.
\newblock {\em Phys. Rept.\/}~{\em 265}, 253.

\bibitem{}{amendolia:86}{\protect\citeauthoryear{Amendolia et~al.}{Amendolia
  et~al.}{1986}}
Amendolia, S.~R. et~al. (1986).
\newblock A measurement of the space--like pion electromagnetic form factor.
\newblock {\em Nucl. Phys.\/}~{\em B277}, 168.

\bibitem{}{antonuccio:97a}{\protect\citeauthoryear{Antonuccio
et~al.}{Antonuccio, Brodsky, and  Dalley}{1997}}
Antonuccio, F., S.~J. Brodsky, and S.~Dalley (1997).
\newblock Light cone wave functions at small $x$.
\newblock {\em Phys. Lett.\/}~{\em B412}, 104.

\bibitem{}{ashery:99}{\protect\citeauthoryear{Ashery}{Ashery}{1999}}
Ashery, D. (1999).
\newblock Diffractive dissociation of high momentum pions.
\newblock hep-ex/9910024.

\bibitem{}{banks:99}{\protect\citeauthoryear{Banks}{Banks}{1999}}
Banks, T. (1999)
\newblock TASI lectures on matrix theory.
\newblock hep-th/9911068.

\bibitem{}{banks:97}{\protect\citeauthoryear{Banks et~al.}{Banks,
Fischler, Shenker, and Susskind}{1997}}
Banks, T., W.~Fischler, S.~Shenker, and L.~Susskind (1997).
\newblock M theory as a matrix model: A conjecture.
\newblock {\em Phys.~Rev.\/}~{\em D 55}, 5112.

\bibitem{}{bardakci:68}{\protect\citeauthoryear{Bardakci and Halpern}{Bardakci and Halpern}{1968}}
Bardakci, K., M.B.~Halpern (1968).
\newblock Theories at infinite momentum.
\newblock {\em Phys.~Rev.\/}~{\em 176}, 1686.

\bibitem{}{bardeen:80}{\protect\citeauthoryear{Bardeen
et~al.}{Bardeen, Pearson, and Rabinovici}{1980}}
Bardeen, W.~A., R.~B. Pearson, and E.~Rabinovici (1980).
\newblock Hadron masses in Quantum Chromodynamics on the transverse lattice.
\newblock {\em Phys. Rev.\/}~{\em D21}, 1037.

\bibitem{}{bassetto:91a}{\protect\citeauthoryear{Bassetto
et~al.}{Bassetto, Nardelli, and Soldati}{1991}}
Bassetto, A., G.~Nardelli, and R.~Soldati (1991).
\newblock Yang--Mills theories in algebraic noncovariant gauges: Canonical
  quantization and renormalization.
\newblock World Scientific, Singapore.

\bibitem{}{bassetto:00}{\protect\citeauthoryear{Bassetto
et~al.}{Bassetto, Vian, and Griguolo}{2000}}
Bassetto, A., F.~Vian, and L.~Griguolo (2000).
\newblock Light--front vacuum and instantons in two dimensions.
\newblock hep-th/0004026.

\bibitem{}{belyaev:98a}{\protect\citeauthoryear{Belyaev and
Johnson}{Belyaev and Johnson}{1998a}}
Belyaev, V. and M.~Johnson (1998a).
\newblock Twist--2 light--cone pion wave function.
\newblock {\em Mod.~Phys.~Lett.\/}~{\em A13}, 2909.

\bibitem{}{belyaev:98b}{\protect\citeauthoryear{Belyaev and Johnson}{Belyaev
  and Johnson}{1998b}}
Belyaev, V. and M.~Johnson (1998b).
\newblock Pion light--cone wave functions and light--front quark model.
\newblock {Phys.~Lett.\/}~{\em B423}, 379.

\bibitem{}{bentz:99}{\protect\citeauthoryear{Bentz et~al.}{Bentz,
Hama, Matsuki, and Yazaki}{1999}}
Bentz, W., T.~Hama, T.~Matsuki, and K.~Yazaki (1999).
\newblock NJL model on the light--cone and pion structure function.
\newblock {\em Nucl. Phys.\/}~{\em A651}, 143.

\bibitem{}{bernard:85}{\protect\citeauthoryear{Bernard
et~al.}{Bernard, Brockmann, and Weise}{1985}}
Bernard, V., R.~Brockmann, and W.~Weise (1985).
\newblock The Goldstone pion and the quark--antiquark pion (II). Pion
size and decay.
\newblock Regensburg Preprint, TPR 85-02.

\bibitem{}{bethe:57}{\protect\citeauthoryear{Bethe and Salpeter}{Bethe and
  Salpeter}{1957}}
Bethe, H. and E.~Salpeter (1957).
\newblock {\em Quantum Mechanics of One- and Two-Electron Atoms}.
\newblock Springer, Berlin, Heidelberg, New York.


\bibitem{}{bjorken:64}{\protect\citeauthoryear{Bjorken and Drell}{Bjorken and
  Drell}{1964}}
Bjorken, J. and S.~Drell (1964).
\newblock {\em Relativistic Quantum Mechanics}.
\newblock McGraw-Hill.

\bibitem{}{blin:88}{\protect\citeauthoryear{Blin et~al.}{Blin, Hiller,
and Schaden}{1988}}
Blin, A.~H., B.~Hiller, and M.~Schaden (1988).
\newblock Electromagnetic form factors in the Nambu--Jona-Lasinio model.
\newblock {\em Z. Phys.\/}~{\em A331}, 75.

\bibitem{}{bogolubov:90}{\protect\citeauthoryear{Bogolubov
et~al.}{Bogolubov, Logunov, Oksak, and Todorov}{1990}}
Bogolubov, N., A.~Logunov, A.~Oksak, and I.~Todorov (1990).
\newblock {\em General Principles of Quantum Field Theory}.
\newblock Kluwer Academic.

\bibitem{}{bogolubov:75}{\protect\citeauthoryear{Bogolubov
et~al.}{Bogolubov, Logunov, and Todorov}{1975}}
Bogolubov, N., A.~Logunov, and I.~Todorov (1975).
\newblock {\em Introduction to Axiomatic Field Theory}.
\newblock Benjamin, New York.

\bibitem{}{borderies:93}{\protect\citeauthoryear{Borderies
et~al.}{Borderies, Grange, and Werner}{1993}}
Borderies, A., P.~Grange, and E.~Werner (1993).
\newblock Light--cone quantization and 1/$N$ expansion.
\newblock {\em Phys. Lett.\/}~{\em B319}, 490.

\bibitem{}{borderies:95}{\protect\citeauthoryear{Borderies
et~al.}{Borderies, Grange, and Werner}{1995}}
Borderies, A., P.~Grange, and E.~Werner (1995).
\newblock Broken phase of the $O(N)$ $\phi^4$ model in light--cone quantization
  and 1/$N$ expansion.
\newblock {\em Phys. Lett.\/}~{\em B345}, 458.

\bibitem{}{brodsky:85}{\protect\citeauthoryear{Brodsky et~al.}{Brodsky, Ji, and
  Sawicki}{1985}}
Brodsky, S., C.-R. Ji, and M.~Sawicki (1985).
\newblock Evolution equation and relativistic bound--state wave functions for
  scalar field models in four and six dimensions.
\newblock {\em Phys.~Rev.\/}~{\em D32}, 1530.

\bibitem{}{brodsky:89}{\protect\citeauthoryear{Brodsky and Lepage}{Brodsky and
  Lepage}{1989}}
Brodsky, S. and G.~Lepage (1989).
\newblock Exclusive processes in Quantum Chromodynamics.
\newblock in: {\em Perturbative Quantum Chromodynamics}, A.H.~Mueller, ed.,
  World Scientific, Singapore.

\bibitem{}{brodsky:98}{\protect\citeauthoryear{Brodsky, Pauli, and
  Pinsky}{Brodsky et~al.}{1998}}
Brodsky, S.~J., H.-C. Pauli, and S.~S. Pinsky (1998).
\newblock Quantum Chromodynamics and other field theories on the light--cone.
\newblock {\em Phys. Rept.\/}~{\em 301}, 299.

\bibitem{}{brodsky:73b}{\protect\citeauthoryear{Brodsky
et~al.}{Brodsky, Roskies, and Suaya}{1973}}
Brodsky, S.~J., R.~Roskies, and R.~Suaya (1973).
\newblock Quantum electrodynamics and renormalization theory in the infinite
  momentum frame.
\newblock {\em Phys. Rev.\/}~{\em D8}, 4574.

\bibitem{}{burkardt:96c}{\protect\citeauthoryear{Burkardt}{Burkardt}{1996}}
Burkardt, M. (1996).
\newblock Parity invariance and effective light--front hamiltonians.
\newblock {\em Phys. Rev.\/}~{\em D54}, 2913.

\bibitem{}{burkardt:98}{\protect\citeauthoryear{Burkardt}{Burkardt}{1998}}
Burkardt, M. (1998).
\newblock Finiteness conditions for light front--hamiltonians.
\newblock {\em Phys. Rev.\/}~{\em D57}, 1136.

\bibitem{}{callan:76b}{\protect\citeauthoryear{Callan
  et~al.}{Callan, Coote, and Gross}{1976}}
Callan, C., N.~Coote, and D.~Gross (1976).
\newblock Two--dimensional Yang--Mills theory: A model of quark
confinement. 
\newblock {\em Phys.~Rev.\/}~{\em D13}, 1649.

\bibitem{}{casher:76}{\protect\citeauthoryear{Casher}{Casher}{1976}}
Casher, A. (1976).
\newblock Gauge fields on the null plane.
\newblock {\em Phys. Rev.\/}~{\em D14}, 452.

\bibitem{}{casher:71}{\protect\citeauthoryear{Casher et~al.}{Casher,
Noskowicz, and Susskind}{1971}}
Casher, A., S.~Noskowicz, and L.~Susskind (1971).
\newblock Goldstone--Nambu bosons in dual and parton theories. 
\newblock {\em Nucl.~Phys.\/}~{\em B32}, 75.

\bibitem{}{casher:74}{\protect\citeauthoryear{Casher and Susskind}{Casher and
  Susskind}{1974}}
Casher, A. and L.~Susskind (1974).
\newblock A quark model of mesons based on chiral symmetry.
\newblock {\em Phys.~Rev.\/}~{\em D9}, 436.

\bibitem{}{chakrabarti:99}{\protect\citeauthoryear{Chakrabarti
  et~al.}{Chakrabarti, Mukherjee, Kundu, and Harindranath}{1999}}
Chakrabarti, D., A.~Mukherjee, R.~Kundu, and A.~Harindranath (1999).
\newblock A numerical experiment in DLCQ: Microcausality, continuum limit and
  all that.
\newblock {\em Phys.Lett.\/}~{\em B480}, 409.

\bibitem{}{chakrabarty:89}{\protect\citeauthoryear{Chakrabarty 
et~al.}{Chakrabarty, Gupta, Singh, and Mitra}{1989}}
Chakrabarty, S., K.~Gupta, N.N.~Singh, and  A.N.~Mitra (1989).
\newblock Hadron spectroscopy and form factors at quark level.
\newblock {\em Prog.~Part.~Nucl.~Phys.\/}~{\em 22}, 43.

\bibitem{}{chang:73a}{\protect\citeauthoryear{Chang
  et~al.}{Chang, Root, and Yan}{1973}}
Chang, S.-J., R.~G. Root, and T.-M. Yan (1973).
\newblock Quantum field theories in the infinite momentum frame. I.
  Quantization of scalar and Dirac fields.
\newblock {\em Phys. Rev.\/}~{\em D7}, 1133.

\bibitem{}{chernyak:84}{\protect\citeauthoryear{Chernyak and
  Zhitnitsky}{Chernyak and Zhitnitsky}{1984}}
Chernyak, V. and A.~Zhitnitsky (1984).
\newblock Asymptotic behaviour of exclusive processes in QCD. 
\newblock {\em Phys.~Rept.\/}~{\em 112}, 173.

\bibitem{}{chung:88b}{\protect\citeauthoryear{Chung et~al.}{Chung, Coester, and
  Polyzou}{1988}}
Chung, P., F.~Coester, and W.~Polyzou (1988).
\newblock Charge form factors of quark--model pions.
\newblock {\em Phys.~Lett.\/}~{\em B205}, 545.

\bibitem{}{coester:92}{\protect\citeauthoryear{Coester}{Coester}{1992}}
Coester, F. (1992).
\newblock Null--plane dynamics of particles and fields.
\newblock {\em Prog. Part. Nucl. Phys.\/}~{\em 29}, 1.

\bibitem{}{coester:82}{\protect\citeauthoryear{Coester and Polyzou}{Coester and
  Polyzou}{1982}}
Coester, F. and W.~Polyzou (1982).
\newblock Relativistic quantum mechanics of particles with direct interaction.
\newblock {\em Phys.~Rev.\/}~{\em D26}, 1348.

\bibitem{}{coleman:73c}{\protect\citeauthoryear{Coleman}{Coleman}{1973}}
Coleman, S. (1973).
\newblock There are no Goldstone bosons in two dimensions.
\newblock {\em Comm. Math. Phys.\/}~{\em 31}, 259.

\bibitem{}{courant:62}{\protect\citeauthoryear{Courant and Hilbert}{Courant and
  Hilbert}{1962}}
Courant, R. and D.~Hilbert (1962).
\newblock {\em Methods of Mathematical Physics}.
\newblock Interscience, New York.

\bibitem{}{dancoff:50}{\protect\citeauthoryear{Dancoff}{Dancoff}{1950}}
Dancoff, S. (1950).
\newblock Nonadiabatic meson theory of nuclear forces.
\newblock {\em Phys.~Rev.\/}~{\em 78}, 382.

\bibitem{}{dewitt:83}{\protect\citeauthoryear{DeWitt}{DeWitt}{1984}}
DeWitt, B. (1984).
\newblock The space-time approach to quantum field theory.
\newblock in: {\em Relativity, Groups and Topology, II}, B.S.~DeWitt and
  R.~Stora, eds., Proceedings Les Houches 1983, Session XL, North-Holland,
  Amsterdam.

\bibitem{}{diakonov:95}{\protect\citeauthoryear{Diakonov}{Diakonov}{1996}}
Diakonov, D. (1996).
\newblock Chiral symmetry breaking by instantons.
\newblock in: {\em Selected Topics in Nonperturbative QCD}, A.~Di Giacomo and
  D.~Diakonov, eds., Proceedings International School of Physics ``Enrico
  Fermi'', Course CXXX, Varenna, Italy, 1995, IOS Press, Amsterdam.

\bibitem{}{dietmaier:89}{\protect\citeauthoryear{Dietmaier
et~al.}{Dietmaier, Heinzl, Schaden, and Werner}{1989}}
Dietmaier, C., T.~Heinzl, M.~Schaden, and E.~Werner (1989).
\newblock The fermion condensate of the Nambu--Jona-Lasinio model in
light--cone quantization.
\newblock {\em Z. Phys.\/}~{\em A334}, 220.

\bibitem{}{dirac:49}{\protect\citeauthoryear{Dirac}{Dirac}{1949}}
Dirac, P. A.~M. (1949).
\newblock Forms of relativistic dynamics.
\newblock {\em Rev. Mod. Phys.\/}~{\em 21}, 392.

\bibitem{}{dirac:64}{\protect\citeauthoryear{Dirac}{Dirac}{1964}}
Dirac, P. A.~M. (1964).
\newblock {\em Lectures on quantum mechanics}.
\newblock Benjamin, New York.

\bibitem{}{domokos:71}{\protect\citeauthoryear{Domokos}{Domokos}{1972}}
Domokos, G. (1972).  
\newblock Introduction to the characteristic initial--value problem in quantum
  field theory.
\newblock in: Boulder Lectures, Vol. XIV,  A.O. Barut, W.E. Brittin, eds.,
  Colorado University Press, Boulder.

\bibitem{}{dosch:98}{\protect\citeauthoryear{Dosch and Narison}{Dosch and
  Narison}{1998}}
Dosch, H. and S.~Narison (1998).
\newblock Direct extraction of the chiral quark condensate and bounds on the
  light quark masses.
\newblock {\em Phys.~Lett.\/}~{\em B417}, 173.

\bibitem{}{drell:70c}{\protect\citeauthoryear{Drell and Yan}{Drell and
  Yan}{1970}}
Drell, S. and T.-M. Yan (1970).
\newblock Connection of elastic electromagnetic nucleon form factors at large
  $q^2$ and deep inelastic structure functions near threshold.
\newblock {\em Phys.~Rev.~Lett.\/}~{\em 24}, 181.

\bibitem{}{dziembowski:88}{\protect\citeauthoryear{Dziembowski}{Dziembowski}{1988}}
Dziembowski, Z. (1988).
\newblock Relativistic model of nucleon and pion structure: Static properties
  and electromagnetic soft form factors.
\newblock {\em Phys.~Rev.\/}~{\em D37}, 778.

\bibitem{}{faddeev:88}{\protect\citeauthoryear{Faddeev and Jackiw}{Faddeev and
  Jackiw}{1988}}
Faddeev, L. and R.~Jackiw (1988).
\newblock Hamiltonian reduction of unconstrained and constrained systems.
\newblock {\em Phys. Rev. Lett.\/}~{\em 60}, 1692.

\bibitem{}{fock:35}{\protect\citeauthoryear{Fock}{Fock}{1935}}
Fock, V. (1935).
\newblock Zur Theorie des Wasserstoffatoms.
\newblock {\em Z.~Phys.\/}~{\em 98}, 145.

\bibitem{}{fubini:73}{\protect\citeauthoryear{Fubini et~al.}{Fubini,
Hanson, and Jackiw}{1973}}
Fubini, S., A.~Hanson, and R.~Jackiw (1973).
\newblock New approach to field theory.
\newblock {\em Phys.~Rev.\/}~{\em D7}, 1732.

\bibitem{}{fuda:90}{\protect\citeauthoryear{Fuda}{Fuda}{1990}}
Fuda, M. (1990).
\newblock A new picture for light--front dynamics.
\newblock {\em Ann.~Phys.~(N.Y.)\/}~{\em 197}, 265.

\bibitem{}{gelfand:64}{\protect\citeauthoryear{Gelfand and Shilov}{Gelfand and
  Shilov}{1964}}
Gelfand, I. and G.~Shilov (1964).
\newblock {\em Generalized Functions}.
\newblock Academic Press, New York.

\bibitem{}{gell-mann:68}{\protect\citeauthoryear{Gell-Mann
et~al.}{Gell-Mann, Oakes, and Renner}{1968}}
Gell-Mann, M., R.~J. Oakes, and B.~Renner (1968).
\newblock Behavior of current divergences under $SU(3) \times SU(3)$.
\newblock {\em Phys. Rev.\/}~{\em 175}, 2195.

\bibitem{}{gitman:90}{\protect\citeauthoryear{Gitman and Tyutin}{Gitman and
  Tyutin}{1990}}
Gitman, D. and I.~Tyutin (1990).
\newblock {\em Quantization of Fields with Constraints}.
\newblock Sprin\-ger, Berlin, Hei\-del\-berg, New York.

\bibitem{}{glueck:99}{\protect\citeauthoryear{Gl{\"u}ck
et~al.}{Gl{\"u}ck, Reya, and Schienbein}{1999}}
Gl{\"u}ck, M., E.~Reya, and I.~Schienbein (1999).
\newblock Pionic parton distributions revisited.
\newblock {\em Eur. Phys. J.\/}~{\em C10}, 313.

\bibitem{}{goldstone:61}{\protect\citeauthoryear{Goldstone}{Goldstone}{1961}}
Goldstone, J. (1961).
\newblock Field theories with 'superconductor' solutions.
\newblock {\em Nuovo Cim.\/}~{\em 19}, 154.

\bibitem{}{green:87}{\protect\citeauthoryear{Green
  et~al.}{Green, Schwarz, and Witten}{1987}}
Green, M., J.~Schwarz, and E.~Witten (1987).
\newblock {\em Superstring Theory}.
\newblock Cambridge University Press, Cambridge.

\bibitem{}{gribov:78}{\protect\citeauthoryear{Gribov}{Gribov}{1978}}
Gribov, V.~N. (1978).
\newblock Quantization of non--Abelian gauge theories.
\newblock {\em Nucl. Phys.\/}~{\em B139}, 1.

\bibitem{}{gross:88}{\protect\citeauthoryear{Gross}{Gross}{1989}}
Gross, F. (1989)
\newblock Relativistic nuclear physics with the spectator model.
\newblock in: {\em Nuclear and Particle Physics on the Light--Cone},
Proceedings, LAMPF Workshop, Los Alamos, 1988, M.B.~Johnson,
L.S.~Kisslinger, eds., World Scientific, Singapore.

\bibitem{}{hanson:76}{\protect\citeauthoryear{Hanson et~al.}{Hanson, Regge, and
  Teitelboim}{1976}}
Hanson, A., T.~Regge, and C.~Teitelboim (1976).
\newblock {\em Constrained Hamiltonian Systems}.
\newblock Accademia Nazionale dei Lincei, Rome.

\bibitem{}{harada:98}{\protect\citeauthoryear{Harada
  et~al.}{Harada, Heinzl, and Stern}{1998}}
Harada, K., T.~Heinzl, and C.~Stern (1998).
\newblock Variational mass perturbation theory for light--front bound--state
  equations.
\newblock {\em Phys. Rev.\/}~{\em D57}, 2460.

\bibitem{}{harada:94}{\protect\citeauthoryear{Harada et~al.}{Harada,
Sugihara, Taniguchi, and Yahiro}{1994}}
Harada, K., T.~Sugihara, M.~Taniguchi, and M.~Yahiro (1994).
\newblock The massive Schwinger model with $SU(2)_f$ on the light--cone.
\newblock {\em Phys. Rev.\/}~{\em D49}, 4226.

\bibitem{}{harindranath:90}{\protect\citeauthoryear{Harindranath et~al.}{Harindranath, Perry and Wilson}{1990}}
Harindranath, A., R.J.~Perry, and K.G.~Wilson (1990).
\newblock Light--front Tamm--Dancoff field theory.
\newblock {\em Phys. Rev. Lett.\/}~{\em 65}, 2959.

\bibitem{}{hatsuda:94}{\protect\citeauthoryear{Hatsuda and Kunihiro}{Hatsuda
  and Kunihiro}{1994}}
Hatsuda, T. and T.~Kunihiro (1994).
\newblock QCD phenomenology based on a chiral effective Lagrangian.
\newblock {\em Phys. Rept.\/}~{\em 247}, 221.

\bibitem{}{heinzl:95b}{\protect\citeauthoryear{Heinzl}{Heinzl}{1996a}}
Heinzl, T. (1996a).
\newblock Hamiltonian formulations of Yang--Mills quantum theory and the Gribov
  problem.
\newblock hep-th/9604018.

\bibitem{}{heinzl:96b}{\protect\citeauthoryear{Heinzl}{Heinzl}{1996b}}
Heinzl, T. (1996b).
\newblock Fermion condensates and the trivial vacuum of light--cone quantum
  field theory.
\newblock {\em Phys. Lett.\/}~{\em B388}, 129.

\bibitem{}{heinzl:98}{\protect\citeauthoryear{Heinzl}{Heinzl}{1998}}
Heinzl, T. (1998).
\newblock Light--cone dynamics of particles and fields.
\newblock hep-th/9812190.

\bibitem{}{heinzl:92c}{\protect\citeauthoryear{Heinzl et~al.}{Heinzl,
Krusche, Simb\"urger, and Werner}{1992}}
Heinzl, T., S.~Krusche, S.~Simb\"urger, and E.~Werner (1992).
\newblock Nonperturbative light--cone quantum field theory beyond the tree
  level.
\newblock {\em Z. Phys.\/}~{\em C56}, 415.

\bibitem{}{heinzl:99}{\protect\citeauthoryear{Heinzl et~al.}{Heinzl, Scheu, and
  Kr{\"o}ger}{1999}}
Heinzl, T., N.~Scheu, and H.~Kr{\"o}ger (1999).
\newblock Loss of causality in discretized light--cone quantization.
\newblock hep-th/9908173. 

\bibitem{}{heinzl:96a}{\protect\citeauthoryear{Heinzl et~al.}{Heinzl,
Stern, Werner, and Zellermann}{1996}}
Heinzl, T., C.~Stern, E.~Werner, and B.~Zellermann (1996).
\newblock The vacuum structure of light--front $\phi^4_{1+1}$ theory.
\newblock {\em Z. Phys.\/}~{\em C72}, 353.

\bibitem{}{heinzl:94a}{\protect\citeauthoryear{Heinzl and Werner}{Heinzl and
  Werner}{1994}}
Heinzl, T. and E.~Werner (1994).
\newblock Light--front quantization as an initial--boundary--value problem.
\newblock {\em Z. Phys.\/}~{\em C62}, 521.

\bibitem{}{hellerman:99}{\protect\citeauthoryear{Hellerman and
  Polchinski}{Hellerman and Polchinski}{1999}}
Hellerman, S. and J.~Polchinski (1999).
\newblock Compactification in the lightlike limit.
\newblock {\em Phys.~Rev.\/}~{\em D 59}, 125002.

\bibitem{}{hiller:00}{\protect\citeauthoryear{Hiller}{Hiller}{2000}}
Hiller, J. (2000).
\newblock Calculations with DLCQ.
\newblock to appear in the Proceedings of 10th International Light Cone 
Meeting on Non-Perturbative QCD and Hadron Phenomenology: From Hadrons
to Strings, Heidelberg, Germany, June 12--16, 2000, hep-ph/0007309.

\bibitem{}{holstein:95}{\protect\citeauthoryear{Holstein}{Holstein}{1995}}
Holstein, B.~R. (1995).
\newblock Chiral perturbation theory: A primer.
\newblock hep-ph/9510344. 

\bibitem{}{holstein:98}{\protect\citeauthoryear{Holstein}{Holstein}{1998}}
Holstein, B.~R. (1998).
\newblock Klein's paradox.
\newblock {\em Am.~J.~Phys.\/}~{\em 66}, 507.

\bibitem{}{hornbostel:92}{\protect\citeauthoryear{Hornbostel}{Hornbostel}{1992}}
Hornbostel, K. (1992).
\newblock Nontrivial vacua from equal time to the light--cone.
\newblock {\em Phys. Rev.\/}~{\em D45}, 3781.

\bibitem{}{hornbostel:88}{\protect\citeauthoryear{Hornbostel
et~al.}{Hornbostel, Brodsky, and Pauli}{1988}}
Hornbostel, K., S.~J. Brodsky, and H.-C. Pauli (1988).
\newblock Quantization on the light--cone: Response to a comment by Hagen.
\newblock {\em Phys. Rev.\/}~{\em D37}, 2363.

\bibitem{}{itakura:99}{\protect\citeauthoryear{Itakura and Maedan}{Itakura and
  Maedan}{2000}}
Itakura, K. and S.~Maedan (2000).
\newblock Dynamical chiral symmetry breaking on the light--front. I: DLCQ
  approach.
\newblock {\em Phys. Rev.\/}~{\em D61}, 045009.

\bibitem{}{jackiw:87}{\protect\citeauthoryear{Jackiw}{Jackiw}{1987}}
Jackiw, R. (1987). 
\newblock Functional representations for quantized fields.
\newblock in: {\em Conformal Field Theory, Anomalies, and Superstrings},
  C.K.~Chew et al., eds., Proceedings of the First Asia Pacific Workshop on
  High Energy Physics, Singapore, 1987, World Scientific, Singapore.

\bibitem{}{jackiw:93}{\protect\citeauthoryear{Jackiw}{Jackiw}{1993}}
Jackiw, R. (1993).
\newblock (Constrained) quantisation without tears.
\newblock hep-th/9306075. 

\bibitem{}{jaen:84}{\protect\citeauthoryear{Jaen
  et~al.}{Jaen, Molina, and Llosa}{1984}}
Jaen, X., A.~Molina, and J.~Llosa (1984).
\newblock Front form predictive mechanics non--interaction theorem.
\newblock in: Lecture Notes in Physics, 212, Springer, Berlin,
Heidelberg, New York.

\bibitem{}{jaus:90}{\protect\citeauthoryear{Jaus}{Jaus}{1990}}
Jaus, W. (1990).
\newblock Semileptonic decays of B and D mesons in the light--front formalism.
\newblock {\em Phys.~Rev.\/}~{\em D41}, 3394.

\bibitem{}{ji:92}{\protect\citeauthoryear{Ji
  et~al.}{Ji, Chung, and Cotanch}{1992}}
Ji, C.-R., P.~Chung, and S.~Cotanch (1992).
\newblock Light--cone quark--model axial--vector wave function.
\newblock {\em Phys.~Rev.\/}~{\em D45}, 4214.

\bibitem{}{ji:90}{\protect\citeauthoryear{Ji and Cotanch}{Ji and
  Cotanch}{1990}}
Ji, C.-R. and S.~Cotanch (1990).
\newblock Simple relativistic quark--model analysis of flavored pseudoscalar
  mesons.
\newblock {\em Phys.~Rev.\/}~{\em D41}, 2319.

\bibitem{}{jordan:28}{\protect\citeauthoryear{Jordan and Pauli}{Jordan and
  Pauli}{1928}}
Jordan, P. and W.~Pauli (1928).
\newblock Zur Quantenelektrodynamik ladungsfreier Felder. 
\newblock {\em Z. Phys.\/}~{\em 47}, 151.

\bibitem{}{kaplan:95}{\protect\citeauthoryear{Kaplan}{Kaplan}{1995}}
Kaplan, D. (1995).
\newblock Effective field theories.
\newblock nucl-th/9506035.

\bibitem{}{kiefer:94}{\protect\citeauthoryear{Kiefer and Wipf}{Kiefer and
  Wipf}{1994}}
Kiefer, C. and A.~Wipf (1994).
\newblock Functional Schr\"odinger equation for fermions in external gauge
  fields.
\newblock {\em Ann.~Phys. (N.Y.)\/}~{\em 236}, 241.

\bibitem{}{klevansky:92}{\protect\citeauthoryear{Klevansky}{Klevansky}{1992}}
Klevansky, S.~P. (1992).
\newblock The Nambu--Jona-Lasinio model of Quantum Chromodynamics.
\newblock {\em Rev. Mod. Phys.\/}~{\em 64}, 649.

\bibitem{}{kroll:96}{\protect\citeauthoryear{Kroll and Raulfs}{Kroll and
  Raulfs}{1996}}
Kroll, P. and M.~Raulfs (1996).
\newblock The $\pi \gamma$ transition form factor and the pion wave function.
\newblock {\em Phys.~Lett.\/}~{\em B387}, 848.

\bibitem{}{lavelle:87}{\protect\citeauthoryear{Lavelle
et~al.}{Lavelle, Werner, and Glazek}{1987}}
Lavelle, M., E.~Werner, and S.~Glazek (1987).
\newblock Hadron wave functions with condensate induced running masses.
\newblock {\em Few Body Systems, Suppl.\/}~{\em 2}, 519.

\bibitem{}{leibbrandt:84}{\protect\citeauthoryear{Leibbrandt}{Leibbrandt}{1984}}
Leibbrandt, G. (1984).
\newblock The light--cone gauge in Yang--Mills theory. 
\newblock {\em Phys.~Rev.\/}~{\em D29}, 1699.

\bibitem{}{lenz:91}{\protect\citeauthoryear{Lenz et~al.}{Lenz, Thies,
Yazaki, and Levit}{1991}}
Lenz, F., M.~Thies, K.~Yazaki, and S.~Levit (1991).
\newblock Hamiltonian formulation of two-dimensional gauge theories on the
  light--cone.
\newblock {\em Ann.~Phys.~(N.Y.)\/}~{\em 208}, 1.

\bibitem{}{lepage:80}{\protect\citeauthoryear{Lepage and Brodsky}{Lepage and
  Brodsky}{1980}}
Lepage, G. and S.~Brodsky (1980).
\newblock Exclusive processes in Quantum Chromodynamics.
\newblock {\em Phys.~Rev.\/}~{\em D22}, 2157.

\bibitem{}{lepage:81}{\protect\citeauthoryear{Lepage et~al.}{Lepage,
Brodsky, Huang, and Mackenzie}{1981}}
Lepage, G., S.~Brodsky, T.~Huang, and P.~Mackenzie (1981).
\newblock Hadronic wave functions in QCD.
\newblock Proceedings of the Banff Summer Institute.

\bibitem{}{lepage:79a}{\protect\citeauthoryear{Lepage and Brodsky}{Lepage and
  Brodsky}{1979a}}
Lepage, G.~P. and S.~J. Brodsky (1979a).
\newblock Exclusive processes in Quantum Chromodynamics: Evolution equations
  for hadronic wave functions and the form factors of mesons.
\newblock {\em Phys. Lett.\/}~{\em B87}, 359.

\bibitem{}{lepage:79b}{\protect\citeauthoryear{Lepage and Brodsky}{Lepage and
  Brodsky}{1979b}}
Lepage, G.~P. and S.~J. Brodsky (1979b).
\newblock Exclusive processes in Quantum Chromodynamics: the form factors of
  baryons at large momentum transfer.
\newblock {\em Phys. Rev. Lett.\/}~{\em 43}, 545.

\bibitem{}{leutwyler:65}{\protect\citeauthoryear{Leutwyler}{Leutwyler}{1965}}
Leutwyler, H. (1965).
\newblock A no--interaction theorem in classical relativistic Hamiltonian
  particle mechanics.
\newblock {\em Nuovo Cim.\/}~{\em 37}, 556.

\bibitem{}{leutwyler:74a}{\protect\citeauthoryear{Leutwyler}{Leutwyler}{1974a}}
Leutwyler, H. (1974a).
\newblock On the average transverse momentum of the quarks within a meson.
\newblock {\em Phys.~Lett.\/}~{\em B48}, 45.

\bibitem{}{leutwyler:74b}{\protect\citeauthoryear{Leutwyler}{Leutwyler}{1974b}}
Leutwyler, H. (1974b).
\newblock Mesons in terms of quarks on a null--plane.
\newblock {\em Nucl.~Phys.\/}~{\em B76}, 413.

\bibitem{}{leutwyler:78}{\protect\citeauthoryear{Leutwyler and Stern}{Leutwyler
  and Stern}{1978}}
Leutwyler, H. and J.~Stern (1978).
\newblock Relativistic dynamics on a null--plane.
\newblock {\em Ann.~Phys.~(N.Y.)\/}~{\em 112}, 94.

\bibitem{}{liu:93}{\protect\citeauthoryear{Liu and Soper}{Liu and Soper}{1993}}
Liu, H.~H. and D.~E. Soper (1993).
\newblock Implementation of the Leibbrandt--Mandelstam gauge
prescription in the null--plane bound--state equation.
\newblock {\em Phys. Rev.\/}~{\em D48}, 1841.

\bibitem{}{lucha:91}{\protect\citeauthoryear{Lucha et~al.}{Lucha,
Sch\"oberl, and Gromes}{1991}}
Lucha, W., F.~Sch\"oberl, and D.~Gromes (1991).
\newblock Bound states of quarks.
\newblock {\em Phys.~Rept.\/}~{\em 200}, 127.

\bibitem{}{lunin:99}{\protect\citeauthoryear{Lunin and Pinsky}{Lunin and Pinsky}{1999}}
Lunin, O. and S..S.~Pinsky (1999).
\newblock SDLCQ: Supersymmetric discrete light--cone quantization.
\newblock in: {\it New Directions in Quantum Chromodynamics}, C.-R.~Ji and D.-P.~Min, eds., AIP Conference Proceedings 494, Melville, NY.   

\bibitem{}{mandelstam:83}{\protect\citeauthoryear{Mandelstam}{Mandelstam}{1983}}
Mandelstam, S. (1983).
\newblock Light--cone superspace and the ultraviolet finiteness of the
$N=4$ model.  
\newblock {\em Nucl.~Phys.\/}~{\em B213}, 149.

\bibitem{}{manohar:96}{\protect\citeauthoryear{Manohar}{Manohar}{1996}}
Manohar, A. (1997).
\newblock Effective field theories.
\newblock in: {\em Perturbative and Nonperturbative Aspects of Quantum Field
  Theory}, H.~Latal and W.~Schweiger, eds., Lecture Notes in Physics, Vol.~479,
  p.~311, Springer, Berlin, Heidelberg, New York.

\bibitem{}{manohar:84}{\protect\citeauthoryear{Manohar and Georgi}{Manohar and
  Georgi}{1984}}
Manohar, A. and H.~Georgi (1984).
\newblock Chiral quarks and the nonrelativistic quark model.
\newblock {\em Nucl.~Phys.\/}~{\em B234}, 189.

\bibitem{}{maris:00}{\protect\citeauthoryear{Maris and Tandy}{Maris
and Tandy}{2000}}
Maris, P. and  P.C.~Tandy (2000).
\newblock The quark photon vertex and the pion charge radius.
\newblock {\em Phys.Rev.\/}~{\em C61}, 045202.

\bibitem{}{maskawa:76}{\protect\citeauthoryear{Maskawa and Yamawaki}{Maskawa
  and Yamawaki}{1976}}
Maskawa, T. and K.~Yamawaki (1976).
\newblock The problem of $p^\p = 0$ mode in the null--plane field
theory and Dirac's method of quantization.
\newblock {\em Prog. Theor. Phys.\/}~{\em 56}, 270.

\bibitem{}{mccartor:92}{\protect\citeauthoryear{McCartor and Robertson}{McCartor and Robertson}{1992}}
McCartor, G. and  D.G.~Robertson (1992).
\newblock Bosonic zero modes in discretized light--cone field theory.
\newblock {\em Z.~Phys.\/}~{\em C53}, 679.

\bibitem{}{melosh:74}{\protect\citeauthoryear{Melosh}{Melosh}{1974}}
Melosh, H. (1974).
\newblock Quarks: Currents and constituents.
\newblock {\em Phys.Rev.\/}~{\em D9}, 1095.

\bibitem{}{michael:82}{\protect\citeauthoryear{Michael and Payne}{Michael and
  Payne}{1982}}
Michael, C. and F.~Payne (1982).
\newblock Bound states of heavy quarks on the light plane.
\newblock {\em Z.~Phys.\/}~{\em C12}, 145.

\bibitem{}{mukherjee:93}{\protect\citeauthoryear{Mukherjee
et~al.}{Mukherjee, Nag, Sanyal, Morii, Morishita, and Tsuge}{1993}}
Mukherjee, S., R. Nag, S. Sanyal, T. Morii, J. Morishita and M. Tsuge (1993). 
\newblock Quark potential approach to baryons and mesons.
\newblock {\em Phys.~Rept.\/}~{\em 231}, 201.

\bibitem{}{nakanishi:77}{\protect\citeauthoryear{Nakanishi and
  Yamawaki}{Nakanishi and Yamawaki}{1977}}
Nakanishi, N. and K.~Yamawaki (1977).
\newblock A consistent formulation of the null--plane quantum field theory.
\newblock {\em Nucl. Phys.\/}~{\em B122}, 15.

\bibitem{}{nambu:61a}{\protect\citeauthoryear{Nambu and Jona-Lasinio}{Nambu and
  Jona-Lasinio}{1961}}
Nambu, Y. and G.~Jona-Lasinio (1961).
\newblock Dynamical model of elementary particles based on an analogy with
  superconductivity. I.
\newblock {\em Phys. Rev.\/}~{\em 122}, 345.

\bibitem{}{namyslowski:85}{\protect\citeauthoryear{Namys{\l}owski}{Namys{\l}ow%
ski}{1985}}
Namys{\l}owski, J. (1985).
\newblock Light--cone perturbation theory and its application to different
  fields.
\newblock {\em Prog.~Part.~Nucl.~Phys.\/}~{\em 14}, 49.

\bibitem{}{neubert:94}{\protect\citeauthoryear{Neubert}{Neubert}{1994}}
Neubert, M. (1994).
\newblock Heavy quark symmetry.
\newblock {\em Phys. Rept.\/}~{\em 245}, 259.

\bibitem{}{neville:71a}{\protect\citeauthoryear{Neville and Rohrlich}{Neville
  and Rohrlich}{1971}}
Neville, R. and F.~Rohrlich (1971).
\newblock Quantum field theory off null--planes.
\newblock {\em Nuovo Cim.\/}~{\em 1A}, 625.

\bibitem{}{newton:49}{\protect\citeauthoryear{Newton and Wigner}{Newton and
  Wigner}{1949}}
Newton, T. and E.~Wigner (1949).
\newblock Localized states for elementary systems.
\newblock {\em Rev.~Mod.~Phys.\/}~{\em 21}, 400.

\bibitem{}{parker:70a}{\protect\citeauthoryear{Parker and Schmieg}{Parker and
  Schmieg}{1970}}
Parker, L. and G.~Schmieg (1970).
\newblock Special relativity and diagonal transformations.
\newblock {\em Am.~J.~Phys.\/}~{\em 38}, 218.

\bibitem{}{pauli:85a}{\protect\citeauthoryear{Pauli and Brodsky}{Pauli and
  Brodsky}{1985a}}
Pauli, H.-C. and S.~J. Brodsky (1985a).
\newblock Solving field theory in one space and one time dimension.
\newblock {\em Phys. Rev.\/}~{\em D32}, 1993.

\bibitem{}{pauli:85b}{\protect\citeauthoryear{Pauli and Brodsky}{Pauli and
  Brodsky}{1985b}}
Pauli, H.-C. and S.~J. Brodsky (1985b).
\newblock Discretized light--cone quantization: Solution to a field theory in
  one space and one time dimension.
\newblock {\em Phys. Rev.\/}~{\em D32}, 2001.

\bibitem{}{peierls:52}{\protect\citeauthoryear{Peierls}{Peierls}{1952}}
Peierls, R. (1952).
\newblock The commutation laws of relativistic field theory.
\newblock {\em Proc.~Roy.~Soc.\/}~{\em A214}, 143.

\bibitem{}{perry:99}{\protect\citeauthoryear{Perry}{Perry}{1999}}
Perry, R.~J. (1999).
\newblock Light--front Quantum Chromodynamics.
\newblock nucl-th/9901080.

\bibitem{}{petrov:97}{\protect\citeauthoryear{Petrov and Pobylitsa}{Petrov and
  Pobylitsa}{1997}}
Petrov, V.~Y. and P.~Pobylitsa (1997).
\newblock Pion wave function from the instanton vacuum model.
\newblock hep-ph/9712203. 

\bibitem{}{petrov:98}{\protect\citeauthoryear{Petrov et~al.}{Petrov,
Polyakov, Ruskov, Weiss, and Goeke}{1999}}
Petrov, V.~Y., M.~V. Polyakov, R.~Ruskov, C.~Weiss, and K.~Goeke (1999).
\newblock Pion and photon light--cone wave functions from the instanton vacuum.
\newblock {\em Phys. Rev.\/}~{\em D59}, 114018.

\bibitem{}{polyakov:87}{\protect\citeauthoryear{Polyakov}{Polyakov}{1987}}
Polyakov, A. (1987).
\newblock {\em Gauge Fields and Strings}.
\newblock Harwood Academic, Chur.

\bibitem{}{prokhvatilov:88}{\protect\citeauthoryear{Prokhvatilov and
  Franke}{Prokhvatilov and Franke}{1988}}
Prokhvatilov, E. and V.~Franke (1988).
\newblock Approximate description of QCD condensates in light--cone
coordinates. 
\newblock {\em Sov.~J,~Nucl.~Phys.\/}~{\em 47}, 559.

\bibitem{}{prokhvatilov:89}{\protect\citeauthoryear{Prokhvatilov and
  Franke}{Prokhvatilov and Franke}{1989}}
Prokhvatilov, E. and V.~Franke (1989).
\newblock Limiting transition to light--like coordinates and the QCD
  Hamiltonian.
\newblock {\em Sov.~J.~Nucl.~Phys.\/}~{\em 49}, 688.

\bibitem{}{radyushkin:95}{\protect\citeauthoryear{Radyushkin}{Radyushkin}{1995}}
Radyushkin, A.V. (1995).
\newblock Quark--hadron duality and intrinsic transverse momentum.
\newblock {\em Acta Phys. Polon.\/}~{\em B26}, 2067.

\bibitem{}{roberts:94}{\protect\citeauthoryear{Roberts and Williams}{Roberts
  and Williams}{1994}}
Roberts, C.~D. and A.~G. Williams (1994).
\newblock Dyson--Schwinger equations and their application to hadronic physics.
\newblock {\em Prog. Part. Nucl. Phys.\/}~{\em 33}, 477.

\bibitem{}{roberts:00}{\protect\citeauthoryear{Roberts and Schmidt}{Roberts
  and Schmidt}{2000}}
Roberts, C.~D. and S.~M. Schmidt (2000).
\newblock Dyson-Schwinger equations: density, temperature and continuum
strong QCD.
\newblock nucl-th/0005064.

\bibitem{}{rohrlich:71}{\protect\citeauthoryear{Rohrlich}{Rohrlich}{1971}}
Rohrlich, F. (1971).
\newblock Null--plane field theory.
\newblock {\em Acta Physica Austriaca, Suppl.\/}~{\em 8}, 277.

\bibitem{}{salmons:99}{\protect\citeauthoryear{Salmons
et~al.}{Salmons, Grange, and Werner}{1999}}
Salmons, S., P.~Grange, and E.~Werner (1999).
\newblock Field dynamics on the light--cone: Compact versus continuum
  quantization.
\newblock {\em Phys.~Rev.\/}~{\em D 60}, 067701.

\bibitem{}{savinov:95}{\protect\citeauthoryear{Savinov}{Savinov}{1995}}
Savinov, V. (1995).
\newblock A measurement of the form factors of light pseudoscalar mesons at a
  large momentum transfer.
\newblock hep-ex/9507005. 

\bibitem{}{savkli:98}{\protect\citeauthoryear{\c{S}avkl{\i} and
  Tabakin}{\c{S}avkl{\i} and Tabakin}{1998}}
\c{S}avkl{\i}, \c{C}. and F.~Tabakin (1998).
\newblock Quark--antiquark bound states within a Dyson--Schwinger
Bethe--Salpeter formalism.
\newblock {\em Nucl.~Phys.\/}~{\em A 628}, 645.

\bibitem{}{scheu:98}{\protect\citeauthoryear{Scheu}{Scheu}{1997}}
Scheu, N. (1997).
\newblock On the computation of structure functions and mass spectra in a
  relativistic Hamiltonian formalism: A lattice point of view.
\newblock Ph.D. Thesis, Universit{\'e} Laval, Qu{\'e}bec.
\newblock hep-th/9804190.  

\bibitem{}{schwinger:49}{\protect\citeauthoryear{Schwinger}{Schwinger}{1949}}
Schwinger, J. (1949).
\newblock Quantum Electrodynamics. II. Vacuum polarization and
self--energy. 
\newblock {\em Phys.~Rev.\/}~{\em 75}, 651.

\bibitem{}{schwinger:51}{\protect\citeauthoryear{Schwinger}{Schwinger}{1951}}
Schwinger, J. (1951).
\newblock The theory of quantized fields. I.
\newblock {\em Phys. Rev.\/}~{\em 82}, 914.

\bibitem{}{schwinger:53a}{\protect\citeauthoryear{Schwinger}{Schwinger}{1953a}}
Schwinger, J. (1953a).
\newblock The theory of quantized fields. II.
\newblock {\em Phys. Rev.\/}~{\em 91}, 713.

\bibitem{}{schwinger:53b}{\protect\citeauthoryear{Schwinger}{Schwinger}{1953b}}
Schwinger, J. (1953b).
\newblock A note on the quantum dynamical principle.
\newblock {\em Phil.~Mag.\/}~{\em 44}, 1171.


\bibitem{}{shifman:79}{\protect\citeauthoryear{Shifman
et~al.}{Shifman, Vainshtein, and Zakharov}{1979}}
Shifman, M.~A., A.~I. Vainshtein, and V.~I. Zakharov (1979).
\newblock QCD and resonance physics. Sum rules.
\newblock {\em Nucl. Phys.\/}~{\em B147}, 385.

\bibitem{}{shigetani:93}{\protect\citeauthoryear{Shigetani
et~al.}{Shigetani, Suzuki, and Toki}{1993}}
Shigetani, T., K.~Suzuki, and H.~Toki (1993).
\newblock Pion structure function in the Nambu and Jona-Lasinio model.
\newblock {\em Phys.~Lett.\/}~{\em B308}, 383.

\bibitem{}{sokolov:79}{\protect\citeauthoryear{Sokolov and Shatnii}{Sokolov and
  Shatnii}{1979}}
Sokolov, S. and A.~Shatnii (1979).
\newblock Physical equivalence of the three forms of relativistic dynamics and
  addition of interactions in the front and instant form.
\newblock {\em Theor.~Math.~Phys.\/}~{\em 37}, 1029.

\bibitem{}{srivastava:99}{\protect\citeauthoryear{Srivastava and
  Brodsky}{Srivastava and Brodsky}{2000}}
Srivastava, P.~P. and S.~J. Brodsky (2000).
\newblock Light--front--quantized QCD in covariant gauge.
\newblock {\em Phys. Rev.\/}~{\em D61}, 025013.

\bibitem{}{stern:99}{\protect\citeauthoryear{Stern}{Stern}{1999}}
C.~Stern (1999)
\newblock Chirale Symmetriebrechung und gebundene Zust\"ande in
Lichtkegelquantenfeldtheorien. 
\newblock PhD thesis, Regensburg, (unpublished).

\bibitem{}{streater:63}{\protect\citeauthoryear{Streater and Wightman}{Streater
  and Wightman}{1963}}
Streater, R. and A.~Wightman (1963).
\newblock {\em PCT, Spin and Statistics, and all that}.
\newblock Benjamin, New York.

\bibitem{}{sundermeyer:82}{\protect\citeauthoryear{Sundermeyer}{Sundermeyer}{1%
982}}
Sundermeyer, K. (1982).
\newblock {\em Constrained Dynamics}.
\newblock Springer, Berlin, Heidelberg, New York.

\bibitem{}{susskind:68}{\protect\citeauthoryear{Susskind}{Susskind}{1968}}
Susskind, L. (1968). 
\newblock Model of self--induced strong interactions.
\newblock {\em Phys. Rev.\/}~{\em 165}, 1535.

\bibitem{}{susskind:97}{\protect\citeauthoryear{Susskind}{Susskind}{1997}}
Susskind, L. (1997). 
\newblock Another conjecture about M(atrix) theory.
\newblock hep-th/9704080. 

\bibitem{}{thooft:74}{\protect\citeauthoryear{'t~Hooft}{'t~Hooft}{1974}}
't~Hooft, G. (1974).
\newblock A two--dimensional model for mesons.
\newblock {\em Nucl. Phys.\/}~{\em B75}, 461.

\bibitem{}{thooft:75}{\protect\citeauthoryear{'t~Hooft}{'t~Hooft}{1975}}
't~Hooft, G. (1975).
\newblock Gauge theory for strong interactions.
\newblock Lecture given at the International School of Subnuclear Physics,
  Erice, Italy, 1975; in: {\em New Phenomena in Subnuclear Physics, Part A},
  A.~Zichichi, ed., Plenum, New York.

\bibitem{}{tamm:45}{\protect\citeauthoryear{Tamm}{Tamm}{1945}}
Tamm, I. (1945).
\newblock Relativistic interaction of elementary particles. 
\newblock {\em J.~Phys.~(Moscow)\/}~{\em 9}, 449.

\bibitem{}{terentev:76}{\protect\citeauthoryear{Terent'ev}{Terent'ev}{1976}}
Terent'ev, M. (1976).
\newblock On the structure of the wave functions of mesons considered as bound
  states of relativistic quarks.
\newblock {\em Sov.~J.~Nucl.~Phys.\/}~{\em 24}, 106.

\bibitem{}{tomboulis:73}{\protect\citeauthoryear{Tomboulis}{Tomboulis}{1973}}
Tomboulis, E. (1973).
\newblock Quantization of the Yang--Mills field in the null--plane frame.
\newblock {\em Phys. Rev.\/}~{\em D8}, 2736.

\bibitem{}{vogl:91}{\protect\citeauthoryear{Vogl and Weise}{Vogl and
  Weise}{1991}}
Vogl, U. and W.~Weise (1991).
\newblock The Nambu and Jona-Lasinio model: Its implications for hadrons and
  nuclei.
\newblock {\em Prog. Part. Nucl. Phys.\/}~{\em 27}, 195.

\bibitem{}{weise:84}{\protect\citeauthoryear{Weise}{Weise}{1984}}
Weise, W. (1984). 
\newblock Quarks, chiral symmetry and dynamics of nuclear constituents.
\newblock in: {\em International Review of Nuclear Physics, Vol.1, Quarks and
  Nuclei}, W.~Weise, ed., World Scientific, Singapore.

\bibitem{}{wilson:94}{\protect\citeauthoryear{Wilson et~al.}{Wilson,
Walhout, Harindranath, Zhang, Perry, and Glazek}{1994}}
Wilson, K.G., T.S.~Walhout, A.~Harindranath, W.-M.~Zhang, R.J.~Perry,
and S.D.~Glazek (1994).
\newblock Nonperturbative QCD: A weak coupling treatment on the light--front.
\newblock {\em Phys. Rev.\/}~{\em D49}, 6720.

\bibitem{}{witten:78}{\protect\citeauthoryear{Witten}{Witten}{1978}}
Witten, E. (1978).
\newblock Chiral Symmetry, the 1/$N$ expansion and the $SU(N)$
Thirring model.
\newblock {\em Nucl.~Phys.\/}~{\em B145}, 110.

\bibitem{}{wittman:89}{\protect\citeauthoryear{Wittman}{Wittman}{1989}}
Wittman, R.S. (1989)
\newblock Symmetry breaking in the $(\phi^4)_2$ theory and the
light--front vacuum. 
\newblock in: {\em Nuclear and Particle Physics on the Light--Cone},
Proceedings, LAMPF Workshop, Los Alamos, 1988, M.B.~Johnson,
L.S.~Kisslinger, eds., World Scientific, Singapore.

\bibitem{}{wu:77}{\protect\citeauthoryear{Wu}{Wu}{1977}}
Wu, T.T. (1977).
\newblock Two--dimensional Yang--Mills theory in the leading 1/$N$ expansion.
\newblock {\em Phys. Lett.\/}~{\em 71B}, 142.

\bibitem{}{yndurain:96}{\protect\citeauthoryear{Yndurain}{Yndurain}{1996}}
Yndurain, F. (1996).
\newblock {\em Relativistic Quantum Mechanics and Introduction to Field
  Theory}.
\newblock Springer, Berlin, Heidelberg, New York.

\bibitem{}{zhitnitsky:86}{\protect\citeauthoryear{Zhitnitsky}{Zhitnitsky}{1986%
}}
Zhitnitsky, A. (1986). 
\newblock On chiral symmetry breaking in QCD$_2$($N_C \to \infty$). 
\newblock {\em Sov. J. Nucl. Phys.\/}~{\em 43}, 999.

\bibitem{}{zinn-justin:96}{\protect\citeauthoryear{Zinn-Justin}{Zinn-Justin}{1%
996}}
Zinn-Justin, J. (1996).
\newblock {\em Quantum Field Theory and Critical Phenomena}.
\newblock Clarendon Press, Oxford.

\end{thebibliography}

\end{document}